\documentclass[10pt, a4paper]{article}

\usepackage{fullpage}
\usepackage{amssymb}
\usepackage{amsmath}
\usepackage{amsfonts}

\usepackage{amsmath}
\usepackage{tabulary,capt-of}
\usepackage{subcaption}
\usepackage{multirow}
\usepackage{tikz}
\usepackage{diagbox}
\usepackage{xcolor}
\usepackage{hyperref}

\newcommand{\avg}[1]{{\left\{\!\left[ #1 \right]\!\right\}}}

\usepackage{csquotes}

\begin{document}

\title{Benchmark for numerical solutions of flow in heterogeneous groundwater formations}

\author{Cristian D. Alecsa$^{1}$, Imre Boros$^{1, 2}$, Florian Frank$^{3}$, Mihai Nechita$^{1,4}$, \\
Alexander Prechtel$^{3}$, Andreas Rupp$^{3, 5}$,  Nicolae Suciu$^{1, 3}\footnote{Corresponding author. \textit{Email adress: suciu@math.fau.de}}$ \\ \\
\smallskip %\vspace{0.3cm}
\textit{$^{1}$Tiberiu Popoviciu Institute of Numerical Analysis, Romanian Academy,}\\
\textit{Fantanele 57, 400320 Cluj-Napoca, Romania}\\
\textit{$^{2}$ Department of Mathematics, Babe\c s-Bolyai University,} \\
\textit{Mihail Kog\u alniceanu, 1, 400084 Cluj-Napoca, Romania}\\
\textit{$^{3}$ Mathematics Department, Friedrich-Alexander University of Erlangen-Nuremberg,}\\
\textit{Cauerstra{\ss}e. 11, 91058 Erlangen, Germany}\\
\textit{$^{4}$ Department of Mathematics, University College London,}\\
\textit{Gower Street, London, WC1E 6BT, United Kingdom}\\
\textit{$^{5}$ Interdisciplinary Center for Scientific Computing, Ruprecht-Karls-University,}\\
\textit{Im Neuenheimer Feld 205, 69120  Heidelberg, Germany}}

%\author[a,b]{Cristian D. Alecsa}\author[a,b]{Imre Boros}\author[c]{Florian Frank}\author[c]{Peter Knabner}\author[e]{...}
%\author[c]{Alexander Prechtel}\author[c,d]{Andreas Rupp}\author[a,c]{Nicolae Suciu\corref{1}} \ead{suciu@math.fau.de}
%\cortext[1]{Corresponding author.}
%\address[a]{Tiberiu Popoviciu Institute of Numerical Analysis, Romanian Academy,\\ Fantanele 57, 400320 Cluj-Napoca, Romania}
%\address[b]{Department of Mathematics, Babe\c s-Bolyai University, \\ Mihail Kog\u alniceanu, 1, 400084 Cluj-Napoca, Romania}
%\address[c]{Mathematics Department, Friedrich-Alexander University of Erlangen-Nuremberg,\\ Cauerstra{\ss}e. 11, 91058 Erlangen, Germany}
%\address[d]{Interdisciplinary Center for Scientific Computing, Ruprecht-Karls-University,\\Im Neuenheimer Feld 205, 69120  Heidelberg, Germany}
%\address[e]{...}

%\author{Ihr Name  \\
%	Ihr Unternehmen / Universit\"at  \\
%	Teststra\ss e -99 \\
%	0123456 Testhausen \\
%	\and
%	Der Andere  \\
%	Sein Unternehmen / Universit\"at \\
%	Musterstra\ss e 00 \\
%	6543210 Musterdorf \\
%	}

\date{}
\maketitle

\begin{abstract}
This article presents numerical investigations on accuracy and
convergence properties of several numerical approaches for simulating steady state flows in heterogeneous aquifers. Finite difference, finite element, discontinuous Galerkin, spectral, and random walk methods are tested on one- and two-dimensional benchmark flow problems. Realizations of log-normal hydraulic conductivity fields are generated by Kraichnan algorithms in closed form as finite sums of random periodic modes, which allow direct code verification by comparisons with manufactured reference solutions. The quality of the methods is assessed for increasing number of random modes and for increasing variance of the log-hydraulic conductivity fields with Gaussian and exponential correlation. Experimental orders of convergence are calculated from successive refinements of the grid. The numerical methods are further validated by comparisons between statistical inferences obtained from Monte Carlo ensembles of numerical solutions and theoretical first-order perturbation results. It is found that while for Gaussian correlation of the log-conductivity field all the methods perform well, in the exponential case their accuracy deteriorates and, for large variance and number of modes, the benchmark problems are practically not tractable with reasonably large computing resources, for all the methods considered in this study.
\end{abstract}



%\begin{keyword}
%Darcy flow \sep accuracy \sep convergence \sep computational tractability
%\sep finite difference \sep finite element \sep discontinuous Galerkin
%\sep spectral methods \sep global random walk
%
%\MSC[2010] 65N06 \sep 65N30 \sep 65N35 \sep 65N12 \sep 76S05 \sep 86A05%
%
%%65N06   Finite difference methods%
%%65N30   Finite elements, Rayleigh-Ritz and Galerkin methods, finite methods%
%%65N35   Spectral, collocation and related methods%
%%65N12   Stability and convergence of numerical methods%
%%76S05   Flows in porous media; filtration; seepage
%%86A05   Hydrology, hydrography, oceanography%
%
%\end{keyword}

\section{Introduction}
\label{intro}

Solving the flow problem is the first step in modeling contaminant transport in natural porous media formations. Since typical parameters for aquifers often lead to advection-dominated transport problems \cite{Dagan1989}, accurate flow solutions are essential for reliable simulations of the effective dispersion of the solute plumes. The numerical feasibility of the flow problem in realistic conditions has been addressed in a pioneering work by \cite{Ababouetal1989}, who pointed out that the discretization
steps must be much smaller than the heterogeneity scale, which in turn is
much smaller than the dimension of the computational domain. Such order relations also served as guidelines for implementing solutions with finite
difference methods (FDM) \cite{deDreuzyetal2007,KurbanmuradovandSabelfeld2010,Trefryetal2003} or finite element methods (FEM) \cite{Bellinetal1992,Raduetal2011,SalandinandFiorotto1998} used in Monte Carlo simulations. In spectral approaches \cite{Gotovacetal2007,LiandMcLaughlin1991} or probabilistic collocation and chaos expansion approaches \cite{LiandZahng2007,Lietal2009} the spatial resolution is also related to the number of terms in the series representation of the coefficients and of the solution of the flow problem. Despite the advances in numerical methods, computing accurate flow solutions in case of highly heterogeneous formations faces computational challenges in terms of code efficiency and computational resources (see e.g., \cite{Gotovacetal2009,LiandZahng2007}).

Field experiments show a broad range of heterogeneities of the aquifer systems characterized by variances $\sigma^2$ of the log-hydraulic conductivity from less than one, for instance, at Borden site, Ontario \cite{RajaramandGelhar1991,RitziandSoltanian2015}, up to variances between $\sigma^2=3.4$ and $\sigma^2=8.7$ at the Macrodispersion Experiment (MADE) site in Columbus, Mississippi \cite{Bohlingetal2016,Rehfeldtetal1992}. The geostatistical interpretation of the measured conductivity data is often based on a log-normal probability distribution with an assumed exponential correlation model \cite{Bohlingetal2016,Gelhar1986,Rehfeldtetal1992}. But, since dispersion coefficients do not depend on the shape of the correlation function \cite{Dagan1989} and are essentially determined by the correlation scale \cite{RajaramandGelhar1991}, a Gaussian model can be chosen as well \cite{KurbanmuradovandSabelfeld2010,Trefryetal2003}. Moreover, experimental data from Borden, MADE, and other sites can be reinterpreted by using a multifractal analysis and power-law correlations \cite{LiuandMolz1997,MolzandBoman1995,Molzetal1997}. However, random fields with power-law correlations can be generated as linear combinations of random modes with either exponential or Gaussian correlation \cite{DiFedericoandNeuman1997,Suciuetal2015}. Therefore, the influence of the correlation structure on the accuracy of the numerical solutions of the flow problem can be generally analyzed by considering Gaussian and exponential correlations of the log-conductivity field.

Numerical simulations of flow and transport in groundwater were carried out for increasing log-conductivity variance from small to moderate values, e.g., up to $\sigma^2=1.6$ in \cite{Bellinetal1992}, to values as high as $\sigma^2=9$ in \cite{deDreuzyetal2007} or $\sigma^2=16$ in \cite{KurbanmuradovandSabelfeld2010}. Though exponential shape of the log-conductivity correlation is considered in most cases, a Gaussian shape which yields smoother fields could be preferred for numerical reasons \cite{Trefryetal2003}. Different approaches were validated through comparisons with linear theoretical approximations of first-order in $\sigma^2$ and the numerical methods were further used to explore the limits of the first-order theory. A direct validation of the flow solver by comparisons with analytical manufactured solutions has been done by \cite{KurbanmuradovandSabelfeld2010}, for both Gaussian and exponential correlations with variance $\sigma^2=1$. Serious challenges are posed by exponential correlation models with high variances, when conductivity differences may span several orders of magnitude. Using a spectral collocation approach based on expansions of the solution in series of compactly supported and infinitely differentiable ``atomic'' functions, \cite{Gotovacetal2009} tested the flow solver for variances up to $\sigma^2=8$. The accuracy of the solution was assessed with mass balance errors between control planes, the criterion of small grid P\'{e}clet number (defined with advection velocities given by the derivatives of the log-hydraulic conductivity and constant unit diffusion coefficient highlighted by expanding the second derivatives in the flow equation), as well as by corrections of the solution induced by the increase of the number of collocation points. It was found that increasing the resolution of the log-hydraulic conductivity requires increasing the resolution of the solution to fulfill the accuracy criteria. While in case of Gaussian correlation of log-hydraulic conductivity the desired accuracy is achieved, for $\sigma^2$ up to 8, with relatively small number of collocation points per integral scale, in case of exponential correlation, which requires increasing resolution with increasing $\sigma^2$, the required resolution of the solution may lead to problem dimensions which exceed the available computational resources \cite[Section 5.2]{Gotovacetal2009}.

In line with the investigations discussed above, we present in this article a comparative study of several numerical schemes for the Darcy flow equation, based on different concepts. The schemes will be tested in terms of accuracy and convergence behavior on benchmark problems designed for a broad range of parameters describing the variance and the spectral complexity of the log-hydraulic conductivity fields with both Gaussian and exponential correlation structures used as models for heterogeneous aquifers.

Comparisons of numerical schemes for flow problems have been made for constant
hydraulic parameters, the goal being to solve nonlinear reactive transport problems \cite{Carrayrouetal2010a,Carrayrouetal2010b}, for constant parameters on sub-domains and complex geometry, in order to simulate flow in fractured
porous media \cite{deDreuzyetal2013,Flemischetal2018}, by considering functional dependence of the hydraulic conductivity on temperature, in case of coupled flow and heat transport \cite{Grenieretal2018}, or in case of heterogeneous hydraulic conductivity fields with low variability
\cite{Raduetal2011}. Such comparative studies use benchmark problems and code intercomparison for the validation of the numerical methods (e.g., \cite{Carrayrouetal2010b}) or, whenever available, comparisons with analytical solutions for the purpose of code verification (e.g., \cite{Grenieretal2018}).

In the present study, we adopt a different strategy. The hydraulic
conductivity is generated as a realization of a Kraichnan random
field \cite{Kraichnan1970} in a closed form, consisting of finite
sums of cosine random modes, with a robust randomization method
\cite{Krameretal2007,KurbanmuradovandSabelfeld2010}. Unlike in turning
bands \cite{Mantoglouetal1982} or HYDRO\_GEN \cite{BellinandRubin1996} methods, the Kraichnan approach allows a direct control of the accuracy of the field, by increasing the number of modes, and of its variability, through the variance of the log-hydraulic conductivity random field. Moreover, the explicit analytical form of the generated field allows the computation of the
source term occurring when a manufactured analytical solution is forced to verify the flow equation \cite{Brunneretal2012,KurbanmuradovandSabelfeld2010,Roache2002,Roy2005}. A realization of the hydraulic conductivity and the source term are computed by summing up cosine modes with parameters given by a fixed realization of the wave numbers and phases produced by the randomization method. These parameters are computed with C codes and stored as data files. Codes written in different programming languages, for instance Matlab, C++, or Python upload the same data files to construct precisely the same manufactured solution to be used in code verifications. One obtains in this way a benchmark to evaluate the performance of different numerical methods for a wide range of variances of the log-hydraulic conductivity and of the number of random modes. While the variance characterizes the spatial heterogeneity of the aquifer system, the number of modes is related to the spatial scale of the problem. The latter is demonstrated by numerical investigations which show that, in order to ensure the self-averaging behavior of the simulated transport process, the number of modes in the Kraihchnan routine has to be of the order of the number of correlation lengths traveled by the solute plume \cite[Fig. 4]{Eberhardetal2007}.

We compare the accuracy of the FDM, the FEM, the discontinuous Galerkin method (DGM), and the Chebyshev collocation spectral method (CSM) by solving two dimensional (2D) problems. A Galerkin spectral method (GSM) and a newly developed ``global random walk'' method (GRW) are tested only in the one-dimensional (1D) case. Code verification is done by evaluating the errors with respect to manufactured analytical hydraulic head solutions for combinations of parameters of the Kraichnan routine, i.e. number of modes between $10^2$ and $10^4$ and variance of the log-hydraulic conductivity between 0.1 and 10. Estimations of the
convergence order are obtained numerically by successive refinements
of the discretization. The 2D codes verified in this way are further
used to solve the homogeneous equation for the hydraulic head and to
compute the Darcy velocity. Ensembles of 100 realizations of the flow solution are computed and used in Monte Carlo inferences of the statistics of the hydraulic head and of the components of the Darcy velocity. The numerical methods are validated by comparisons with first-order theoretical results \cite{Bakretal1978,Dagan1989,Mizelletal1982}.

In order to test the numerical methods in conditions of hydraulic conductivity with increasing variability, comparisons are done for log-hydraulic conductivities with exponential and Gaussian correlations. While a Gaussian correlation ensures the smoothness of the generated hydraulic conductivity samples, in case of exponential correlation the smoothness significantly deteriorates with the increase of the number of modes, the generated samples approaching non-differentiable functions in the limit of an infinite number of modes \cite{Suciu2010,Suciu2019}.

The remainder of this paper is organized as follows. Section, \ref{problem}
presents the benchmark problems, the Kraichnan randomization method, and some considerations of the sample-smoothness of the generated hydraulic conductivity fields with Gaussian and exponential correlation. Sections \ref{fdm}, \ref{fem}, \ref{dgm}, \ref{spectral}, and \ref{grw}
contain results obtained with the numerical approaches compared in this
study. Section \ref{stat} presents Monte Carlo inferences and comparisons with first-order theory. Finally, some conclusions of the present study are drawn in Section \ref{concl}. Appendix~\ref{appendixA} presents the computation of the manufactured solutions. Some details and analysis of the numerical results are presented in Appendix~\ref{appendixB} and Appendix~\ref{appendixB}.

\section{Benchmark flow problems}
\label{problem}

We consider the two-dimensional steady-state velocity field in
saturated porous media with constant porosity determined by the
gradient of the hydraulic head $h(x,y)$ according to Darcy's law \cite{Dagan1989}
\begin{equation}\label{eq2.1}
V_{x}=-K\frac{\partial h}{\partial x}, \quad V_{y}=-K\frac{\partial
h}{\partial y},
\end{equation}
where $K(x,y)$ is an isotropic hydraulic conductivity consisting of a
fixed realization of a spatial random function. The velocity solves a conservation equation $\partial V_x / \partial_x+\partial V_y / \partial_y = 0$, which after the substitution of the velocity components (\ref{eq2.1}) gives the following equation for the hydraulic head $h$,
\begin{equation}\label{eq2.2}
-\left[\frac{\partial}{\partial x}\left(  K\frac{\partial h}{\partial
x}\right) +\frac{\partial}{\partial y}\left(K\frac{\partial
h}{\partial y}\right)\right] = 0.
\end{equation}
We are looking for numerical solutions of the equations
(\ref{eq2.1})--(\ref{eq2.2}) in a rectangular domain $\Omega=[0,L_{x}]\times[0,L_{y}]$, subject to
boundary conditions given by
\begin{align}
h(0,y)  & =H, \;\;h(L_{x},y)=0, \;\; \forall y\in[0,L_y]\label{eq2.3}\\
\frac{\partial h}{\partial y}(x,0)  & =\frac{\partial
h}{\partial y}(x,L_{y})=0, \;\; \forall x\in[0,L_x]\label{eq2.4}
\end{align}
where $H$ is a constant head.

The hydraulic conductivity fields $K$ considered in this study are isotropic, log-normally distributed, with exponential and Gaussian correlations. The random log-hydraulic conductivity field $Y=\ln(K)$ is specified by the ensemble mean $\langle Y\rangle$, the variance $\sigma^2$, and the correlation length
$\lambda$. The Kraichnan algorithm
\cite{Kraichnan1970} used to generate samples of fluctuating part
of the log-hydraulic conductivity field $Y'=Y-\langle Y\rangle$ is a
randomized spectral representation of statistically homogeneous
Gaussian fields \cite{Krameretal2007}. A general randomization
formula, valid for both exponential and Gaussian correlation types,
is given by a finite sum of $N$ cosine functions,
\begin{equation}\label{eq2.5}
Y'(x,y) =\sigma\sqrt{\frac{2}{N}} \sum_{i=1}^N \cos\left(\phi_i+
2\pi(k_{i,1}x+k_{i,2}y)\right).
\end{equation}
The probability density of the random vector $(k_{i,1},k_{i,2})$ is
given by the Fourier transform of the correlation function of the
statistically homogeneous random field $Y'$ divided by the variance
$\sigma^2$ and the phases $\phi_i$ are random numbers uniformly distributed in $[0,2\pi]$ \cite[Eq. (22)]{KurbanmuradovandSabelfeld2010}.

The smoothness of the $K$ field is essentially determined by the
correlation of the random field $Y'$. Sufficient conditions for
sample smoothness are given by the existence of higher order
derivatives of the correlation function $C(\mathbf{r})=\langle
Y'(\mathbf{x}) Y'(\mathbf{x}+\mathbf{r})\rangle$. For instance, in
the 1D case, if $C(r)$ has a second order derivative at
$r=0$, then the 1D field $Y'(x)$ is equivalent to a field
with samples that are continuous with probability one in every
finite interval. Moreover, if the fourth derivative $C^{(4)}(0)$ exists, $Y' (x)$ is equivalent to a field the sample functions of which have, with
probability one, continuous derivatives in every finite interval.
These sufficient conditions can hardly be relaxed and one expects
that they are very close to the necessary conditions
\cite{CramerandLeadbetter1967,Yaglom1987}.

\begin{figure}[h]
\centering \begin{minipage}[t]{0.45\linewidth} \centering
\vspace*{0in}
\includegraphics[width=\linewidth]{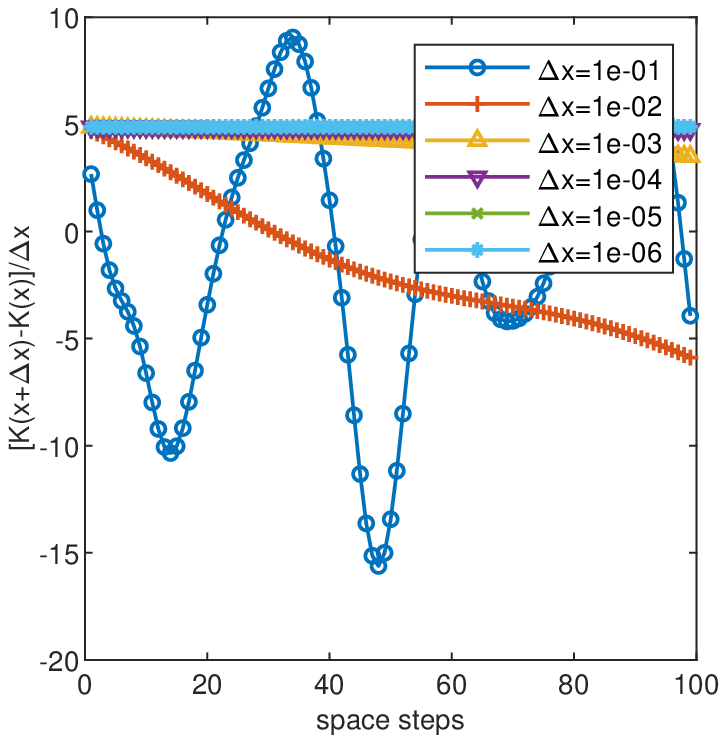}
\caption{\label{fig:derivativeGss}Estimations of the derivative for a
realization of the log-normal field $K$ with Gaussian correlation
for decreasing space step and fixed number of modes $N=10^6$.}
\end{minipage}
\hspace{0.2cm} \centering \begin{minipage}[t]{0.45\linewidth}
\centering \vspace*{0in}
\includegraphics[width=\linewidth]{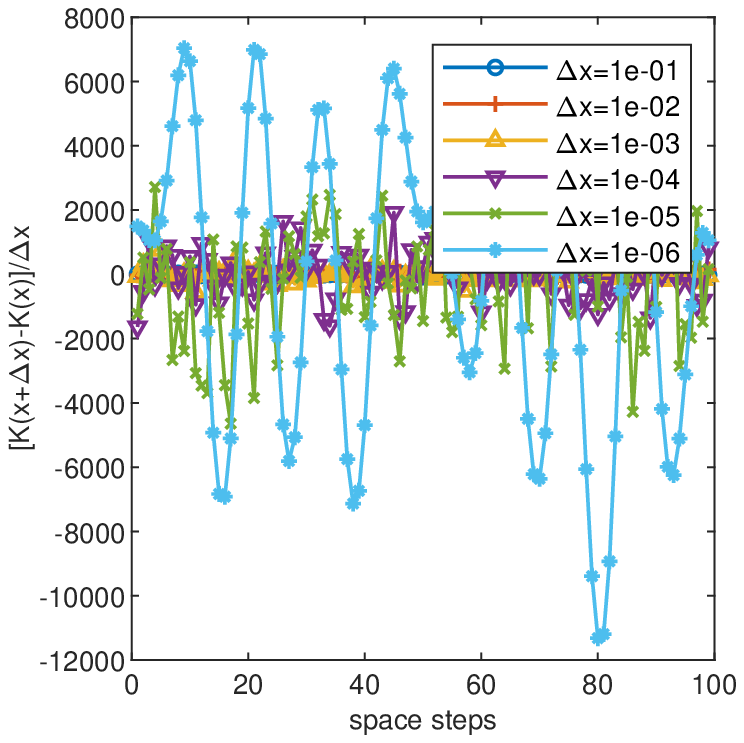}
\caption{\label{fig:derivativeExp}Estimations of the derivative for a
realization of the log-normal field $K$ with exponential correlation
for decreasing space step and fixed number of modes $N=10^6$.}
\end{minipage}
\end{figure}

The smoothness conditions of \cite{CramerandLeadbetter1967} are
fulfilled for instance if the correlation of $Y'$ has a Gaussian
shape $C(r)\sim e^{-r^{2}}$ and is infinitely differentiable, which results
in a random field (\ref{eq2.5}) with continuous sample derivatives.
A counterexample is the field $Y'$ with an exponential correlation $C(r)\sim e^{-r}$, which is not differentiable at the origin. Since, rigorously
speaking, the differentiability of $C(r)$ at the origin is only a
sufficient condition, we proceed with numerical investigations on
the sample-smoothness of the $K$ field in case of Gaussian and
exponential correlations of the log-hydraulic conductivity $Y$, for fixed variance $\sigma^2=0.1$. The estimations of the spatial derivative $d K/ d x$ of a given realization of the numerically generated 1D $K$ field, computed at $x=1+i\Delta x,\;\; i=1\ldots 100$, are presented in Figs.~\ref{fig:derivativeGss}~and~\ref{fig:derivativeExp}. One can see that, while in case of Gaussian correlation the estimations approach the exact value of $d K/ d x$ at $x=1$, according to (\ref{Flow:Eq3}), for exponential correlation of $Y$, the realization of the $K$ field apparently approaches a non-differentiable function for the large number $N=10^6$ random modes in (\ref{eq2.5}).

\begin{figure}[h]
\centering \begin{minipage}[t]{0.45\linewidth} \centering
\vspace*{0in}
\includegraphics[width=\linewidth]{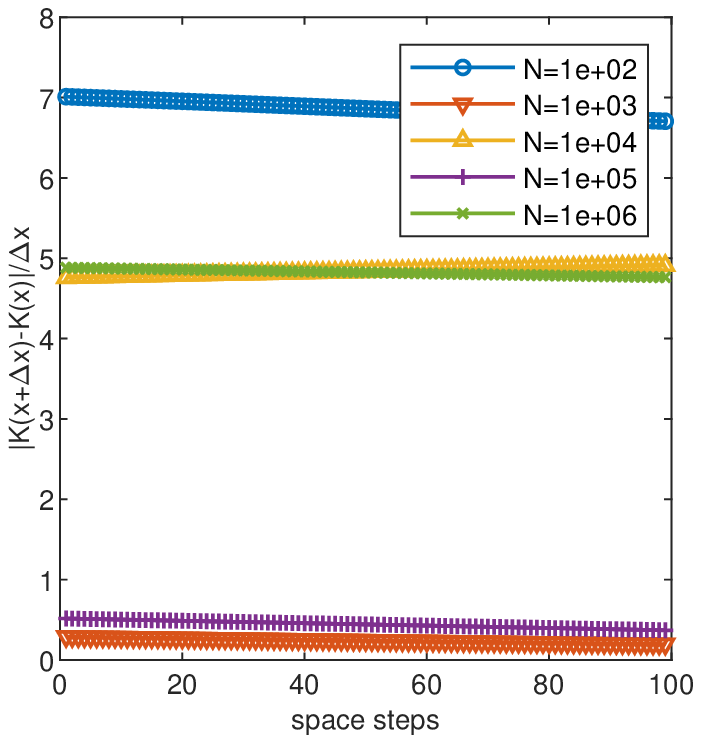}
\caption{\label{fig:LipaschitzGss}Estimations of the Lipschitz constant
for a realization of the log-normal field $K$ with Gaussian
correlation for a fixed space step $\delta x=10^{-4}$ and increasing number of modes $N$.}
\end{minipage}
\hspace{0.2cm} \centering \begin{minipage}[t]{0.45\linewidth}
\centering \vspace*{0in}
\includegraphics[width=\linewidth]{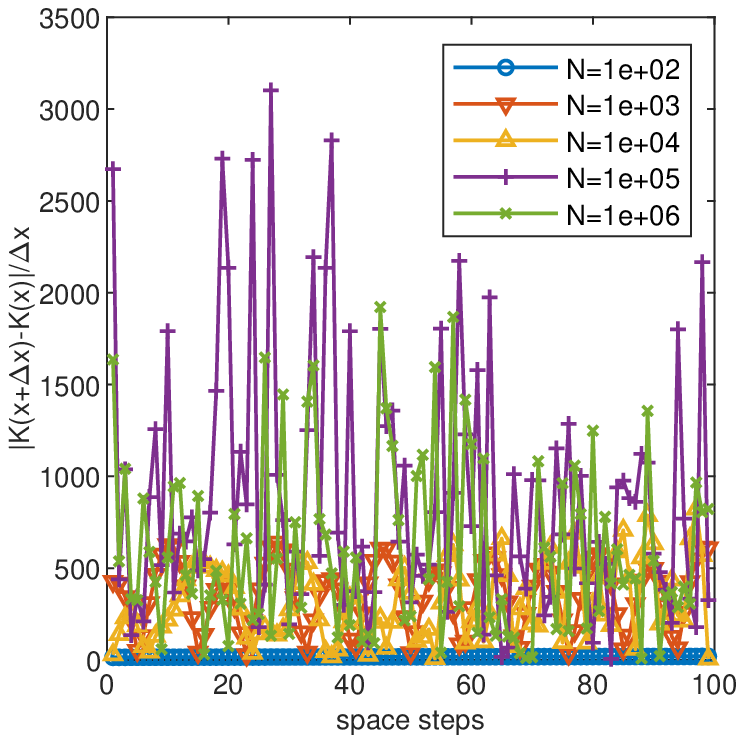}
\caption{\label{fig:LipschitzExp}Estimations of the Lipschitz constant
for a realization of the log-normal field $K$ with exponential
correlation for a fixed space step $\delta x=10^{-4}$ and increasing number of modes $N$.}
\end{minipage}
\end{figure}

Lipschitz continuity, required in many circumstances, can still hold if the sample of the random field is not differentiable. Since Lipschitz continuity of the coefficients is often assumed in formulations of the flow problem \cite{Arbogastetal1996,KnabnerandAngermann2003,NochettoandVerdi1988,Raduetal2004}, we estimate the Lipschitz constant of the same samples of the 1D $K$ fields for a fixed space step and increasing $N$. The results presented in Figs.~\ref{fig:LipaschitzGss}~and~\ref{fig:LipschitzExp} show that while in case of Gaussian correlation, the estimated Lipschitz constant is bounded, in case of exponential correlation it increases with the number of modes $N$, indicating the lack of Lipschitz continuity. It should however be noticed that the samples (\ref{eq2.5}) of the random function $Y'$ are analytical functions that are differentiable and Lipschitz continuous. The apparent lack of smoothness shown by these numerical simulations is an interplay between the number of modes and the space step. Therefore, the results presented
in Figs.~\ref{fig:derivativeGss}~-~\ref{fig:LipschitzExp} rather illustrate
the challenges that numerical methods face if the hydraulic conductivity is modeled as a log-normal random field with exponential correlation.

In order to investigate the robustness and the limits of applicability of the
numerical schemes tested in this study, we consider highly
oscillating coefficients $K$ generated by (\ref{eq2.5}) for isotropic
correlations of the random field $Y'$, $C(r)=\langle Y'(x,y) Y'(x+r_x,y+r_y)\rangle$,
where $r=(r_{x}^{2}+r_{y}^{2})^{1/2}$.
As correlation functions, we consider the two extreme cases presented above, i.e., the Gaussian correlation,
\begin{equation}\label{eq2.6}
C(r)=\sigma^2\exp\left(-\frac{r^2}{\lambda^2}\right),
\end{equation}
which yields analytical samples of $Y'$, and, respectively, the exponential correlation,
\begin{equation}\label{eq2.7}
C(r)=\sigma^2\exp\left(-\frac{|r|}{\lambda}\right),
\end{equation}
which, in the limit of infinite number of modes $N$, produces non-smooth samples.

To compute the hydraulic conductivity $K=\exp(Y)=\exp(\langle
Y\rangle)\exp(Y')$, one needs a relation between the geometric mean $K_g=\exp(\langle Y\rangle)$ and the arithmetic mean $\langle K\rangle$ of the random field $K$ \cite{Dengetal1995,GelharandAxness1983,Zhangetal2007}.
According to the boundary conditions (\ref{eq2.3}-\ref{eq2.4}), the mean velocity $(U,0)$ is aligned to the longitudinal axis $x$. The mean $\langle K\rangle$ is the effective hydraulic conductivity related by Darcy's law
(\ref{eq2.1}) to the mean slope of the hydraulic head,
$J=-H/L_x$, and the mean velocity $U$ by $U=\langle K\rangle J$. We adopt here, as far as we do not compare velocity fields with different variances, a first order approximation in $\sigma^2$ given by $K_g=\langle K\rangle\exp(-\sigma^2/2)$, which has been used
for variances $\sigma^2>1$ in both analytical \cite{Attinger2003} and numerical \cite{Bellinetal1992} computations. Higher order corrections
were also derived as asymptotic approximations for small $\sigma^2$ \cite{Dengetal1995}. But our numerical tests have shown that neither of these corrections is appropriate for statistical inferences, because they lead to underestimations of the mean velocity. Therefore, to compare results obtained with different values of $\sigma^{2}$ presented in Section~\ref{stat}, we use dimensionless velocities, in units of $\langle K\rangle J$.

The algorithms for Gaussian and exponential correlation functions
described in Sections 5.2 and 5.3 of
\cite{KurbanmuradovandSabelfeld2010} were implemented as
C-functions. These C-functions were used to compute the wavenumbers
$k_{i,1}$, $k_{i,2}$ and phases $\phi_i$ entering as numerical
parameters in Eq. (\ref{eq2.5}). The same numerical parameters can be used in either Matlab or C++ codes, to ensure identical values of the $K$ fields, in the limit of double precision, irrespective of programming language.

The mean hydraulic conductivity is fixed to $\langle K\rangle=15$~m/day, a value representative for gravel or coarse sand aquifers \cite{Dagan1989}. The correlation length $\lambda$ in (\ref{eq2.6}) and (\ref{eq2.7}) is set to $\lambda=1$~m and the dimensions of the spatial domain are given in $\lambda$ units. For a given realization of $K$, the set of wavenumbers and phases is computed for the maximum number of modes considered in this study, $N=10^4$. Samples of the $K$ field are further constructed for values of $\sigma^2$ between 0.1 and 10, and for $N$ between $10^2$ and $10^4$.

The longitudinal dimension of the domain, $L_x$, has to be tens of times larger than $\lambda$ (for instance, \cite{Ababouetal1989} recommend $L_x/\lambda =25$). Constrained by considerations of computational tractability, we fix the dimensions of the domain to $L_x=20$ and $L_y=10$. The constant hydraulic head at the inflow boundary is fixed to $H=1$. To enable comparisons of the 2D solutions obtained with the FDM, FEM, DGM approaches tested in the present benchmark, the number of discretization points is set to $n_x=10^3+1$, in the longitudinal direction, and to $n_y=5\cdot 10^2+1$ in the transverse direction. Consequently, the space steps are fixed at $\Delta x=\Delta y=2\cdot 10^{-2}$, which, for unit correlation length considered in this study, gives $\lambda/\Delta x=50$, i.e., one of the highest resolutions of the log-conductivity field used in the past in similar finite difference/element approaches (see e.g., \cite[Table 1]{deDreuzyetal2007}).

Before proceeding to solve the 2D problem, we test the codes for the 1D problem
\begin{equation}\label{eq2.8}
	\begin{cases}
    	-(K h')' = 0, \quad \forall x \in(0,L) \\[2mm]
    	h(0) = H, \quad h(L) = 0,
    \end{cases}
\end{equation}
where the length of the domain is fixed to $L=200$. The FDM, FEM, and GRW schemes are tested for fixed number of grid points $n=2\cdot 10^5+1$ (of the same order of magnitude as in the 2D case) and $\Delta x=10^{-3}$. GRW solutions of the 1D problem (\ref{eq2.8}) were also computed for $\Delta x=10^{-1}$ and $\Delta x=4\cdot 10^{-4}$.

In case of spectral approaches, the number of collocation points, for both CSM and GSM schemes, is chosen after preliminary tests as large as necessary, possible optimal, to ensure the desired accuracy (see Section~\ref{spectral}).

The codes based on the numerical schemes considered in this study are
first verified by comparisons with manufactured solutions. The
approach consists of choosing analytical functions of the dependent
variables, $h(x,y)$ and $h(x)$, on which one applies the differential
operator from the left hand side of Eqs.~(\ref{eq2.2})~and~(\ref{eq2.8}), the result being a source term $f$. Hence, $h(x,y)$ and $h(x)$ are exact solutions of Eqs.~(\ref{eq2.2})~and~(\ref{eq2.8}), respectively, with a source term $f$ added in the right hand side and modified boundary functions according to the chosen manufactured solution \cite{Brunneretal2012,Oberkampfetal1998,Raduetal2011,Roache1998,Roache2002,Roy2005}.

An example of choosing the appropriate manufactured solution for the flow solver within a numerical setting similar to that of the present study can be found in \cite[Appendix B]{KurbanmuradovandSabelfeld2010}. In order to test the numerical approximations of the differential operators in the flow equation (\ref{eq2.2}), the manufactured solutions used in this study are simple, but not trivial analytical functions with nonzero derivatives up to the highest order of differentiation. The construction of the manufactured solutions and of the source terms $f$ for the 1D and 2D problems solved in the present benchmark is presented in Appendix~\ref{appendixA}. For the purpose of code verification we chose 21 parameter pairs formed with $\sigma^2=0.1,\; 1,\; 2,\; 4,\; 6,\; 8,\; 10$, and $N=10^2,\; 10^3,\; 10^4$.

The code verification is completed by convergence tests aiming at comparing the computational order of convergence with the theoretical one. In case of FDM, FEM, DGM, and GRW approaches, computational orders of convergence are estimated by computing error norms of the solutions obtained by successively halving the spatial discretization step with respect to a reference solution obtained with the finest discretization (see e.g. \cite{Roy2003,Lordetal2014}). The approach is different in case of spectral methods, where the convergence is indicated by the decrease towards the round-off plateau of the coefficients of the spectral representation \cite{AurentzandTrefethen2017}. Convergence tests are performed for $\sigma^2=0.1,\; 4,\; 10$, and $N=10^2,\; 10^3,\; 10^4$.

Validations tests are finally conducted by comparing mean values and variances of the velocity components inferred from ensembles of 100 realizations of the solution of the boundary value problem (\ref{eq2.1}-\ref{eq2.4}), with $f=0$, for $\sigma^2=0.1,\; 0.5,\; 1,\; 1.5,\; 2$ and fixed $N=10^2$, with theoretical and numerical results given in the literature \cite{Ababouetal1989,Bellinetal1992,Dagan1989,KurbanmuradovandSabelfeld2010,SalandinandFiorotto1998,Trefryetal2003}.

The formulation of the benchmark problems, data files, numerical codes and functions are given in a Git repository at
\href{https://github.com/PMFlow/FlowBenchmark}{https://github.com/PMFlow/FlowBenchmark}, which can help interested readers to test their numerical methods for flows in highly heterogeneous aquifers.

\section{Finite difference method}
\label{fdm}

The first discretization approach tested in this benchmark study is the FDM, known to be one of the most feasible and easily implementable numerical methods for partial differential equations \cite{Larsson2009}. In this section we present the discretization error of the hydraulic head for various parameters $N$ and $\sigma^2$ estimated by comparisons of the 1D and 2D FDM solutions with manufactured analytical solutions. Validation tests through Monte Carlo inferences and comparisons with results given by theoretical perturbation approaches are deferred to Section~\ref{stat}.

\subsection{One-dimensional case}

The non-homogeneous 1D flow problem (\ref{eq2.8}) with right hand side $-f$, with the coefficient $K$ and the source term $f$ given by (\ref{Flow:Eq3}) and (\ref{Flow:Eq4}), respectively, is solved with the following FDM scheme \cite{Leveque2007,Lietal2017}
\begin{equation}\label{FDM_EQ1}
\begin{split}
K \left( x_i-\dfrac{\Delta x}{2} \right)  h_{i-1} - \left[ K \left( x_i + \dfrac{\Delta x}{2} \right) + K \left( x_i - \dfrac{\Delta x}{2} \right) \right]  h_i + K \left(x_i + \dfrac{\Delta x}{2} \right)  h_{i+1} = \Delta x^2 f_i ,
\end{split}
\end{equation}
where $f_i$, $i \in \lbrace 1 , \ldots, n \rbrace$, represents the value of the data $f$ at the grid point $x_i$ and $h_i$ is the value of the numerical solution at the same grid point.

The deviation from the analytical manufactured solution $h(x_i)$ given by (\ref{Flow:Eq1}) of the numerical solutions $h_i$ for different parameters $N$ and $\sigma^2$ computed with the FDM scheme (\ref{FDM_EQ1}), implemented in a Matlab code, is measured with the discrete $L^2$ norm
\begin{equation}\label{FDM_EQ3}
L^2_{\Delta x} = \left( \Delta x \sum\limits_{i=1}^{n} ( h_i - h(x_i) )^2 \right)^{1/2}.
\end{equation}
The $L^2_{\Delta x}$ errors of the numerical solutions for Gaussian and exponential correlation of the $\ln(K)$ field are given in Tables~\ref{tab:titleGauss1D_NonHomogenuous} and \ref{tab:titleExp1D_NonHomogenuous}, respectively. In the case of Gaussian correlation, the benchmark problems are accurately solved for all $N$ and $\sigma^2$. In case of exponential correlation, the 1D FDM scheme fails to solve the problems with $N = 10^4$ for the prescribed step size $\Delta x=10^{-3}$. The total computational time, for both Gaussian and exponential cases, increases, in the same way for all $\sigma^2$, from about one second, for $N = 10^2$, to about one minute, for $N = 10^4$. The time to solve the linear system of equations is about 0.005 seconds in all cases. The computers used in this study are comparable in performance. Moreover, the linear systems of equations are solved with the {\it mldivide} Matlab function in FDM and DGM schemes and with the LU~solver in FEM schemes which both call the UMFPACK's Unsymmetric MultiFrontal method. Thus, the corresponding computing time allows a fair comparison of these three related approaches from the point of view of the computational demand.

\begin{table}[!ht]
    \centering
    \captionof{table}{Numerical errors of the 1D FDM solutions for Gaussian correlation of the $\ln(K)$ field.}
    \label{tab:titleGauss1D_NonHomogenuous}
    \begin{tabular}{|| c || c | c | c | c | c | c | c || c ||}
\hline\hline
\diagbox[width=3em]{$N$}{$\sigma^2$} & 0.1 & 1 & 2 & 4 & 6 & 8 & 10  \\ \hline
$10^2$ & 1.64e-6  & 4.00e-6 & 6.64e-6 & 1.24e-5 & 7.42e-5 & 9.70e-5 & 8.49e-4 \\ \hline
$10^3$ & 1.31e-6 & 5.86e-6 & 1.11e-5 & 2.29e-5 & 8.24e-5 & 1.93e-4 & 8.42e-4 \\ \hline
$10^4$ & 1.57e-6 & 3.88e-6 & 8.83e-6 & 4.74e-5 & 2.13e-4 & 1.15e-3 & 4.86e-3 \\ \hline\hline
\end{tabular}
\end{table}

\begin{table}[!ht]
    \centering
        \captionof{table}{Numerical errors of the 1D FDM solutions for exponential correlation of the $\ln(K)$ field.}
    \label{tab:titleExp1D_NonHomogenuous}
   \begin{tabular}{|| c || c | c | c | c | c | c | c || c ||}
\hline\hline
\diagbox[width=3em]{$N$}{$\sigma^2$} & 0.1 & 1 & 2 & 4 & 6 & 8 & 10  \\ \hline
$10^2$ & 4.60e-5 & 4.57e-4 & 9.08e-4 & 1.81e-3 & 2.77e-3 & 3.78e-3 & 4.35e-3 \\ \hline
$10^3$ & 9.00e-6 & 9.42e-5 & 1.87e-4 & 3.69e-4 & 6.46e-4 & 8.43e-4 & 1.53e-3 \\ \hline
$10^4$ & 1.03e+1 & 3.19e+2 & 5.94e+3 & 1.75e+5 & 2.19e+6 & 1.86e+7 & 1.24e+8 \\
\hline\hline
\end{tabular}
\end{table}

In the following, we investigate numerically the convergence of the FDM scheme by solving the homogeneous problem (\ref{eq2.8}) ($f=0$) for successive refinements of the grid (see e.g. \cite{Lordetal2014}). We start with $\Delta x=10^{-1}$ and obtain $5$ refinements of the grid by successively halving the space steps up to $\Delta x=3.125\cdot 10^{-3}$. The corresponding solutions are denoted by $h^{(k)}$, $k=1,\ldots,6$. The estimated order of convergence (EOC) that describes the decrease in logarithmic scale of the error, quantified by the vector norm $\varepsilon_k = \|h^{(k)}-h^{(6)}\|_{l^{2}}$, is computed according to
\begin{equation}\label{FDM_eqEOC}
EOC=\log\left(\frac{\varepsilon_k}{\varepsilon_{k+1}}\right)/\log(2), \quad\quad k=1,\ldots,4.
\end{equation}
The EOC values corresponding to Gaussian and exponential correlations for different parameters $N$ and $\sigma^2$ are presented in Tables~\ref{tab:titleGauss1D_Homogeneous} and \ref{tab:titleExp1D_Homogeneous}. It is observed that the computational order of convergence in case of Gaussian correlation with $\sigma^2\leq 4$ is similar to the predicted theoretical order of convergence $2$. For the largest value of the variance, $\sigma^2 = 10$, one obtains negative EOC values, which correspond to non-monotonous decrease of the error. In the case of the exponential correlation, negative EOC values occur for all values of $N$ and $\sigma^2$ and a trend of the errors is hardly observable.

\begin{table}[!ht]
\centering
\caption{Computational order of convergence of the 1D FDM scheme for Gaussian correlation of the $\ln(K)$ field.}
\label{tab:titleGauss1D_Homogeneous}
\begin{tabular}{ |c|c|c|c|c|c|c|c|c|c|c|c|c| }
\hline\hline
$N$ & $\sigma^2$ & $\varepsilon_1$ & EOC & $\varepsilon_2$ & EOC &$\varepsilon_3$ & EOC & $\varepsilon_4$ & EOC & $\varepsilon_5$ \\
\hline
\multirow{3}{1em}{$10^2$} & $0.1$ & 5.75e-5 &2.01 & 1.43e-5 & 2.02& 3.53e-6 & 2.08 & 8.34e-7 & 2.36& 1.63e-7 \\
& $4$ & 4.21e-4 & 2.01& 1.04e-4 & 2.02& 2.57e-5 & 2.08& 6.09e-6 &2.28 & 1.25e-6\\
& $10$ & 9.42e-4 & 2.01& 2.35e-4 & 1.93& 6.18e-5 &1.36 & 2.40e-5 & -0.73& 3.96e-5\\
\hline\hline
\multirow{3}{1em}{$10^3$} & $0.1$ &7.13e-5 &2.01 & 1.78e-5 &2.02  & 4.39e-6 &2.07 & 1.05e-6 &2.30 & 2.13e-7\\
& $4$ &8.81e-4 & 2.01& 2.19e-4 & 2.02& 5.41e-5 & 2.06& 1.29e-5 &2.29 & 2.64e-6\\
& $10$ &2.22e-3 &2.02 & 5.46e-4 &2.00  & 1.36e-4 &1.60 & 4.50e-5 & 1.20& 1.96e-5 \\
\hline\hline
\multirow{3}{1em}{$10^4$} & $0.1$ &4.17e-5 &2.01 & 1.04e-5 &2.02 & 2.56e-6 & 2.07 & 6.08e-7 & 2.34& 1.20e-7 \\
& $4$  &6.12e-4 &2.01 & 1.52e-4 &1.99 & 3.84e-5 &1.89  & 1.03e-5 & 1.53& 3.57e-6 \\
&  $10$ &1.86e-3 & 1.95& 4.81e-4 &1.57  & 1.63e-4 &0.87 & 8.89e-5 &-1.07 & 1.87e-4  \\
\hline\hline
\end{tabular}
\end{table}

\begin{table}[!ht]
\centering
\caption{Computational order of convergence of the 1D FDM scheme for exponential correlation of the $\ln(K)$ field.}
\label{tab:titleExp1D_Homogeneous}
\begin{tabular}{ |c|c|c|c|c|c|c|c|c|c|c|c|c| }
\hline\hline
$N$ & $\sigma^2$ & $\varepsilon_1$ & EOC & $\varepsilon_2$ & EOC &$\varepsilon_3$ & EOC & $\varepsilon_4$ & EOC & $\varepsilon_5$ \\
\hline
\multirow{3}{1em}{$10^2$} & $0.1$ & 1.39e-3 & 0.8 & 8.07e-4 & -0.09 & 8.57e-4 & 5.12 & 2.46e-5 & 2.83 & 3.46e-6 \\
& $4$ & 2.56e-1 & 1.82 & 7.26e-2 & -0.02 & 7.35e-2 & 4.88 & 2.51e-3 & 4.67 & 9.89e-5\\
& $10$ & 8.73e-1 & 2.33 & 1.73e-1 & -0.66& 2.73e-1 & 5.19 & 7.46e-3 & 4.18 & 4.13e-4\\
\hline\hline
\multirow{3}{1em}{$10^3$} & $0.1$ & 9.73e-3 & 5.23 & 2.59e-4 & -0.26 & 3.05e-4 & 5.12 & 8.80e-6 & 2.64 & 1.41e-6\\
& $4$ &4.27e-1 & 1.31 & 1.72e-1 & 1.18 & 7.56e-2 & 5.44 & 1.73e-3 & 6.06 & 2.59e-5\\
& $10$ &1.30e+0 & 0.36 & 1.01e+0 & 1.22 & 4.34e-1 & 5.6 & 8.92e-3 & 4.1 & 5.19e-4 \\
\hline\hline
\multirow{3}{1em}{$10^4$} & $0.1$ &2.31e-2 & 0.26 & 2.27e-2 & -0.04 & 2.34e-2 & 4.71 & 8.93e-4 & 4.88 & 3.05e-5 \\
& $4$  &3.29e-1 & 1.50 & 1.16e-1 & -0.18 & 1.31e-1 & 1.01 & 6.49e-2 & 4.39 & 3.09e-3 \\
&  $10$ &1.00e+0 & 1.86 & 2.76e-1 & 0.14 & 2.50e-1 & 1.15 & 1.13e-1 & 4.76 & 4.18e-3  \\
\hline\hline
\end{tabular}
\end{table}

\subsection{Two-dimensional case}

To solve the non-homogeneous 2D flow problem (\ref{eq2.2}) with right hand side $-f$, we employ the extended form of the numerical scheme (\ref{FDM_EQ1}) for the two-dimensional case, which leads to a linear system with a band matrix \cite{Lietal2017},

\begin{equation}\label{FDM_EQ6}
\begin{split}
A_{i,j}  h_{i-1,j} + B_{i,j}  h_{i,j-1} + C_{i,j}  h_{i,j} + D_{i,j}  h_{i+1,j} + E_{i,j}  h_{i,j+1} = f_{i,j},
\end{split}
\end{equation}
where $f_{i,j}$, $i \in \lbrace 1, \ldots, n_x \rbrace, \; j \in \lbrace 1, \ldots, n_y \rbrace$, represents the data $f$ at the grid points $(x_i,y_j)$ and $h_{i,j}$ is the value of the numerical solution at the same grid points.
In formula (\ref{FDM_EQ6}), we have used the following notations :
\begin{equation}\label{FDM_EQ7}
\begin{split}
\begin{cases}
& A_{i,j} := \dfrac{1}{(\Delta x)^2}  K \left( x_i-\dfrac{\Delta x}{2} , y_j \right) , \\[4mm]
& B_{i,j} := \dfrac{1}{(\Delta y)^2}  K \left( x_i, y_j-\dfrac{\Delta y}{2} \right) , \\[4mm]
& D_{i,j} := \dfrac{1}{(\Delta x)^2}  K \left( x_i+\dfrac{\Delta x}{2} , y_j \right) , \\[4mm]
& E_{i,j} := \dfrac{1}{(\Delta y)^2}  K \left( x_i,y_j+\dfrac{\Delta y}{2} \right) , \\[4mm]
& C_{i,j} := - \left[ A_{i,j} + B_{i,j} + D_{i,j} + E_{i,j}\right] .
\end{cases}
\end{split}
\end{equation}

The non-homogeneous 2D problem (\ref{eq2.2}), with source term given by (\ref{Flow:Eq8}) and modified Dirichlet and Neumann boundary conditions (\ref{Flow:Eq6}), is solved with a Matlab code for the FDM scheme (\ref{FDM_EQ6}-\ref{FDM_EQ7}) on the 2D domain described in Section~\ref{problem}, with uniform step sizes $\Delta x=\Delta y=2\cdot 10^{-2}$. Errors of the numerical solutions $h_{i,j}$ with respect to the manufactured solution $h(x_i,y_j)$ given by (\ref{Flow:Eq5}) are quantified by the norm
\begin{equation}\label{FDM_EQ9}
L^2_{\Delta x, \Delta y} = \left( \Delta x  \Delta y  \sum\limits_{i=1}^{N_x} \sum\limits_{i=1}^{N_y} (h_{i,j} - h(x_i,y_j))^2 \right)^{1/2}.
\end{equation}

\begin{table}[!ht]
    \centering
    \captionof{table}{Numerical errors of the 2D FDM solutions for Gaussian correlation of the $\ln(K)$ field.}
    \begin{tabular}{|| c || c | c | c | c | c | c | c || c ||}
\hline\hline
\diagbox[width=3em]{$N$}{$\sigma^2$} & 0.1 & 1 & 2 & 4 & 6 & 8 & 10  \\ \hline
$10^2$ & 1.03e-3 & 2.00e-3 & 7.95e-3 & 4.34e-2 & 1.45e-1 & 3.88e-1 & 9.12e-1 \\ \hline
$10^3$ & 1.09e-3 & 8.91e-3 & 4.23e-2 & 3.65e-1 & 1.88e+0 & 7.43e+0 & 2.47e+1 \\ \hline
$10^4$ & 1.03e-3 & 1.16e-3 & 1.81e-3 & 4.52e-3 & 1.00e-2 & 2.04e-2 & 3.88e-2 \\ \hline\hline
\end{tabular}
    \label{tab:titleGauss2D_NonHomogeneous}
\end{table}

\begin{table}[!ht]
    \centering
    \captionof{table}{Numerical errors of the 2D FDM for exponential correlation of the $\ln(K)$ field.}
    \begin{tabular}{|| c || c | c | c | c | c | c | c || c ||}
\hline\hline
\diagbox[width=3em]{$N$}{$\sigma^2$} & 0.1 & 1 & 2 & 4 & 6 & 8 & 10  \\ \hline
$10^2$ & 1.17e-1 & 2.60e+0 & 3.61e+0 & 9.41e+1 & 7.15e+2 & 3.02e+3 & 9.46e+3 \\ \hline
$10^3$ & 3.58e-2 & 3.09e+0 & 1.42e+1 & 9.79e+1 & 1.45e+3 & 9.54e+3 & 3.91e+4 \\ \hline
$10^4$ & 1.57e+0 & 5.65e+1 & 5.74e+2 & 6.74e+3 & 3.41e+4 & 1.22e+5 & 3.54e+5 \\ \hline\hline
\end{tabular}
    \label{tab:titleExp2D_NonHomogeneous}
\end{table}

The errors of the 2D FDM scheme in case of Gaussian correlation shown in Table~\ref{tab:titleGauss2D_NonHomogeneous} are acceptable even for large values $N$ and $\sigma^2$, excepting the range ($N=10^3, \sigma^2\geq 6$). Instead, in exponential case (Table~\ref{tab:titleExp2D_NonHomogeneous}), the errors progressively increase with increasing  $N$ and $\sigma^2$, the only acceptable values smaller than one being obtained only for ($N\leq 10^3,\; \sigma^2=0.1$). As concerning the computational time, it remains relatively low, with values increasing with $N$ between about 10 seconds and 7 minutes, for both correlation models and all $\sigma^2$. The linear system of equations alone is solved in about 5 seconds in case of exponential correlation with $N=10^4$ and $\sigma^2=10$ and in about 3 seconds in all the other cases.

As in the 1D case, the EOC for the homogeneous version of the problem (\ref{eq2.2}-\ref{eq2.4}) is computed by successively halving the step size five times, starting with $\Delta x=\Delta y=10^{-1}$ and ending with $\Delta x=\Delta y=3.125\cdot 10^{-3}$. The results are presented in Tables~\ref{tab:titleGauss2D_Homogeneous} and \ref{tab:titleExp2D_Homogeneous}. In the Gaussian case, the EOC values are close to the theoretical convergence order 2, but in the exponential case, one obtains negative EOC values and a decreasing trend of the errors is not observable, for almost all combinations of $N$ and $\sigma^2$.

\begin{table}[!ht]
\centering
\caption{Computational order of convergence of the 2D FDM scheme for Gaussian correlation of the $\ln(K)$ field.}
\label{tab:titleGauss2D_Homogeneous}
\begin{tabular}{ |c|c|c|c|c|c|c|c|c|c|c|c|c| }
\hline\hline
$N$ & $\sigma^2$ & $\varepsilon_1$ & EOC & $\varepsilon_2$ & EOC &$\varepsilon_3$ & EOC & $\varepsilon_4$ & EOC & $\varepsilon_5$ \\
\hline
\multirow{3}{1em}{$10^2$} & $0.1$ & 2.28e-3 & 2.00 & 5.71e-4 & 2.02 & 1.41e-4 & 2.07 & 3.36e-5 & 2.32 & 6.72e-6 \\
& $4$ & 3.03e-2 & 1.98 & 7.68e-3 & 2.01 & 1.90e-3 & 2.07 & 4.54e-4 & 2.32 & 9.09e-5\\
& $10$ & 4.10e-2 & 1.98 & 1.04e-2 & 2.01 & 2.58e-3 & 2.07 & 6.16e-4 & 2.32 & 1.23e-4\\
\hline\hline
\multirow{3}{1em}{$10^3$} & $0.1$ & 4.92e-3 & 2.00 & 1.23e-3 & 2.02 & 3.04e-4 & 2.07 & 7.24e-5 &	2.32 & 1.45e-5\\
& $4$ &1.43e-1 & 1.99 & 3.59e-2 & 2.01 & 8.90e-3 & 2.07 & 2.12e-3 & 2.32 & 4.24e-4 \\
& $10$ &3.58e-1 & 1.98 & 9.09e-2 & 2.00 & 2.26e-2 & 2.07 & 5.38e-3 & 2.32 & 1.08e-3 \\
\hline\hline
\multirow{3}{1em}{$10^4$} & $0.1$ &1.11e-3 & 2.00 & 2.77e-4 & 2.01 & 6.84e-5 & 2.07 & 1.63e-5 & 2.32 & 3.26e-6 \\
& $4$  &1.98e-2 & 2.00 & 4.94e-3 & 2.01 & 1.22e-3 & 2.07 & 2.90e-4 & 2.32 & 5.80e-5  \\
&  $10$ &5.76e-2 & 2.01 & 1.43e-2 & 2.02 & 3.51e-3 & 2.07 & 8.34e-4 & 2.32 & 1.67e-4 \\
\hline\hline
\end{tabular}
\end{table}

\begin{table}[!ht]
\centering
\caption{Computational order of convergence of the 2D FDM scheme for exponential correlation of the $\ln(K)$ field.}
\label{tab:titleExp2D_Homogeneous}
\begin{tabular}{ |c|c|c|c|c|c|c|c|c|c|c|c|c| }
\hline\hline
$N$ & $\sigma^2$ & $\varepsilon_1$ & EOC & $\varepsilon_2$ & EOC &$\varepsilon_3$ & EOC & $\varepsilon_4$ & EOC & $\varepsilon_5$ \\
\hline
\multirow{3}{1em}{$10^2$} & $0.1$ &1.00e-2 & 0.00 & 1.00e-2 & 2.27 & 2.07e-3 & -1.75 & 6.95e-3 & 0.30 & 5.65e-3 \\
& $4$ &1.06e+0 & 1.14 & 4.82e-1 & 0.72 & 2.93e-1 & 0.72 & 1.78e-1 & 0.25 & 1.50e-1 \\
& $10$ &2.80e+0 & 1.47 & 1.01e+0 & 0.46 & 7.33e-1 & 1.18 & 3.24e-1 & 0.12 & 2.98e-1  \\
\hline\hline
\multirow{3}{1em}{$10^3$} & $0.1$ &9.31e-3 & 1.47 & 3.36e-3 & 1.01 & 1.67e-3 & 0.59 & 1.11e-3 & 0.38 & 8.52e-4 \\
& $4$ &1.45e-1 & 0.08 & 1.37e-1 & 0.40 & 1.04e-1 & 0.59 & 6.89e-2 & 2.14 & 1.561e-2 \\
& $10$ &2.64e-1 & -1.30 & 6.49e-1 & 0.60 & 4.28e-1 & 0.41 & 3.23e-1 & 2.98 & 4.09e-2 \\
\hline\hline
\multirow{3}{1em}{$10^4$} & $0.1$ &2.61e-2 & 2.69 & 4.04e-3 & 1.89 & 1.09e-3 & 1.61 & 3.58e-4 & 0.81 & 2.05e-4 \\
& $4$  &4.63e-1 & 1.68 & 1.45e-1 & 1.38 & 5.56e-2 & -1.04 & 1.14e-1 & 2.81 & 1.63e-2\\
&  $10$ &8.45e-1 & 0.31 & 6.83e-1 & 2.08 & 1.62e-1 & -1.50 & 4.59e-1 & 2.60 & 7.55e-2  \\
\hline\hline
\end{tabular}
\end{table}

\newpage
\section{Finite element method}
\label{fem}
We consider the conforming piecewise linear finite element method (FEM), see e.g. \cite{KnabnerandAngermann2003}, implemented in FEniCS \cite{Alnaes2015}. The conductivity $K$ and the source term $f$ are interpolated into finite element spaces as well. We consider linear ($\mathbb{P}_1$), quadratic ($\mathbb{P}_2$) and cubic ($\mathbb{P}_3$) elements for this. Better approximations can be obtained by taking higher order finite element spaces for $K$ and $f$. The assembled linear system is solved with the default solver in FEniCS, which is the Unsymmetric MultiFrontal method in UMFPACK.

\subsection{One-dimensional case}\label{fem1D}

We solve the non-homogeneous version of the 1D problem (\ref{eq2.8}) with right hand side $-f$, for the exact analytical solution provided in Appendix~\ref{appendixA}. For this, we take $2\cdot 10^{5}$ elements (intervals) and a constant mesh size $\Delta x=10^{-3}$. We look for a continuous function $h_{\Delta x}$ that is piecewise affine on each element. We take test functions that are continuous, piecewise affine on each element and which vanish at the endpoints of the domain. We multiply the equation with a test function $\varphi_{\Delta x}$, integrate by parts and obtain that $h_{\Delta x}$ satisfies
$$
\int_0^L K h_{\Delta x}' \varphi_{\Delta x}' \,\textrm{dx} = - \int_0^L f \varphi_{\Delta x} \,\textrm{dx},
$$
for any test function $\varphi_{\Delta x}$. Tables \ref{tab:Gauss1D_P1_FEM}--\ref{tab:Exp1D_P3_FEM} contain the $L^2_{\Delta x}$ errors, computed by (\ref{FDM_EQ3}), for both Gaussian and exponential correlations and the same values of the parameters $N$ and $\sigma^2$ as in Section \ref{fdm}. The time required to solve the linear system increases with $N$ and $\sigma^2$, from about 0.01 seconds to about 0.1 seconds. In the case of Gaussian correlation, all the benchmark problems are solved accurately. In the case of exponential correlation, we can see that the 1D linear FEM with the mesh size $\Delta x=10^{-3}$ fails to solve the problems when $N = 10^4$.
\begin{table}[!ht]
	\centering
	\captionof{table}{$\mathbb{P}_1$ approximation for $K$ and $f$. Numerical errors of the 1D FEM solutions for Gaussian correlation of the $\ln(K)$ field.}
	\label{tab:Gauss1D_P1_FEM}
	\begin{tabular}{|| c || c | c | c | c | c | c | c || c ||}
		\hline\hline
		\diagbox[width=3em]{$N$}{$\sigma^2$} & 0.1 & 1 & 2 & 4 & 6 & 8 & 10  \\ \hline
		$10^2$ & 4.49e-6 & 1.45e-5 & 2.03e-5 & 1.11e-4 & 4.98e-4 & 2.55e-3 & 7.78e-3 \\ \hline
		$10^3$ & 1.00e-5 & 2.92e-5 & 5.14e-5 & 1.51e-4 & 5.58e-4 & 1.76e-3 & 8.13e-3 \\ \hline
		$10^4$ & 3.47e-6 & 1.32e-05 & 1.33e-05 & 1.38e-4 & 9.76e-4 & 4.71e-3 & 2.56e-2 \\ \hline\hline
	\end{tabular}
\end{table}

\begin{table}[!ht]
	\centering
	\captionof{table}{$\mathbb{P}_2$ approximation for $K$ and $f$. Numerical errors of the 1D FEM solutions for Gaussian correlation of the $\ln(K)$ field.}
	\label{tab:Gauss1D_FEM}
	\begin{tabular}{|| c || c | c | c | c | c | c | c || c ||}
		\hline\hline
		\diagbox[width=3em]{$N$}{$\sigma^2$} & 0.1 & 1 & 2 & 4 & 6 & 8 & 10  \\ \hline
		$10^2$ & 3.31e-6 & 7.03e-6 & 1.80e-5 & 9.03e-5 & 6.12e-4 & 1.70e-3 & 9.90e-3 \\ \hline
		$10^3$ & 6.34e-6 & 1.63e-5 & 3.34e-5 & 1.23e-4 & 4.79e-4 & 1.80e-3 & 6.36e-3 \\ \hline
		$10^4$ & 2.49e-6 & 3.80e-06 & 1.52e-05 & 1.50e-4 & 9.64e-4 & 6.02e-3 & 2.18e-2 \\ \hline\hline
	\end{tabular}
\end{table}

\begin{table}[!h]
	\centering
	\captionof{table}{$\mathbb{P}_1$ approximation for $K$ and $f$. Numerical errors of the 1D FEM solutions for exponential correlation of the $\ln(K)$ field.}
	\label{tab:Exp1D_P1_FEM}
	\begin{tabular}{|| c || c | c | c | c | c | c | c || c ||}
		\hline\hline
		\diagbox[width=3em]{$N$}{$\sigma^2$} & 0.1 & 1 & 2 & 4 & 6 & 8 & 10  \\ \hline
		$10^2$ & 6.59e-6 & 1.59e-5 & 3.00e-5 & 1.32e-4 & 6.03e-4 & 2.58e-3 & 1.43e-2 \\ \hline
		$10^3$ & 4.92e-6 & 1.28e-5 & 2.84e-5 & 1.77e-4 & 1.23e-3 & 5.54e-3 & 3.79e-2 \\ \hline
		$10^4$ & 8.88e+0 & 2.06e+3 & 2.10e+4 & 4.85e+5 & 5.46e+6 & 4.30e+7 & 2.67e+8 \\
		\hline\hline
	\end{tabular}
\end{table}

\begin{table}[!h]
	\centering
	\captionof{table}{$\mathbb{P}_2$ approximation for $K$ and $f$. Numerical errors of the 1D FEM solutions for exponential correlation of the $\ln(K)$ field.}
	\label{tab:Exp1D_FEM}
	\begin{tabular}{|| c || c | c | c | c | c | c | c || c ||}
		\hline\hline
		\diagbox[width=3em]{$N$}{$\sigma^2$} & 0.1 & 1 & 2 & 4 & 6 & 8 & 10  \\ \hline
		$10^2$ & 5.25e-6 & 1.25e-5 & 2.64e-5 & 1.16e-4 & 6.31e-4 & 2.05e-3 & 1.05e-2 \\ \hline
		$10^3$ & 4.52e-6 & 1.07e-5 & 2.83e-5 & 1.63e-4 & 1.10e-3 & 5.78e-3 & 2.77e-2 \\ \hline
		$10^4$ & 3.42e+0 & 1.05e+2 & 1.96e+3 & 5.78e+4 & 7.20e+5 & 6.09e+6 & 4.03e+7 \\
		\hline\hline
	\end{tabular}
\end{table}

\begin{table}[!h]
	\centering
	\captionof{table}{$\mathbb{P}_3$ approximation for $K$ and $f$. Numerical errors of the 1D FEM solutions for exponential correlation of the $\ln(K)$ field.}
	\label{tab:Exp1D_P3_FEM}
	\begin{tabular}{|| c || c | c | c | c | c | c | c || c ||}
		\hline\hline
		\diagbox[width=3em]{$N$}{$\sigma^2$} & 0.1 & 1 & 2 & 4 & 6 & 8 & 10  \\ \hline
		$10^2$ & 5.20e-6 & 1.24e-5 & 2.70e-5 & 1.32e-4 & 5.44e-4 & 2.21e-3 & 1.40e-2 \\ \hline
		$10^3$ & 4.48e-6 & 1.07e-5 & 2.94e-5 & 1.69e-4 & 1.13e-3 & 5.68e-3 & 4.00e-2 \\ \hline
		$10^4$ & 1.28e+0 & 3.90e+1 & 7.29e+2 & 2.13e+4 & 2.63e+5 & 2.21e+6 & 1.41e+7 \\
		\hline\hline
	\end{tabular}
\end{table}

Next, we consider the homogeneous 1D problem (\ref{eq2.8}) with $f=0$ and take 5 refinements of the grid by successively halving the mesh size from  $\Delta x=10^{-1}$ to $\Delta x=3.125\cdot 10^{-3}$, as in the previous section.
The $L^2$ error norms and the estimated orders of convergence (EOC) are presented in Tables \ref{tab:Gauss1D_P1_FEM_Homogeneous}--\ref{tab:Exp1D_FEM_Homogeneous}.
In the case of Gaussian correlation, the EOC values are close to the expected theoretical value 2, with the exception of the parameter combination ($N=10^4, \sigma^2=10)$, where the EOC decreases below 1. In the case of exponential correlation, the convergence behaviour degenerates and the errors oscillate a lot.

\begin{table}[!ht]
	\centering
	\caption{$\mathbb{P}_1$ approximation for $K$ and $f$. Computational order of convergence of the 1D FEM for Gaussian correlation of the $\ln(K)$ field.}
	\label{tab:Gauss1D_P1_FEM_Homogeneous}
	\begin{tabular}{ |c|c|c|c|c|c|c|c|c|c|c|c|c| }
		\hline\hline
		$N$ & $\sigma^2$ & $\varepsilon_1$ & EOC & $\varepsilon_2$ & EOC &$\varepsilon_3$ & EOC & $\varepsilon_4$ & EOC & $\varepsilon_5$ \\
		\hline
		\multirow{3}{1em}{$10^2$} & $0.1$ & 2.67e-4 & 2.00 & 6.69e-5 & 2.01 & 1.65e-5 & 2.05 & 3.98e-6 & 2.27 & 8.25e-7 \\
		& $4$ & 1.62e-2 & 1.99 & 4.07e-3 & 2.01 & 1.00e-3 & 2.06 & 2.41e-4 & 2.28 & 4.94e-5\\
		& $10$ & 3.79e-2 & 1.97 & 9.65e-3 & 1.95 & 2.48e-3 & 1.87 & 6.76e-4 & 1.73 & 2.03e-4\\
		\hline\hline
		\multirow{3}{1em}{$10^3$} & $0.1$ & 1.34e-4 & 1.99 & 3.35e-5 & 2.01 & 8.32e-6 & 2.05 & 1.99e-6 & 2.26 & 4.16e-7\\
		& $4$ &1.45e-2 & 1.99 & 3.64e-3 & 2.01 & 9.03e-4 & 2.06 & 2.16e-4 & 2.29 & 4.42e-5 \\
		& $10$ &2.35e-2 & 1.98 & 5.95e-3 & 1.98 & 1.50e-3 & 1.94 & 3.90e-4 & 1.82 & 1.10e-4 \\
		\hline\hline
		\multirow{3}{1em}{$10^4$} & $0.1$ &3.52e-4 & 1.99 & 8.83e-5 & 2.01 & 2.18e-5 & 2.05 & 5.25e-6 & 2.27 & 1.08e-6 \\
		& $4$  &7.89e-3 & 1.98 & 1.99e-3 & 2.01 & 4.94e-4 & 2.07 & 1.17e-4 & 2.31 & 2.36e-5  \\
		&  $10$ &1.61e-2 & 2.05 & 3.88e-3 & 2.27 & 8.05e-4 & 0.94 & 4.17e-4 & 0.41 & 3.13e-4 \\
		\hline\hline
	\end{tabular}
\end{table}

\begin{table}[!ht]
	\centering
	\caption{$\mathbb{P}_2$ approximation for $K$ and $f$. Computational order of convergence of the 1D FEM for Gaussian correlation of the $\ln(K)$ field.}
	\label{tab:Gauss1D_FEM_Homogeneous}
	\begin{tabular}{ |c|c|c|c|c|c|c|c|c|c|c|c|c| }
		\hline\hline
		$N$ & $\sigma^2$ & $\varepsilon_1$ & EOC & $\varepsilon_2$ & EOC &$\varepsilon_3$ & EOC & $\varepsilon_4$ & EOC & $\varepsilon_5$ \\
		\hline
		\multirow{3}{1em}{$10^2$} & $0.1$ & 9.63e-5 & 2.00 & 2.40e-5 & 2.00 & 5.98e-6 & 2.03 & 1.46e-6 & 2.18 & 3.22e-7 \\
		& $4$ & 5.48e-3 & 2.00 & 1.36e-3 & 2.01 & 3.38e-4 & 2.05 & 8.16e-5 & 2.25 & 1.71e-5\\
		& $10$ & 1.29e-2 & 1.97 & 3.30e-3 & 1.88 & 8.92e-4 & 1.66 & 2.82e-4 & 1.21 & 1.21e-4\\
		\hline\hline
		\multirow{3}{1em}{$10^3$} & $0.1$ & 5.46e-5 & 2.00 & 1.36e-5 & 2.00 & 3.40e-6 & 2.02 & 8.35e-7 & 2.10 & 1.94e-7\\
		& $4$ &4.99e-3 & 2.00 & 1.24e-3 & 2.01 & 3.08e-4 & 2.04 & 7.46e-5 & 2.22 & 1.59e-5 \\
		& $10$ &8.14e-3 & 1.99 & 2.04e-3 & 1.92 & 5.40e-4 & 1.64 & 1.72e-4 & 1.20 & 7.45e-5 \\
		\hline\hline
		\multirow{3}{1em}{$10^4$} & $0.1$ &1.17e-4 & 1.99 & 2.95e-5 & 2.00 & 7.33e-6 & 2.03 & 1.78e-7 & 2.18 & 3.92e-7 \\
		& $4$  &2.70e-3 & 2.00 & 6.73e-4 & 2.01 & 1.66e-4 & 2.07 & 3.95e-5 & 2.24 & 8.33e-6  \\
		&  $10$ &5.40e-3 & 2.27 & 1.11e-3 & 0.99 & 5.58e-4 & -0.16 & 6.26e-4 & 0.17 & 5.53e-4 \\
		\hline\hline
	\end{tabular}
\end{table}

\begin{table}[!ht]
	\centering
	\caption{$\mathbb{P}_1$ approximation for $K$ and $f$. Computational order of convergence of the 1D FEM for exponential correlation of the $\ln(K)$ field.}
	\label{tab:Exp1D_P1_FEM_Homogeneous}
	\begin{tabular}{ |c|c|c|c|c|c|c|c|c|c|c|c|c| }
		\hline\hline
		$N$ & $\sigma^2$ & $\varepsilon_1$ & EOC & $\varepsilon_2$ & EOC &$\varepsilon_3$ & EOC & $\varepsilon_4$ & EOC & $\varepsilon_5$ \\
		\hline
		\multirow{3}{1em}{$10^2$} & $0.1$ &8.86e-4 & 1.54 & 3.03e-4 & 0.13 & 2.76e-4 & 4.52 & 1.20e-5 & 1.83 & 3.35e-6 \\
		& $4$ &4.42e-2 & 0.67 & 2.76e-2 & -0.07 & 2.90e-2 & 5.28 & 7.48e-4 & 1.72 & 2.25e-4 \\
		& $10$ &1.09e-1 & 0.11 & 1.01e-1 & 0.12 & 9.24e-2 & 4.79 & 3.33e-3 & 1.66 & 1.04e-3  \\
		\hline\hline
		\multirow{3}{1em}{$10^3$} & $0.1$ &2.47e-3 & 2.47 & 4.45e-4 & 1.51 & 1.55e-4 & 2.05 & 3.73e-5 & 2.29 & 7.60e-6 \\
		& $4$ &5.05e-2 & -0.03 & 5.17e-2 & 1.12 & 2.36e-2 & 3.62 & 1.91e-3 & 2.15 & 4.301e-4 \\
		& $10$ &2.05e-1 & -0.66 & 3.26e-1 & 1.68 & 1.01e-1 & 4.09 & 5.91e-3 & 2.26 & 1.23e-3 \\
		\hline\hline
		\multirow{3}{1em}{$10^4$} & $0.1$ &2.66e-3 & -2.01 & 1.07e-2 & 1.37 & 4.15e-3 & 3.69 & 3.21e-4 & 4.96 & 1.03e-5 \\
		& $4$  &2.00e-1 & -0.01 & 2.01e-1 & 1.48 & 7.23e-2 & 1.01 & 3.57e-2 & 4.37 & 1.72e-3\\
		&  $10$ &6.06e-1 & 0.31 & 4.88e-1 & 1.73 & 1.46e-1 & 0.63 & 9.48e-2 & 4.85 & 3.26e-3  \\
		\hline\hline
	\end{tabular}
\end{table}

\begin{table}[!ht]
	\centering
	\caption{$\mathbb{P}_2$ approximation for $K$ and $f$. Computational order of convergence of the 1D FEM for exponential correlation of the $\ln(K)$ field.}
	\label{tab:Exp1D_FEM_Homogeneous}
	\begin{tabular}{ |c|c|c|c|c|c|c|c|c|c|c|c|c| }
		\hline\hline
		$N$ & $\sigma^2$ & $\varepsilon_1$ & EOC & $\varepsilon_2$ & EOC &$\varepsilon_3$ & EOC & $\varepsilon_4$ & EOC & $\varepsilon_5$ \\
		\hline
		\multirow{3}{1em}{$10^2$} & $0.1$ &3.99e-4 & 0.85 & 2.20e-4 & 1.24 & 3.19e-5 & 3.46 & 8.42e-6 & 2.08 & 1.99e-6 \\
		& $4$ &3.38e-2 & 1.08 & 1.59e-2 & 1.10 & 7.45e-3 & 4.51 & 3.26e-4 & 1.74 & 9.73e-5 \\
		& $10$ &1.19e-1 & 0.80 & 6.88e-2 & 1.09 & 3.23e-2 & 4.61 & 1.31e-3 & 1.31 & 5.28e-4  \\
		\hline\hline
		\multirow{3}{1em}{$10^3$} & $0.1$ &1.25e-3 & 2.90 & 1.67e-4 & 1.28 & 6.89e-5 & 2.41 & 1.29e-5 & 2.31 & 2.59e-6 \\
		& $4$ &4.07e-2 & -0.01 & 4.11e-2 & 2.36 & 8.00e-3 & 3.02 & 9.81e-4 & 2.72 & 1.48e-4 \\
		& $10$ &3.95e-1 & 0.59 & 2.61e-1 & 3.14 & 2.96e-2 & 3.22 & 3.17e-3 & 2.79 & 4.57e-4 \\
		\hline\hline
		\multirow{3}{1em}{$10^4$} & $0.1$ &7.24e-3 & -0.01 & 7.29e-2 & 1.55 & 2.48e-3 & 4.80 & 8.88e-5 & 4.09 & 5.21e-6 \\
		& $4$  &9.41e-2 & 0.80 & 5.37e-2 & 1.83 & 1.50e-2 & 1.23 & 6.40e-3 & 3.55 & 5.46e-4\\
		&  $10$ &2.45e-1 & 0.61 & 1.60e-1 & 3.17 & 1.76e-2 & -0.03 & 1.81e-2 & 3.29 & 1.84e-3  \\
		\hline\hline
	\end{tabular}
\end{table}

\subsection{Two-dimensional case}\label{fem2D}
To write the finite element formulation, let us denote by $\Gamma_1 := \{0,L_x\}\times [0,L_y]$ the part of the boundary where Dirichlet conditions are imposed, and by $\Gamma_2 := [0,L_x] \times \{0,L_y\}$ the part of the boundary where Neumann conditions are given.

Consider a triangulation $\mathcal{T}_{\Delta x}$ with the mesh size $\Delta x$. The trial space contains continuous functions that are piecewise affine on each triangle and that satisfy the Dirichlet boundary condition on $\Gamma_1$. The test space consists in continuous piecewise affine functions that vanish on $\Gamma_1$.

For the non-homogeneous equation (\ref{eq2.2}) with right hand side $-f$, the linear FEM solves for a trial function $h_{\Delta x}$ such that for any test function $\varphi_{\Delta x}$ the following weak formulation is satisfied:
$$
\int_{\mathcal{T}_{\Delta x}} K \nabla h_{\Delta x} \cdot \nabla \varphi_{\Delta x} \,\textrm{dx}\textrm{dy} = - \int_{\mathcal{T}_{\Delta x}} f \varphi_{\Delta x} \,\textrm{dx}\textrm{dy} + \int_{\Gamma_2} K \varphi_{\Delta x} \nabla h_{\Delta x} \cdot \nu \,\textrm{ds},
$$
where $\nu$ is the unit outward normal.

For the numerical tests, we first triangulate the domain by isosceles-right triangles with a short side of $\Delta x = \Delta y = 2\cdot 10^{-2}$ and we aim to approximate the exact solution given by (\ref{Flow:Eq5}--\ref{Flow:Eq6}). Tables \ref{tab:Gauss2D_P1_FEM}--\ref{tab:Exp2D_FEM} show the errors computed using the norm (\ref{FDM_EQ9}) (the same as in the finite difference case). One notices that the magnitude of the errors is in between $10^{-3}$ and $10^{-2}$ in the case of Gaussian correlation when using $\mathbb{P}_2$ approximations for $K$ and $f$. For the exponential correlation, the method cannot provide a good approximation of the solution: the errors increase substantially with $\sigma^2$, by about five or six orders of magnitude from $\sigma^2=0.1$ to $\sigma^2=10$ for each $N$. In this case, the roughness of the conductivity $K$ (highly oscillating) and the multiscale nature of the problem have a crucial impact. The linear system is solved in around 15 seconds in each case.
\begin{table}[!h]
	\centering
	\captionof{table}{$\mathbb{P}_1$ approximation for $K$ and $f$. Numerical errors of the 2D FEM solutions for Gaussian correlation of the $\ln(K)$ field.}
	\label{tab:Gauss2D_P1_FEM}
	\begin{tabular}{|| c || c | c | c | c | c | c | c || c ||}
		\hline\hline
		\diagbox[width=3em]{$N$}{$\sigma^2$} & 0.1 & 1 & 2 & 4 & 6 & 8 & 10  \\ \hline
		$10^2$ & 4.23e-3 & 7.58e-3 & 1.60e-2 & 8.51e-2 & 2.93e-1 & 7.96e-1 & 1.87e+0 \\ \hline
		$10^3$ & 5.81e-3 & 2.73e-2 & 9.37e-2 & 7.29e-1 & 3.74e+0 & 1.49e+1 & 4.99e+2 \\ \hline
		$10^4$ & 4.67e-3 & 1.19e-2 & 1.68e-2 & 2.40e-2 & 3.27e-2 & 4.97e-2 & 8.36e-2 \\ \hline\hline
	\end{tabular}
\end{table}

\begin{table}[!h]
	\centering
	\captionof{table}{$\mathbb{P}_2$ approximation for $K$ and $f$. Numerical errors of the 2D FEM solutions for Gaussian correlation of the $\ln(K)$ field.}
	\label{tab:Gauss2D_FEM}
	\begin{tabular}{|| c || c | c | c | c | c | c | c || c ||}
		\hline\hline
		\diagbox[width=3em]{$N$}{$\sigma^2$} & 0.1 & 1 & 2 & 4 & 6 & 8 & 10  \\ \hline
		$10^2$ & 1.61e-3 & 2.97e-3 & 3.93e-3 & 5.60e-3 & 7.06e-3 & 8.28e-3 & 9.20e-3 \\ \hline
		$10^3$ & 2.17e-3 & 5.60e-3 & 7.60e-3 & 1.01e-2 & 1.16e-2 & 1.23e-2 & 1.41e-2 \\ \hline
		$10^4$ & 2.04e-3 & 5.78e-3 & 8.14e-3 & 1.08e-2 & 1.23e-2 & 1.33e-2 & 1.41e-2 \\ \hline\hline
	\end{tabular}
\end{table}

\begin{table}[!h]
	\centering
	\captionof{table}{$\mathbb{P}_1$ approximation for $K$ and $f$. Numerical errors of the 2D FEM solutions for exponential correlation of the $\ln(K)$ field.}
	\label{tab:Exp2D_P1_FEM}
	\begin{tabular}{|| c || c | c | c | c | c | c | c || c ||}
		\hline\hline
		\diagbox[width=3em]{$N$}{$\sigma^2$} & 0.1 & 1 & 2 & 4 & 6 & 8 & 10  \\ \hline
		$10^2$ & 2.84e-2 & 1.51e+0 & 1.07e+0 & 1.26e+2 & 7.06e+2 & 2.68e+3 & 7.91e+3 \\ \hline
		$10^3$ & 1.17e-1 & 2.15e+0 & 9.49e+0 & 1.64e+2 & 1.85e+3 & 1.09e+4 & 4.32e+4 \\ \hline
		$10^4$ & 1.66e+0 & 5.67e+1 & 5.70e+2 & 6.68e+3 & 3.36e+4 & 1.19e+5 & 3.47e+5 \\ \hline\hline
	\end{tabular}
\end{table}

\begin{table}[!h]
	\centering
	\captionof{table}{$\mathbb{P}_2$ approximation for $K$ and $f$. Numerical errors of the 2D FEM solutions for exponential correlation of the $\ln(K)$ field.}
	\label{tab:Exp2D_FEM}
	\begin{tabular}{|| c || c | c | c | c | c | c | c || c ||}
		\hline\hline
		\diagbox[width=3em]{$N$}{$\sigma^2$} & 0.1 & 1 & 2 & 4 & 6 & 8 & 10  \\ \hline
		$10^2$ & 9.08e-3 & 5.07e-1 & 3.57e+0 & 4.28e+1 & 2.41e+2 & 9.24e+2 & 2.74e+3 \\ \hline
		$10^3$ & 1.29e-2 & 8.46e-1 & 3.66e+0 & 4.87e+1 & 6.06e+2 & 3.66e+3 & 1.42e+4 \\ \hline
		$10^4$ & 4.60e-1 & 3.56e+1 & 2.40e+2 & 2.22e+3 & 1.03e+4 & 3.67e+4 & 1.13e+5 \\ \hline\hline
	\end{tabular}
\end{table}

We then consider the homogeneous problem (\ref{eq2.2}--\ref{eq2.4}) with no exact solution being given. We refine the mesh five times starting from $\Delta x = \Delta y = 4\cdot 10^{-1}$ by halving both $\Delta x$ and $\Delta y$. The finest approximation with $\Delta x = \Delta y = 1.25\cdot10^{-2}$ serves as a reference solution for estimating the convergence order. The respective errors $\varepsilon_i$ are then computed with the $L^2$ norm and the results are given in Tables \ref{tab:Gauss2D_P1_FEM_Homogeneous}--\ref{tab:exp2D_FEM_Homogeneous}. The estimated order of convergence is close to the expected order 2 for the Gaussian correlation, but it deteriorates in the case of exponential correlation.
\begin{table}[!ht]
	\centering
	\caption{$\mathbb{P}_1$ approximation for $K$ and $f$. Computational order of convergence of the 2D FEM for Gaussian correlation of the $\ln(K)$ field.}
	\label{tab:Gauss2D_P1_FEM_Homogeneous}
	\begin{tabular}{ |c|c|c|c|c|c|c|c|c|c|c|c|c| }
		\hline\hline
		$N$ & $\sigma^2$ & $\varepsilon_1$ & EOC & $\varepsilon_2$ & EOC &$\varepsilon_3$ & EOC & $\varepsilon_4$ & EOC & $\varepsilon_5$ \\
		\hline
		\multirow{3}{1em}{$10^2$} & $0.1$ & 9.88e-3 & 1.89 & 2.65e-3 & 1.97 & 6.73e-4 & 2.05 & 1.62e-4 & 2.30 & 3.28e-5 \\
		& $4$ & 1.25e-1 & 1.90 & 3.36e-2 & 1.98 & 8.53e-3 & 2.05 & 2.05e-3 & 2.31 & 4.13e-4\\
		& $10$ & 3.13e-1 & 2.06 & 7.50e-2 & 2.00 & 1.86e-2 & 2.05 & 4.47e-3 & 2.31 & 9.00e-4\\
		\hline\hline
		\multirow{3}{1em}{$10^3$} & $0.1$ & 9.63e-3 & 1.81 & 2.72e-3 & 1.95 & 7.03e-4 & 2.04 & 1.70e-4 & 2.30 & 3.45e-5\\
		& $4$ &1.13e-1 & 1.74 & 3.41e-2 & 1.92 & 8.97e-3 & 2.04 & 2.18e-3 & 2.30 & 4.42e-4 \\
		& $10$ &2.60e-1 & 1.93 & 6.79e-2 & 1.97 & 1.72e-2 & 2.04 & 4.18e-3 & 2.30 & 8.47e-4 \\
		\hline\hline
		\multirow{3}{1em}{$10^4$} & $0.1$ &9.85e-3 & 1.83 & 2.77e-3 & 1.95 & 7.12e-4 & 2.04 & 1.72e-4 & 2.30 & 3.49e-5 \\
		& $4$  &1.21e-1 & 1.84 & 3.39e-2 & 1.96 & 8.70e-3 & 2.05 & 2.10e-3 & 2.30 & 4.24e-4  \\
		&  $10$ &2.34e-1 & 1.88 & 6.34e-2 & 1.96 & 1.62e-2 & 2.04 & 3.91e-3 & 2.30 & 7.92e-4 \\
		\hline\hline
	\end{tabular}
\end{table}

\begin{table}[!ht]
	\centering
	\caption{$\mathbb{P}_2$ approximation for $K$ and $f$. Computational order of convergence of the 2D FEM for Gaussian correlation of the $\ln(K)$ field.}
	\label{tab:Gauss2D_FEM_Homogeneous}
	\begin{tabular}{ |c|c|c|c|c|c|c|c|c|c|c|c|c| }
		\hline\hline
		$N$ & $\sigma^2$ & $\varepsilon_1$ & EOC & $\varepsilon_2$ & EOC &$\varepsilon_3$ & EOC & $\varepsilon_4$ & EOC & $\varepsilon_5$ \\
		\hline
		\multirow{3}{1em}{$10^2$} & $0.1$ & 5.02e-3 & 1.94 & 1.30e-3 & 1.99 & 3.27e-4 & 2.05 & 7.85e-5 & 2.28 & 1.61e-5 \\
		& $4$ & 6.30e-2 & 1.98 & 1.58e-2 & 2.01 & 3.93e-3 & 2.06 & 9.395e-4 & 2.30 & 1.90e-4\\
		& $10$ & 1.48e-1 & 2.04 & 3.60e-2 & 2.03 & 8.81e-3 & 2.06 & 2.10e-3 & 2.30 & 4.25e-4\\
		\hline\hline
		\multirow{3}{1em}{$10^3$} & $0.1$ & 5.96e-3 & 1.90 & 1.59e-3 & 1.99 & 4.00e-4 & 2.05 & 9.61e-5 & 2.28 & 1.96e-5\\
		& $4$ &7.59e-2 & 1.90 & 0.02e-2 & 2.00 & 5.05e-3 & 2.06 & 1.20e-3 & 2.30 & 2.45e-4 \\
		& $10$ &1.62e-1 & 2.04 & 3.95e-2 & 2.04 & 9.57e-3 & 2.07 & 2.27e-3 & 2.30 & 4.62e-4 \\
		\hline\hline
		\multirow{3}{1em}{$10^4$} & $0.1$ &5.99e-3 & 1.90 & 1.59e-3 & 1.99 & 4.01e-4 & 2.05 & 9.64e-5 & 2.28 & 1.97e-5 \\
		& $4$  &6.90e-2 & 1.96 & 1.76e-2 & 2.00 & 4.38e-3 & 2.06 & 1.04e-3 & 2.30 & 2.12e-4  \\
		&  $10$ &1.27e-1 & 1.99 & 3.19e-2 & 2.02 & 7.85e-3 & 2.06 & 1.87e-3 & 2.29 & 3.81e-4 \\
		\hline\hline
	\end{tabular}
\end{table}

\begin{table}[!ht]
	\centering
	\caption{$\mathbb{P}_1$ approximation for $K$ and $f$. Computational order of convergence of the 2D FEM for exponential correlation of the $\ln(K)$ field.}
	\label{tab:exp2D_P1_FEM_Homogeneous}
	\begin{tabular}{ |c|c|c|c|c|c|c|c|c|c|c|c|c| }
		\hline\hline
		$N$ & $\sigma^2$ & $\varepsilon_1$ & EOC & $\varepsilon_2$ & EOC &$\varepsilon_3$ & EOC & $\varepsilon_4$ & EOC & $\varepsilon_5$ \\
		\hline
		\multirow{3}{1em}{$10^2$} & $0.1$ & 1.29e-2 & 1.30 & 5.23e-3 & 1.54 & 1.79e-3 & 1.27 & 7.41e-4 & 1.16 & 3.29e-4 \\
		& $4$ & 2.64e-1 & 1.44 & 9.71e-2 & 1.68 & 3.01e-2 & 0.94 & 1.56e-2 & 0.91 & 8.26e-3\\
		& $10$ & 5.63e-1 & 1.19 & 2.45e-1 & 1.60 & 8.10e-2 & 0.99 & 4.07e-2 & 1.02 & 1.99e-2\\
		\hline\hline
		\multirow{3}{1em}{$10^3$} & $0.1$ & 1.67e-2 & 1.28 & 6.85e-3 & 1.79 & 1.98e-3 & 1.26 & 8.27e-4 & 2.17 & 1.82e-4\\
		& $4$ & 1.92e-1 & 1.05 & 9.29e-2 & 1.44 & 3.40e-2 & 0.93 & 1.78e-2 & 2.45 & 3.24e-3 \\
		& $10$ & 3.56e-1 & 0.89 & 1.91e-1 & 1.47 & 6.89e-2 & 0.52 & 4.80e-2 & 2.71 & 7.31e-3 \\
		\hline\hline
		\multirow{3}{1em}{$10^4$} & $0.1$ &1.94e-2 & 1.52 & 6.77e-3 & 0.49 & 4.81e-3 & 2.59 & 7.96e-4 & 1.21 & 3.43e-4 \\
		& $4$  &2.89e-1 & 0.76 & 1.70e-1 & 1.80 & 4.86e-2 & 1.50 & 1.71e-2 & 1.35 & 6.66e-3  \\
		&  $10$ &4.18e-1 & 0.39 & 3.18e-1 & 1.59 & 1.05e-1 & 1.70 & 3.23e-2 & 0.95 & 1.66e-2 \\
		\hline\hline
	\end{tabular}
\end{table}

\begin{table}[!ht]
	\centering
	\caption{$\mathbb{P}_2$ approximation for $K$ and $f$. Computational order of convergence of the 2D FEM for exponential correlation of the $\ln(K)$ field.}
	\label{tab:exp2D_FEM_Homogeneous}
	\begin{tabular}{ |c|c|c|c|c|c|c|c|c|c|c|c|c| }
		\hline\hline
		$N$ & $\sigma^2$ & $\varepsilon_1$ & EOC & $\varepsilon_2$ & EOC &$\varepsilon_3$ & EOC & $\varepsilon_4$ & EOC & $\varepsilon_5$ \\
		\hline
		\multirow{3}{1em}{$10^2$} & $0.1$ & 8.88e-3 & 1.38 & 3.39e-3 & 1.26 & 1.40e-3 & 1.43 & 5.22e-4 & 1.73 & 1.57e-4 \\
		& $4$ & 1.95e-1 & 1.78 & 5.65e-2 & 1.01 & 2.79e-2 & 0.84 & 1.55e-2 & 1.59 & 5.16e-3\\
		& $10$ & 4.38e-1 & 1.86 & 1.19e-1 & 0.74 & 7.15e-2 & 0.65 & 4.54e-2 & 1.53 & 1.57e-2\\
		\hline\hline
		\multirow{3}{1em}{$10^3$} & $0.1$ & 1.40e-2 & 1.76 & 4.14e-3 & 1.53 & 1.43e-3 & 1.45 & 5.22e-4 & 2.09 & 1.21e-4\\
		& $4$ & 2.09e-1 & 1.89 & 5.64e-2 & 1.41 & 2.10e-2 & 1.20 & 9.16e-3 & 1.90 & 2.44e-3 \\
		& $10$ & 5.03e-1 & 2.12 & 1.11e-1 & 1.39 & 4.38e-2 & 1.20 & 1.89e-2 & 1.83 & 5.32e-3 \\
		\hline\hline
		\multirow{3}{1em}{$10^4$} & $0.1$ &1.21e-2 & 1.41 & 4.55e-3 & 1.62 & 1.48e-3 & 1.43 & 5.48e-4 & 1.81 & 1.56e-4 \\
		& $4$  &2.21e-1 & 1.75 & 6.56e-2 & 1.68 & 2.04e-2 & 0.96 & 1.04e-2 & 1.88 & 2.82e-3  \\
		&  $10$ &4.09e-1 & 1.43 & 1.51e-1 & 1.69 & 4.66e-2 & 0.97 & 2.37e-2 & 1.70 & 7.28e-3 \\
		\hline\hline
	\end{tabular}
\end{table}

\section{Discontinuous Galerkin method}
\label{dgm}

In this section, we use the non-symmetric interior penalty DGM to solve 2D problem \eqref{eq2.2}--\eqref{eq2.4} with source term $-f$. We consider a Friedrichs--Keller triangulation ${\mathcal T}_{\Delta x}$ where $\Delta x = \Delta y$ denotes the length of a short side of each isosceles-right triangle. Assuming that the test and trial functions are element-wise polynomials of degree at most~$p$, element-wise multiplication of Eq.~\eqref{eq2.2} by local test functions $\varphi_{\Delta x}$ with support only on~$T\in {\mathcal T}_{\Delta x}$, integration over~$T$, integration by parts, and selection of suitable numerical fluxes yield the following equation for all $T \in {\mathcal T}_{\Delta x}$ that are not adjacent to the domain's boundary:
\begin{align*}
 \int_T \! K \nabla h_{\Delta x} \cdot \nabla \varphi_{\Delta x} \, \textrm{dx}\textrm{dy} + \sum_{F \in \mathcal F_T} \int_F \left[ \avg{ K \nabla \varphi_{\Delta x} } h_{\Delta x} \! \cdot \nu - \avg{ K \nabla h_{\Delta x} } \varphi_{\Delta x} \! \cdot \nu + \frac{\eta}{\Delta x} \avg{h_{\Delta x} \nu} \varphi_{\Delta x} \! \cdot \nu \right] \text ds \\
 = - \int_T \! f \varphi_{\Delta x} \, \textrm{dx}\textrm{dy}.
\end{align*}
Here, $\mathcal F_T$ denotes the set of faces of $T$, $\nu$ is the unit outward normal with respect to $T$, and $\eta \ge 0$ is a stabilization parameter. Since the test and trial functions are discontinuous, the numerical fluxes contain averages $\avg{\cdot}$ of functions. The average of a function at a common interface $F=T^+\cap T^-$ of elements $T^+$ and $T^-$ is (component-wise) given as the arithmetic mean of the traces of this function. The correct way to employ boundary conditions and precise definitions of the corresponding terms can be found in~\cite{DiPietroErn2012,Riviere2008,Rupp2019}. The following simulations are carried out utilizing piecewise linear test and trial functions, i.e., $p=1$.

We consider the 2D spatial domain specified in Section~\ref{problem}, partitioned into isosceles-right triangles with a short side of $\Delta x = \Delta y = 2\cdot 10^{-2}$. Thus, we receive $10^6$ triangles with 3~degrees of freedom (DoF) on each triangle. Similar to the previous sections, an analytical test scenario is evaluated for a scalar~$K$, the values of which are log-normally distributed according to a Gaussian correlation (cf.~Table~\ref{Gauss2Ddg}) and an exponential correlation (cf.~Table~\ref{Exp2Ddg}). The $L^2$ errors for the respective numerical tests with a manufactured analytical solution and varying $\sigma^2$ and $N$ (cf.~Section~\ref{problem}) are outlined below. The implementation was made using the Matlab/Octave toolbox~FESTUNG~\cite{FESTUNG1}.
\begin{table}[!ht] \centering
 \caption{Numerical errors of the 2D DGM solutions for Gaussian correlation of the $\ln(K)$ field.}
 \begin{tabular}{|| l || c | c | c | c | c | c | c || c ||}
  \hline\hline
  \diagbox[width=3em]{$N$}{$\sigma^2$} & 0.1 & 1 & 2 & 4 & 6 & 8 &10\\
  \hline
  $10^2$ & 1.11e-3 & 1.15e-3 & 1.41e-3 & 2.10e-3 & 2.76e-3 & 3.35e-3 & 3.86e-3 \\
  \hline
  $10^3$ & 1.11e-3 & 1.19e-3 & 1.41e-3 & 1.84e-3 & 2.29e-3 & 2.75e-3 & 3.17e-3 \\
  \hline
  $10^4$ & 1.11e-3 & 1.05e-3 & 1.10e-3 & 1.18e-3 & 1.42e-3 & 1.75e-3 & 2.16e-3 \\
  \hline\hline
 \end{tabular}\label{Gauss2Ddg}
\end{table}
\begin{table}[!ht] \centering
 \caption{Numerical errors of the 2D DGM for exponential correlation of the $\ln(K)$ field.}
 \begin{tabular}{|| l || c | c | c | c | c | c | c || c ||}
  \hline\hline
  \diagbox[width=3em]{$N$}{$\sigma^2$} & 0.1 & 1 & 2 & 4 & 6 & 8 &10\\
  \hline
  $10^2$ & 4.58e-2 & 2.59e-1 & 5.31e-1 & 2.91e+0 & 1.53e+1 & 6.08e+1 & 2.00e+2 \\
  \hline
  $10^3$ & 1.37e-2 & 1.72e-1 & 1.20e+0 & 1.32e+1 & 6.87e+1 & 2.52e+2 & 7.61e+2 \\
  \hline
  $10^4$ & 1.74e-1 & 5.27e+0 & 2.39e+1 & 3.42e+2 & 2.26e+3 & 9.48e+3 & 3.05e+4 \\
  \hline\hline
 \end{tabular}\label{Exp2Ddg}
\end{table}

One observes that the DGM scheme works well for the Gaussian correlation where the $L^2$ errors have a magnitude of $10^{-3}$ independently from the choice of $N$ and $\sigma^2$. Moreover, the errors for large values of $N$ and $\sigma^2$ are one order of magnitude smaller than for classical FEM and between one and four orders of magnitude smaller than for FDM. However, for the exponential correlation, the $L^2$ errors increase significantly with increasing $\sigma^2$ and $N$. Even though generally one order of magnitude smaller than for FDM and FEM, the extremely large errors, up to $10^4$, of the numerical solutions in the case of exponential correlation with $N$ and $\sigma^2$ show that these test problems are practically computationally intractable for DGM as well.

\begin{figure}[ht!] \centering
 \begin{tikzpicture}
  \draw (-2,-1) -- (2,-1) -- (2,1) -- (-2,1) -- (-2,-1);
  \draw (0,-1) -- (0,1);
  \draw (-2,0) -- (2,0);
  \draw (-2,1) -- (2,-1);
  \draw (-2,0) -- (0,-1);
  \draw (0,1) -- (2,0);
  \node at (-1.9,-0.9) [circle,fill,inner sep=1pt]{};
  \node at (-0.3,-0.925) [circle,fill,inner sep=1pt]{};
  \node at (-1.9,-0.2) [circle,fill,inner sep=1pt]{};
  \node at (-1.7,-0.075) [circle,fill,inner sep=1pt]{};
  \node at (-0.1,-0.8) [circle,fill,inner sep=1pt]{};
  \node at (-0.1,-0.1) [circle,fill,inner sep=1pt]{};
  \node at (0.1,-0.9) [circle,fill,inner sep=1pt]{};
  \node at (1.7,-0.925) [circle,fill,inner sep=1pt]{};
  \node at (0.1,-0.2) [circle,fill,inner sep=1pt]{};
  \node at (0.3,-0.075) [circle,fill,inner sep=1pt]{};
  \node at (1.9,-0.8) [circle,fill,inner sep=1pt]{};
  \node at (1.9,-0.1) [circle,fill,inner sep=1pt]{};
  \node at (-1.9,0.1) [circle,fill,inner sep=1pt]{};
  \node at (-0.3,0.075) [circle,fill,inner sep=1pt]{};
  \node at (-1.9,0.8) [circle,fill,inner sep=1pt]{};
  \node at (-1.7,0.925) [circle,fill,inner sep=1pt]{};
  \node at (-0.1,0.2) [circle,fill,inner sep=1pt]{};
  \node at (-0.1,0.9) [circle,fill,inner sep=1pt]{};
  \node at (0.1,0.1) [circle,fill,inner sep=1pt]{};
  \node at (1.7,0.075) [circle,fill,inner sep=1pt]{};
  \node at (0.1,0.8) [circle,fill,inner sep=1pt]{};
  \node at (0.3,0.925) [circle,fill,inner sep=1pt]{};
  \node at (1.9,0.2) [circle,fill,inner sep=1pt]{};
  \node at (1.9,0.9) [circle,fill,inner sep=1pt]{};
 \end{tikzpicture}
 \qquad \qquad
 \begin{tikzpicture}
  \draw (-2,-1) -- (2,-1) -- (2,1) -- (-2,1) -- (-2,-1);
  \draw (0,-1) -- (0,1);
  \draw (-2,0) -- (2,0);
  \draw (-2,1) -- (2,-1);
  \draw (-2,0) -- (0,-1);
  \draw (0,1) -- (2,0);
  \node at (-2,-1) [circle,fill,inner sep=1pt]{};
  \node at (0,-1) [circle,fill,inner sep=1pt]{};
  \node at (2,-1) [circle,fill,inner sep=1pt]{};
  \node at (-2,0) [circle,fill,inner sep=1pt]{};
  \node at (0,0) [circle,fill,inner sep=1pt]{};
  \node at (2,0) [circle,fill,inner sep=1pt]{};
  \node at (-2,1) [circle,fill,inner sep=1pt]{};
  \node at (0,1) [circle,fill,inner sep=1pt]{};
  \node at (2,1) [circle,fill,inner sep=1pt]{};
 \end{tikzpicture}
 \caption{Illustration of the degrees of freedom for DGM (left) and FEM (right) for non-isosceles triangles. }\label{Fig:DoFDG}
\end{figure}
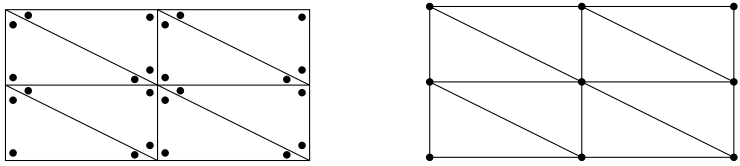

Note that the DGM approach induces significantly larger linear systems of equations than a classical FEM approximation of equal order on the same mesh. This is due to the fact that for DGM schemes, all elements contain three degrees of freedom (DoFs), while for FEM, the corresponding DoFs are shared across element boundaries. This is illustrated in Fig.~\ref{Fig:DoFDG}, where a coarse mesh is illustrated: the center point is associated with six DoFs in DGM, while it is associated with a single DoF in a FEM approximation. Hence, the size of the corresponding linear system of equations for DGM is approximately six times the size of the FEM approximation (when neglecting the boundary DoFs where the ratio is only 3:1 and the corner DoFs where the ratio is 2:1 or 1:1). As a consequence, DGM approximations are more costly than FEM approximations when performed on the same mesh, but they also pose some advantages such as local mass conservation, feasibility for local mesh and $p$ adaptivity, as well as an intuitive treatment of hanging nodes. The computation times to solve each of the linear systems of equations are between 42 and 46 seconds in the aforementioned scenario.

Next, we consider the 2D case in which no analytical solution is given and $f$ is set to zero. For the precise definition of the simulation scenario, we refer to Section~\ref{problem}. Starting with $\Delta x = \Delta y = 4\cdot 10^{-1}$, as for 2D FEM solutions analyzed in the previous section, we refine the mesh five times by dividing both $\Delta x$ and $\Delta y$ by two. This yields a series of successively refined approximate solutions, where the finest approximation with $\Delta x = \Delta y = 1.25\cdot10^{-2}$ serves as reference solution. The respective errors $\varepsilon_i$ are defined as the $L^2$ difference between the $i$th and the 6th approximate solution. The estimated orders of convergence (EOC) for scenarios with values of $K$ which are Gaussian correlated are listed in Table~\ref{Gauss2DdgHom}. The analogous scenario with exponential correlation is evaluated in Table~\ref{Exp2DdgHom}.
\begin{table}[!ht] \centering
 \caption{Computational order of convergence of the 2D DGM scheme for Gaussian correlation of the $\ln(K)$ field.}
 \begin{tabular}{ |c|c|c|c|c|c|c|c|c|c|c|c|c| }
  \hline\hline
  $N$ & $\sigma^2$ & $\varepsilon_1$ & EOC & $\varepsilon_2$ & EOC &$\varepsilon_3$ & EOC & $\varepsilon_4$ & EOC & $\varepsilon_5$ \\
  \hline
  \multirow{3}{1em}{$10^2$} & $0.1$ & 7.43e-3 & 2.02 & 1.83e-3 & 2.01 & 4.52e-4 & 1.92 & 1.19e-4 & 1.34 & 4.69e-5 \\
			    & $4$   & 1.27e-1 & 2.00 & 3.16e-2 & 2.02 & 7.74e-3 & 2.08 & 1.83e-3 & 2.31 & 3.69e-4 \\
			    & $10$  & 1.57e-1 & 2.03 & 3.83e-2 & 2.01 & 9.50e-3 & 2.07 & 2.26e-3 & 2.30 & 4.56e-4 \\
  \hline\hline
  \multirow{3}{1em}{$10^3$} & $0.1$ & 8.15e-3 & 2.10 & 1.89e-3 & 2.07 & 4.49e-4 & 1.94 & 1.17e-4 & 1.39 & 4.45e-5 \\
			    & $4$   & 7.53e-2 & 2.09 & 1.76e-2 & 2.07 & 4.18e-3 & 2.10 & 9.74e-4 & 2.29 & 1.99e-4 \\
			    & $10$  & 2.76e-1 & 2.07 & 6.55e-2 & 2.05 & 1.58e-2 & 2.08 & 3.73e-3 & 2.31 & 7.48e-4 \\
  \hline\hline
  \multirow{3}{1em}{$10^4$} & $0.1$ & 8.20e-3 & 2.04 & 1.99e-3 & 2.02 & 4.89e-4 & 1.95 & 1.26e-4 & 1.38 & 4.83e-5 \\
			    & $4$   & 8.85e-2 & 1.94 & 2.30e-2 & 2.01 & 5.69e-3 & 2.07 & 1.35e-3 & 2.29 & 2.76e-4 \\
			    & $10$  & 2.12e-1 & 1.91 & 5.63e-2 & 1.99 & 1.41e-2 & 2.06 & 3.37e-3 & 2.31 & 6.77e-4 \\
  \hline\hline
 \end{tabular}\label{Gauss2DdgHom}
\end{table}
\begin{table}[!ht] \centering
 \caption{Computational order of convergence of the 2D DGM scheme for exponential correlation of the $\ln(K)$ field.}
 \begin{tabular}{ |c|c|c|c|c|c|c|c|c|c|c|c|c| }
  \hline\hline
  $N$ & $\sigma^2$ & $\varepsilon_1$ & EOC & $\varepsilon_2$ & EOC &$\varepsilon_3$ & EOC & $\varepsilon_4$ & EOC & $\varepsilon_5$ \\
  \hline
  \multirow{3}{1em}{$10^2$} & $0.1$ & 2.17e-2 & 1.02 & 1.07e-2 & 1.85 & 2.96e-3 & 1.69 & 9.16e-4 & 0.14 & 8.29e-4 \\
			    & $4$   & 2.35e-1 & 0.86 & 1.29e-1 & 0.84 & 7.19e-2 & 1.10 & 3.34e-2 & 1.60 & 1.10e-2 \\
			    & $10$  & 2.56e-1 & 1.57 & 8.59e-2 & 0.12 & 7.86e-2 & 0.74 & 4.68e-2 & 1.71 & 1.43e-2 \\
  \hline\hline
  \multirow{3}{1em}{$10^3$} & $0.1$ & 2.32e-2 & 1.42 & 8.66e-3 & 1.74 & 2.58e-3 & 1.35 & 1.01e-3 & 1.90 & 2.70e-4 \\
			    & $4$   & 1.83e-1 & 1.01 & 9.03e-2 & 0.84 & 5.02e-2 & 1.29 & 2.04e-2 & 1.33 & 8.07e-3 \\
			    & $10$  & 5.03e-1 & 2.27 & 1.04e-1 & 1.09 & 4.88e-2 & 1.02 & 2.39e-2 & 1.03 & 1.17e-2 \\
  \hline\hline
  \multirow{3}{1em}{$10^4$} & $0.1$ & 2.36e-2 & 1.21 & 1.02e-2 & 1.49 & 3.62e-3 & 1.62 & 1.17e-3 & 1.61 & 3.82e-4 \\
			    & $4$   & 4.06e-1 & 0.92 & 2.14e-1 & 1.02 & 1.05e-1 & 1.35 & 4.10e-2 & 1.61 & 1.34e-2 \\
			    & $10$  & 3.82e-1 & 1.71 & 1.16e-1 & 1.41 & 4.36e-2 & 0.95 & 2.25e-2 & 1.78 & 6.52e-3 \\
  \hline\hline
 \end{tabular}\label{Exp2DdgHom}
\end{table}
The used DGM is known to be of convergence order~$2$ if the mesh is regular (which holds in our case). Our numerical results support this result for the random $\ln(K)$ fields with Gaussian correlation (cf. Table~\ref{Gauss2DdgHom}), whereas the EOC for exponential correlation seems to be deteriorating to $1$ or $1.5$ in most cases (cf. Table~\ref{Exp2DdgHom}). This could for example be caused by the roughness of the solution if the hydraulic conductivity $K$ corresponds to exponentially correlated $\ln(K)$ fields.

Ensembles of hydraulic head solutions and Darcy velocities obtained with the DGM approach, according to (\ref{eq2.1}), are analyzed within a Monte Carlo framework in Section~\ref{stat} below.

\section{Spectral methods}
\label{spectral}

As an alternative to finite difference/element methods used in the previous sections, we propose in the following the technique of collocation and Galerkin spectral methods in order to solve the flow problems (\ref{eq2.2}-\ref{eq2.4}) and (\ref{eq2.8}). We refer to \cite{AurentzandTrefethen2017,Weideman2000} for a detailed description of the Chebyshev collocation spectral method. In our implementation of the CSM schemes, we use the explicit analytical expressions (\ref{Flow:Eq3}) and (\ref{Flow:Eq7}) of the hydraulic conductivity for the 1D and 2D cases, $K(x)$ and $K(x,y)$, respectively, to compute their derivatives $K'(x)$, $\frac{\partial}{\partial x}K(x,y)$, and $\frac{\partial}{\partial y}K(x,y)$. To avoid the need to know the analytical expression of the coefficients $K$, we also adapt the algorithm of \cite{Shen2016} to construct an GSM approach for our 1D benchmark problems.

\subsection{Chebyshev collocation spectral method}

For the beginning, we homogenize the 1D problem (\ref{eq2.8}) by using the transformation
\begin{equation}\label{homogen}
v(x):=h(x) - \left(\frac{h(L)-h(0)}{L}x + h(0)\right), \quad x\in [0,L].
\end{equation}
Replacing $h(x)$ in (\ref{eq2.8}) by (\ref{homogen}), the 1D flow problem becomes
\begin{equation}\label{final1d}
	\begin{cases}
    	K'(x)v'(x) + K(x)v''(x) = f(x) -\dfrac{h(L)-h(0)}{L}K'(x) , \quad \forall x \in(0,L) ,\\[2mm]
    	v(0) = 0, v(L) = 0.
    \end{cases}
\end{equation}

Further, in order to apply the Chebyshev spectral collocation we use the change of variables $t=\frac{2}{L}x-1$ to transform the flow problem from $[0,L]$ to $[-1,1]$ as follows:
\begin{equation}\label{transformed}
	\begin{cases}
    	\dfrac{2}{L}\tilde{K}'(t)\tilde{v}'(t) + \left(\dfrac{2}{L}\right)^2\tilde{K}(t)\tilde{v}''(t) =
    	\tilde{f}(t)-\dfrac{h(L)-h(0)}{L}\tilde{K}'(x), \quad \forall t \in(-1,1) \\[2mm]
    	\tilde{v}(-1) = 0, \quad \tilde{v}(1) = 0  \ ,
    \end{cases}
\end{equation}
where
$$\tilde{v}(t) = v \left( \frac{L}{2}(t+1) \right) \ , \tilde{K}(t) = K \left( \frac{L}{2}(t+1) \right) \quad\mbox{and}\quad \tilde{K^\prime}(t) = K^\prime \left( \frac{L}{2}(t+1) \right) \ , \tilde{f}(t) = f \left( \frac{L}{2}(t+1) \right).$$

The flow equation written in matrix form is given by
\begin{equation}\label{matrixeq}
    \frac{2}{L}\left(\text{diag}(\tilde{\mathbf{K}}')\tilde{D}^{(1)} + \frac{2}{L}\text{diag}(\tilde{\mathbf{K}})\tilde{D}^{(2)}\right)\tilde{v} = \tilde{\mathbf{f}}-\frac{h(L)-h(0)}{L}\tilde{\mathbf{K}}',
\end{equation}
where $\tilde{D}^{(1)}$ and $\tilde{D}^{(2)}$ are the first and second order Chebyshev differentiation matrices where we deleted the first and last rows and columns in order to incorporate the boundary conditions. The matrices  $D^{(1)}, D^{(2)}$ were generated with {\it chebdif} function given in \cite{Weideman2000}. In the equation (\ref{matrixeq}), $\tilde{\mathbf{K}} \ , \tilde{\mathbf{K}}' \text{ and } \tilde{\mathbf{f}}$ are column vectors that contain the values of $\tilde{K} \ , \tilde{K}' \text{ and } \tilde{f}$ on the Chebyshev nodes, with the exception of the first and last points.

Using (\ref{matrixeq}), we solve the 1D problem (\ref{eq2.8}), with source term given by (\ref{Flow:Eq4}), for different combinations of $\sigma^2$ and $N$, and we compare the numerical solutions with the exact manufactured solution (\ref{Flow:Eq1}) by computing the $l_{\infty}$ vector norm at collocation points.

\begin{figure}[h]
\centering \begin{minipage}[t]{0.45\linewidth} \centering
\vspace*{0in}
\includegraphics[width=\linewidth]{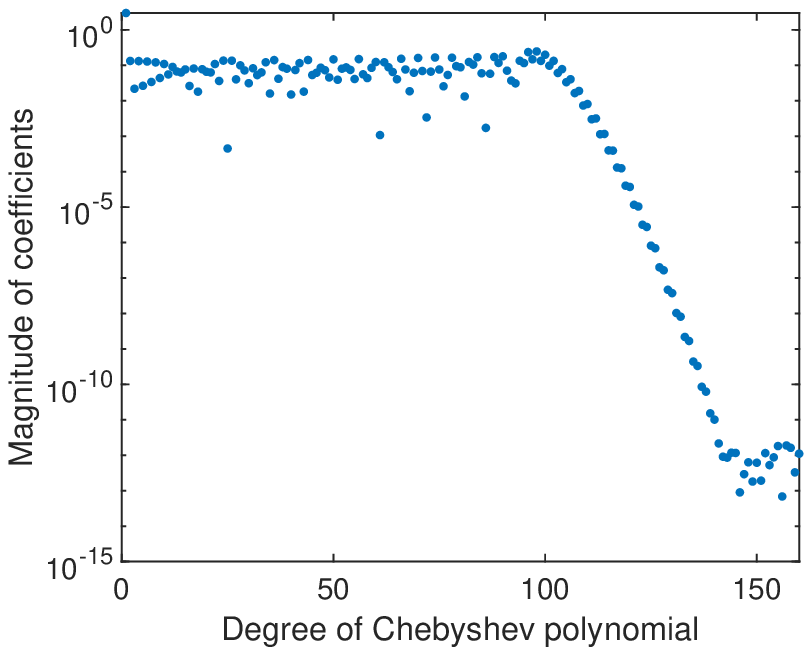}
\caption{\label{fig:Chebyshev1}Non-homogeneous 1D case for Gaussian correlation with $\sigma^2 = 10$ and $N = 10000$. }
\end{minipage}
\hspace{0.2cm} \centering \begin{minipage}[t]{0.45\linewidth}
\centering \vspace*{0in}
\includegraphics[width=\linewidth]{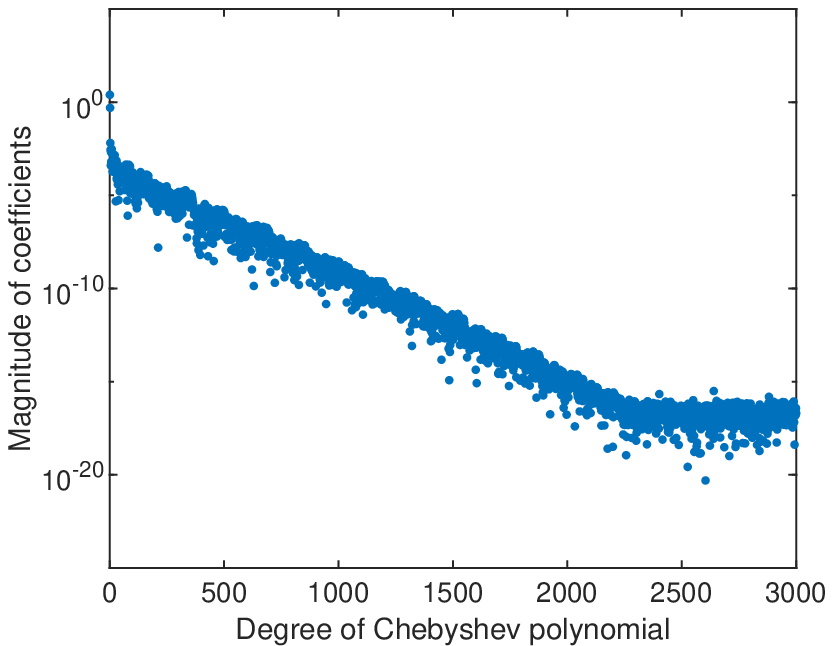}
\caption{\label{fig:Chebyshev2}Homogeneous 1D case for Gaussian correlation with $\sigma^2 = 0.1$ and $N = 100$.}
\end{minipage}
\end{figure}

To establish the necessary number of collocation points, we first investigate the structure of the numerical solution in the worst case scenario, ($N=10^4$, $\sigma^2=10$), for Gaussian correlation of the $\ln(K)$ field. In Fig.~\ref{fig:Chebyshev1} we represent the absolute values of the coefficients of the solution's expansion in the phase space, computed with the \emph{ fast Chebyshev transform}, versus the degree of the Chebyshev polynomial. We observe that for $n \approx 150$ Chebyshev points the coefficients reach the \emph{roundoff plateau}. Since, due to the accumulation of the truncation errors, large numbers of points could result in an increased error, the $l_\infty$ norm should be also computed to establish the optimal value of $n$. Note that the analysis differs from case to case. Therefore, we search for the lowest $l_\infty$ error for all combinations of parameters $N$ and $\sigma^2$ considered in this study, by varying $n$ between $140$ to $200$. The optimal numbers of points identified in this way for Gaussian and exponential correlation of the $\ln(K)$ field are presented in Tables~\ref{tab:SpectralGauss1D_optimal_N} and \ref{tab:SpectralExponential1D_optimal_N}, respectively.

The results of verification tests by comparisons with the manufactured solution (\ref{Flow:Eq1}) are presented in Tables~\ref{tab:SpectralGauss1D_optimal} and \ref{tab:SpectralExponential1D_optimal}. The $l_\infty$ errors correspond to numerical CSM solutions for different combinations of parameters ($N$, $\sigma^2$) computed with the optimal numbers of points from Tables~\ref{tab:SpectralGauss1D_optimal_N} and \ref{tab:SpectralExponential1D_optimal_N}.
The computation time varies between $0.05$ seconds in the test simplest case, ($N=10^2$, $\sigma^2=0.1$), and $1.7$ seconds in the worst case, ($N=10^4$, $\sigma^2=10$).

\begin{table}[!ht]
\centering
\captionof{table}{Gaussian correlation: 1D case. Estimation of optimal $n$.}
\label{tab:SpectralGauss1D_optimal_N}
\begin{tabular}{|| c || c | c | c | c | c | c | c ||}
\hline\hline
\diagbox[width=3em]{N}{$\sigma^2$} & 0.1 & 1 & 2 & 4 & 6 & 8 & 10  \\ \hline
$10^2$   & 150 &  161 &  171 &  167 &  179 &  155 &  146 \\ \hline
$10^3$  & 193 &  161 &  158 &  150 &  192 &  152 &  156 \\ \hline
$10^4$ & 183 &  177 &  154 &  175 &  150 &  169 &  176 \\ \hline
\hline
\end{tabular}
\end{table}

\begin{table}[!ht]
\centering
\captionof{table}{Exponential correlation: 1D case. Estimation of optimal $n$.}
\label{tab:SpectralExponential1D_optimal_N}
\begin{tabular}{|| c || c | c | c | c | c | c | c ||}
\hline\hline
\diagbox[width=3em]{N}{$\sigma^2$} & 0.1 & 1 & 2 & 4 & 6 & 8 & 10  \\ \hline
$10^2$   &   166 &  152 &  167 &  189 &  156 &  155 &  141 \\ \hline
$10^3$  &   182 &  159 &  165 &  176 &  146 &  176 &  164 \\ \hline
$10^4$ &   181 &  156 &  149 &  181 &  164 &  175 &  164 \\ \hline
\hline
\end{tabular}
\end{table}

\begin{table}[!ht]
\centering
\captionof{table}{Gaussian correlation: 1D case. Errors obtained for optimal $n$.}\label{tab:SpectralGauss1D_optimal}
\begin{tabular}{|| c || c | c | c | c | c | c | c ||}
\hline\hline
\diagbox[width=3em]{N}{$\sigma^2$} & 0.1 & 1 & 2 & 4 & 6 & 8 & 10  \\ \hline
$10^2$   & 1.30e-12 &	 1.60e-12 &	 1.67e-12 &	 2.97e-12 &	 1.32e-11 &	 3.00e-11 &	 3.96e-11 \\ \hline
$10^3$  & 2.12e-12 &	 1.09e-12 &	 8.58e-13 &	 1.51e-12 &	 3.75e-12 &	 1.53e-11 &	 2.65e-11 \\ \hline
$10^4$ & 1.06e-12 &	 1.42e-12 &	 9.41e-13 &	 2.18e-12 &	 7.81e-12 &	 1.89e-11 &	 4.91e-11 \\ \hline
\hline
\end{tabular}
\end{table}

\begin{table}[!ht]
\centering
\captionof{table}{Exponential correlation: 1D case. Errors obtained for optimal $n$.}\label{tab:SpectralExponential1D_optimal}
\begin{tabular}{|| c || c | c | c | c | c | c | c ||}
\hline\hline
\diagbox[width=3em]{N}{$\sigma^2$} & 0.1 & 1 & 2 & 4 & 6 & 8 & 10  \\ \hline
$10^2$   & 1.11e-13 &	 1.03e-13 &	 1.99e-13 &	 6.50e-13 &	 1.88e-12 &	 9.77e-12 &	 2.07e-11 \\ \hline
$10^3$  & 3.75e-13 &	 1.09e-13 &	 2.16e-13 &	 1.18e-12 &	 2.56e-12 &	 9.48e-12 &	 2.63e-11 \\ \hline
$10^4$ & 7.15e-14 &	 1.05e-13 &	 1.85e-13 &	 1.70e-12 &	 4.21e-12 &	 3.91e-11 &	 1.44e-10 \\ \hline
\hline
\end{tabular}
\end{table}

We also investigate the structure of the solution and the convergence for the homogeneous 1D problem ($f=0$ in (\ref{eq2.8})) for Gaussian correlation of the $\ln(K)$ field (see Fig.~\ref{fig:Chebyshev2}). We remark that at least $n=2500$ collocation points are needed to reach the roundoff plateau of the coefficients. The much larger value of $n$ is due to the spatial variability of the homogeneous solution which cannot be represented by relatively small $n$, as in case of the simple manufactured solution of the non-homogeneous problem. In absence of an analytical solution of the homogeneous boundary value problem, reaching the roundoff plateau is a guarantee that the numerical solution is highly accurate \cite{AurentzandTrefethen2017}.

In order to apply the Chebyshev collocation method in the 2D case, we reformulate the flow problem (\ref{eq2.2}-\ref{eq2.4}) into the square $[-1,1]^2$, similarly to the procedure used in the 1D case, and we transform the boundary conditions by following the approach of \cite{Hoepffner2007}.

The optimal number of Chebyshev points is established, similarly to the 1D case, by increasing $n$ from $44$ to $64$ in both directions and by computing the $l_\infty$ error for combinations of parameters $N$ and $\sigma^2$. The results are given in Tables~\ref{tab:SpectralGauss2D_optimal_N} and \ref{tab:SpectralExponential2D_optimal_N}, for Gaussian and exponential correlations, respectively.

Using the optimal $n$-values, we verify the 2D code by comparisons with the manufactured solution (\ref{Flow:Eq5}). The $l_\infty$ errors for combinations of parameters $N$ and $\sigma^2$ are given in Tables~\ref{tab:SpectralGauss2D_optimal} and \ref{tab:SpectralExponential2D_optimal}. The average execution time with a grid consisting of $64^2$ collocation points is about 10 seconds.
\begin{table}[!ht]
\centering
\captionof{table}{Gaussian correlation: 2D case. Estimation of optimal $n$.}\label{tab:SpectralGauss2D_optimal_N}
\begin{tabular}{|| c || c | c | c | c | c | c | c ||}
\hline\hline
\diagbox[width=3em]{N}{$\sigma^2$} & 0.1 & 1 & 2 & 4 & 6 & 8 & 10  \\ \hline
$10^2$   & 49 &	 52 &	 50 &	 60 &	 64 &	 45 &	 46 \\ \hline
$10^3$  & 63 &	 51 &	 50 &	 53 &	 48 &	 47 &	 47 \\ \hline
$10^4$ & 49 &	 52 &	 58 &	 50 &	 52 &	 45 &	 48 \\ \hline
\hline
\end{tabular}
\end{table}

\begin{table}[!ht]
\centering
\captionof{table}{Exponential correlation: 2D case. Estimation of optimal $n$.}\label{tab:SpectralExponential2D_optimal_N}
\begin{tabular}{|| c || c | c | c | c | c | c | c ||}
\hline\hline
\diagbox[width=3em]{N}{$\sigma^2$} & 0.1 & 1 & 2 & 4 & 6 & 8 & 10  \\ \hline
$10^2$   & 56 &	 50 &	 48 &	 55 &	 44 &	 55 &	 50 \\ \hline
$10^3$  & 49 &	 48 &	 48 &	 48 &	 46 &	 48 &	 45 \\ \hline
$10^4$ & 61 &	 47 &	 46 &	 46 &	 56 &	 46 &	 46 \\ \hline
\hline
\end{tabular}
\end{table}

\begin{table}[!ht]
\centering
\captionof{table}{Gaussian correlation: 2D case. Errors obtained for optimal $n$.}\label{tab:SpectralGauss2D_optimal}
\begin{tabular}{|| c || c | c | c | c | c | c | c ||}
\hline\hline
\diagbox[width=3em]{N}{$\sigma^2$} & 0.1 & 1 & 2 & 4 & 6 & 8 & 10  \\ \hline
$10^2$   & 3.58e-13 &	 6.75e-13 &	 9.00e-12 &	 1.06e-10 &	 6.54e-10 &	 1.40e-09 &	 7.28e-09 \\ \hline
$10^3$  & 3.71e-13 &	 4.50e-12 &	 1.25e-11 &	 2.50e-10 &	 2.23e-09 &	 1.15e-08 &	 2.53e-08 \\ \hline
$10^4$ & 2.54e-13 &	 6.32e-13 &	 3.10e-12 &	 5.64e-11 &	 3.11e-10 &	 1.26e-09 &	 1.59e-09 \\ \hline
\hline
\end{tabular}
\end{table}

\begin{table}[!ht]
\centering
\captionof{table}{Exponential correlation: 2D case. Errors obtained for optimal $n$.}\label{tab:SpectralExponential2D_optimal}
\begin{tabular}{|| c || c | c | c | c | c | c | c ||}
\hline\hline
\diagbox[width=3em]{N}{$\sigma^2$} & 0.1 & 1 & 2 & 4 & 6 & 8 & 10  \\ \hline
$10^2$   & 8.11e-12 &	 2.15e-11 &	 7.24e-11 &	 3.58e-10 &	 1.00e-09 &	 4.95e-09 &	 3.46e-09 \\ \hline
$10^3$  & 1.63e-11 &	 3.42e-11 &	 3.38e-11 &	 2.47e-10 &	 1.67e-09 &	 5.93e-09 &	 5.75e-09 \\ \hline
$10^4$ & 2.15e-12 &	 1.52e-11 &	 5.16e-11 &	 2.76e-10 &	 1.19e-09 &	 4.31e-09 &	 3.87e-09 \\ \hline
\hline
\end{tabular}
\end{table}

\begin{figure}[h]
\centering \begin{minipage}[t]{0.45\linewidth} \centering
\vspace*{0in}
\includegraphics[width=\linewidth]{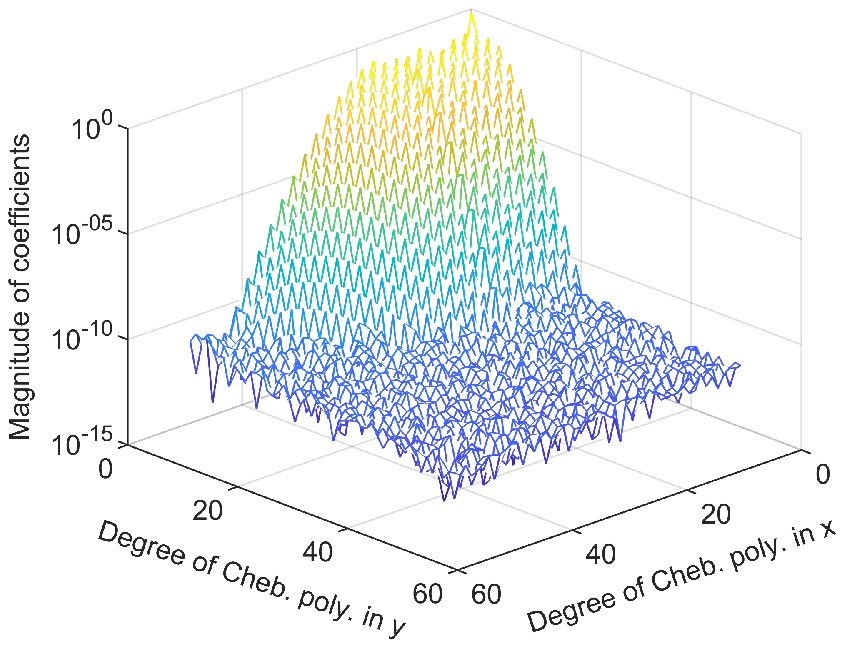}
\caption{\label{fig:Chebyshev_coeff_2d_Gauss}Non-homogeneous 2D case for Gaussian correlation with $\sigma^2 = 10$ and $N = 10000$. }
\end{minipage}
\hspace{0.2cm}
\centering \begin{minipage}[t]{0.45\linewidth}
\centering \vspace*{0in}
\includegraphics[width=\linewidth]{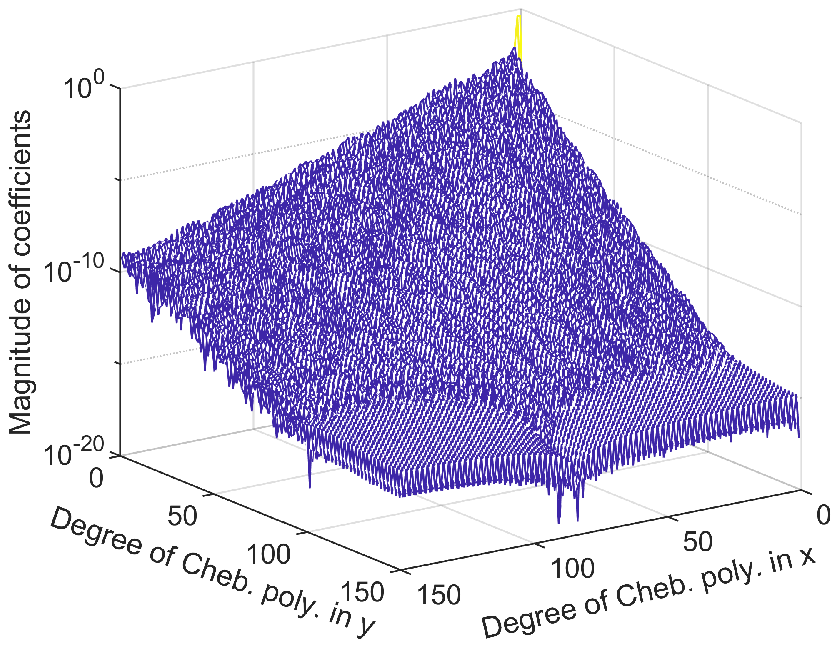}
\caption{\label{fig:Chebyshev_coeff_2d_Gauss_homogen}Homogeneous 2D case for Gaussian correlation with $\sigma^2 = 0.1$ and $N = 100$.}
\end{minipage}
\end{figure}

We conclude this section with a discussion about the convergence and the necessary number of Chebyshev points. The dependence of the spectral expansion of the solution on the degree of the Chebyshev polynomial presented in Fig.~\ref{fig:Chebyshev_coeff_2d_Gauss} indicate that in case of non-homogenous 2D problem the number of about $n_x=50$ collocation points in $x$-direction is about two times larger than the number of points in $y$-direction necessary to reach the roundoff plateau. Therefore, choosing $n_x=n_y=50$ points in both directions ensures the high accuracy of the solutions shown in Tables~\ref{tab:SpectralGauss2D_optimal} and \ref{tab:SpectralExponential2D_optimal}. In case of the homogeneous 2D problem, the roundoff plateau is reached with about $n_y=130$ points in $y$-direction but a significantly larger number $n_x$ of points would be necessary in $x$-direction (see Fig.~\ref{fig:Chebyshev_coeff_2d_Gauss_homogen}). Since, according to the discretization scheme (\ref{matrixeq}), the dimension of the problem increases as $n_{x}^{2}\times n_{y}^{2}$ solving homogeneous 2D problems with CSM schemes requires much larger memory usage than finite difference/element schemes.

\subsection{Chebyshev-Galerkin spectral method for 1D case}
The 1D flow equation in (\ref{eq2.8}) is a particular case of that considered in the GSM approach of \cite{Shen2016},
\begin{equation*}\label{Shen_problem}
\alpha(x)u-(\beta(x)u')' = f(x), \quad x \in (a,b),
\end{equation*}
corresponding to $\alpha(x) \equiv 0$ and $\beta(x) = -K(x)$. In the following, we use the Chebyshev-Galerkin spectral method of \cite{Shen2016} to solve the 1D test cases for the manufactured solution (\ref{Flow:Eq1}). Unlike the CSM schemes used in the previous section, the GSM approach avoids the need to compute the derivative of $K$, by using the weak formulation of the problem.
% and, and can be applided to more general situations where an analytical form of the $K$ field is not available.

\begin{table}[!ht]
\centering
\captionof{table}{Gaussian correlation: 1D case. Errors of the GSM scheme for $n = 4500$ and $L = 200 $.}
\label{tab:Galerkin_Gauss_1D}
\begin{tabular}{|| c || c | c | c | c | c | c | c ||}
\hline\hline
\diagbox[width=3em]{N}{$\sigma^2$} & 0.1 & 1 & 2 & 4 & 6 & 8 & 10  \\ \hline
$10^2$ & 8.87e-12 &   3.46e-11 &	 2.45e-10 &	 1.81e-09 &	 1.11e-08 &	 7.38e-08 &	 2.96e-07  \\ \hline
$10^3$ & 1.05e-11 &	 4.53e-11 &	 2.32e-10 &	 3.43e-09 &	 2.48e-08 &	 5.28e-08 &	 2.32e-07 \\ \hline
$10^4$ & 8.36e-12 &	 6.34e-11 &	 4.07e-10 &	 2.75e-09 &	 3.10e-08 &	 5.87e-08 &	 1.25e-07 \\ \hline
\hline
\end{tabular}
\end{table}

\begin{table}[!ht]
\centering
\captionof{table}{Exponential correlation: 1D case.  Errors of the GSM scheme for $n = 4500$, $L  = 200 $.}
\label{tab:Galerkin_Exp_1D_non_hom_big_L}
\begin{tabular}{|| c || c | c | c | c | c | c | c ||}
\hline\hline
\diagbox[width=3em]{N}{$\sigma^2$} & 0.1 & 1 & 2 & 4 & 6 & 8 & 10  \\ \hline
$10^2$ & 2.55e+02 &	 5.75e+03 &	 3.70e+04 &	 4.17e+05 &	 3.01e+06 &	 1.47e+07 &	 8.49e+07 \\ \hline
$10^3$ & 1.59e+02 &	 3.07e+03 &	 1.42e+04 &	 1.66e+05 &	 3.12e+06 &	 2.68e+07 &	 1.69e+08 \\ \hline
$10^4$ & 9.11e+03 &	 1.19e+05 &	 2.74e+05 &	 4.59e+06 &	 2.78e+07 &	 1.32e+08 &	 5.42e+08
\\ \hline
\hline
\end{tabular}
\end{table}

\begin{table}[!ht]
\centering
\captionof{table}{Exponential correlation: 1D case. Errors of the GSM scheme for $n = 4500$, $L  = 10 $.}
\label{tab:Galerkin_Exp_1D_non_hom_smaller_L}
\begin{tabular}{|| c || c | c | c | c | c | c | c ||}
\hline\hline
\diagbox[width=3em]{N}{$\sigma^2$} & 0.1 & 1 & 2 & 4 & 6 & 8 & 10  \\ \hline
$10^2$ &   5.43e-11 &	 3.12e-08 &	 1.01e-06 &	 4.55e-05 &	 4.46e-04 &	 2.39e-03 &	 9.54e-03 \\ \hline
$10^3$ &  2.59e-12 &	 3.74e-10 &	 5.10e-09 &	 1.89e-07 &	 2.27e-06 &	 1.65e-05 &	 9.28e-05  \\ \hline
$10^4$ & 4.36e+01 &	 2.42e+02 &	 7.58e+02 &	 3.91e+03 &	 1.30e+04 &	 3.49e+04 &	 8.26e+04 \\ \hline
\hline
\end{tabular}
\end{table}

The $l_\infty$ errors of the GSM scheme used to solve the 1D problems with $n = 4500$ collocation points in the domain of length $L  = 200 $ specified in Section~\ref{problem} are presented in Tables~\ref{tab:Galerkin_Gauss_1D} and \ref{tab:Galerkin_Exp_1D_non_hom_big_L}. The average computing time for solving individual cases is of the order of 20 seconds. We can see that, while in case of Gaussian correlation the GSM approximation is almost as accurate as for CSM approximations, in the exponential case the errors are extremely large, from about $10^2$ to the order $10^8$.

By restricting the length of the domain to $L=10$, in exponential case, the manufactured solution is approximated with high accuracy for all values of $\sigma^2$ and $N\leq 10^3$ but the GSM fails to reproduce the exact solution, with large errors again, for all $\sigma^2$ in case of the largest number of modes, $N=10^3$. The situation is somewhat similar to that presented by \cite{Gotovacetal2009}, where the possibility to obtain accurate approximations for high resolution of the $\ln(K)$ field and large $\sigma^2$ is limited by the need to increase the solution's resolution beyond the limits of the available computing resources.

\section{Global random walk method}
\label{grw}

\subsection{One-dimensional case}
\label{grw1D}

GRW algorithms solve parabolic partial differential equations by moving computational particles on regular lattices \cite{Suciu2014,Suciu2019,Vamosetal2003}. Solutions of the 1D elliptic equation for the hydraulic head with time independent boundary conditions can be obtained by solving the associated nonstationary equation with GRW algorithms on staggered grids (C. Vamo\c{s}, 2016, unpublished). Following this idea, we look for a solution of the flow problem (\ref{eq2.8}) given by the explicit steady-state limit of the solution of the explicit staggered FD scheme
\begin{equation*}
\frac{1}{a\Delta t}[h(i,k+1)-h(i,k)]=\frac{1}{{\Delta x}^2}\{[K(i+1/2,k)(h(i+1,k)-h(i,k))]-[K(i-1/2,k)(h(i,k)-h(i-1,k))]\}-f,
\end{equation*}
where $a$ is a constant equal to a unit length.

With the hydraulic head further approximated by the distribution $n(i,k)$ at lattice sites $i$ and times $k$ of a system of $\mathcal{N}$ random walkers, $h(i\Delta x,k\Delta t)\approx n(i,k)a/\mathcal{N}$, the staggered FD scheme becomes
\begin{eqnarray}\label{eq6.1}
n(i,k+1)
&=&\{1-[r(i-1/2,k)+r(i+1/2,k)]\}n(i,k)\nonumber\\
& & + r(i-1/2,k)n(i-1,k)+r(i+1/2,k)n(i+1,k)-\lfloor \mathcal{N}f\Delta t\rfloor,
\end{eqnarray}
where $\lfloor \cdot\rfloor$ is the floor function and the dimensionless parameter $r=Ka\Delta t/{\Delta x}^2$, $r\leq 1$, defines jump probabilities. The contributions to lattice sites $i$ from neighboring sites summed up in (\ref{eq6.1}) are obtained with the GRW algorithm which moves particles from sites $j$ to neighboring sites $i=j\mp 1$ according to
\begin{eqnarray}\label{eq6.2}
n(j,k)=\delta(j,j,k)+\delta(j-1,j,k)+\delta(j+1,j,k).
\end{eqnarray}
For consistency with the staggered scheme (\ref{eq6.1}), the stochastic averages of the quantities $\delta$ in (\ref{eq6.2}) verify
\begin{equation}\label{eq6.3}
\overline{\delta(j,j,k)}=\{1-[r(j-1/2)+r(j+1/2)]\}\overline{n(j,k)}\;, \quad
\overline{\delta(j\mp 1,j,k)}=r(j\mp 1/2)\overline{n(j)}.
\end{equation}
The numbers $r(i\pm 1/2,k)n(i,k)$ of jumping particles are binomial random variables. They are efficiently approximated by summing reminders of multiplication by $r$ and of the floor function applied to $a\mathcal{N}f\Delta t$ and by allocating a particle to the lattice site where the sum of reminders reaches the unity. The steady-state solution corresponds to a constant difference of incoming and outgoing flux of particles \cite[Sect. 3.3.4.1]{Suciu2019}.

Giving up the particle indivisibility and representing $n$ by real numbers, one obtains a deterministic GRW algorithm, where the numbers of particles jumping left/right are exactly given by $r(i\pm 1/2,k)n(i,k)$. Preliminary tests indicate that the GRW algorithm with particles requires fewer time iterations to reach the steady state, but the deterministic version, which preforms fewer arithmetical operations per iteration step, is overall faster. Therefore, we used the deterministic GRW implemented in Matlab codes (with a single exception where a C++ code was used) to solve the 1D benchmark problems.

To solve the flow equation by the GRW algorithm (\ref{eq6.1}-\ref{eq6.3}), the boundary conditions in (\ref{eq2.8}) have to be completed by appropriate initial conditions. To test the stability and the accuracy of the algorithm by comparison with the analytical manufactured solutions (\ref{Flow:Eq1}), we use an initial condition given by the analytical solution itself. With this choice, the $L^{2}_{\Delta x}$ error norm first increases, then approaches a plateau, which corresponds to the steady state deviation from the manufactured solution.

For Gaussian correlation of $\ln(K)$, the number of iterations to approach the steady state, with a space step $\Delta x=10^{-1}$, should be at least of the order $T=10^8$ in the most difficult test case ($N=10^4$, $\sigma^2=10$). With these fixed values of $T$ and $\Delta x$, the computing time needed to estimate the $L^{2}_{\Delta x}$ error norms presented in Table~\ref{tab:grw1} was between 50 and 55 minutes. %32 and 55 minutes. %
By increasing the number of iterations to $T=5\cdot 10^8$, the computing time increases to about 2.30 hours and the numerical solutions get closer to the stationary state (almost constant total flux of particles) in the cases ($N=10^3$, $\sigma^2=4$) and ($N=10^4$, $\sigma^2=10$), but the error norms, of $1.19\cdot 10^{-1}$ and $1.00\cdot 10^{-1}$, respectively, remain within the same order of magnitude as in Table~\ref{tab:grw1}.

\begin{table}[!ht]
    \centering
    \captionof{table}{Numerical errors of 1D GRW solutions for Gaussian correlation of the $\ln(K)$ field.}
    \label{tab:grw1}
    \begin{tabular}{|| c || c | c | c | c | c | c | c || c ||}
\hline\hline
\diagbox[width=3em]{$N$}{$\sigma^2$} & 0.1 & 1 & 2 & 4 & 6 & 8 & 10  \\ \hline
$10^2$ & 1.60e-2 & 3.93e-2 & 5.42e-2 & 6.75e-2 & 7.43e-2 & 8.58e-2 & 9.33e-2 \\ \hline
$10^3$ & 1.28e-2 & 6.01e-2 & 7.71e-2 & 8.70e-2 & 9.70e-2 & 1.15e-1 & 1.19e-1 \\ \hline
$10^4$ & 1.46e-2 & 3.85e-2 & 7.17e-2 & 8.04e-2 & 8.07e-2 & 7.81e-2 & 8.29e-2 \\ \hline\hline
\end{tabular}
\end{table}

In case of the exponential correlation of $\ln(K)$, a finer discretization, with $\Delta x=10^{-3}$ is required to obtain acceptable $L^{2}_{\Delta x}$ errors in the less favorable cases. Since the increase of the number of grid points, $L/\Delta x+1$, leads to very large computation times, numerical GRW solutions were computed only for the three cases presented in Table~\ref{tab:grw2}.
\begin{table}[!ht]
    \centering
    \captionof{table}{Numerical errors of 1D GRW solutions for \\
    exponential correlation of the $\ln(K)$ field.}
    \label{tab:grw2}
    \begin{tabular}{|| c || c | c | c | c | c | c | c || c ||}
\hline\hline
\diagbox[width=3em]{$N$}{$\sigma^2$} & 0.1 & 4 & 10  \\ \hline
$10^2$ & 3.45e-5 &          &          \\ \hline
$10^3$ &          & 4.39e-5 &          \\ \hline
$10^4$ &          &          & 6.06e+0 \\ \hline\hline
\end{tabular}
\end{table}

The easiest exponential case ($N=10^2$, $\sigma^2=0.1$) required $T=2\cdot 10^7$ time iterations and lasted 17.41 hours, the case of intermediate difficulty ($N=10^3$, $\sigma^2=4$) was solved with $T=5\cdot 10^7$ iterations in 30.74 hours, and in the most difficult case ($N=10^4$, $\sigma^2=10$) with $T=5\cdot 10^7$ iterations in 41.46 hours. In the latter case the error is of the order of the maximum value of the analytical solution (\ref{Flow:Eq1})) but, however, much smaller than the errors of order $10^8$ obtained for the same exponential 1D case by FDM, FEM, and GSM approaches. A small error norm of the order $10^{-2}$, was obtained in this case with $\Delta x=4\cdot 10^{-4}$ and $T=10^8$ iterations by consuming a huge amount of computing time of 7.31 days, with a C++ GRW code which is about 1.5 faster than the Matlab code.

The increased robustness of the GRW solutions is obtained with the price of a strong increase of the computing time, mainly when solving non-homogeneous problems. For comparison, accurately constant differences of incoming and outgoing flux of particles, indicating steady-state solutions of the homogeneous problem ($f=0$ in Eq.~\ref{eq2.8}), are obtained for the same combinations of parameters as in Table~\ref{tab:grw2} and for $\Delta x=10^{-1}$ with $T=2\cdot 10^7,\;\;5\cdot 10^7,$ and $5\cdot 10^9$ iterations in 4.51 min, 11.54 min, and 24.09 hours, respectively.

The estimated order of convergence (\ref{FDM_eqEOC}) was computed from solutions of the homogeneous problem for successively halved space steps from $\Delta x=10^{-2}$ to $\Delta x=3.125\cdot 10^{-4}$, for both Gaussian and exponential 1D cases. For these estimations, the initial conditions for the homogeneous flow problem was given by the mean slope of the hydraulic head and the number of iteration was fixed to $T=2\cdot 10^{4}$. The EOC shown in Tables~\ref{tab:grw3}~and~\ref{tab:grw4} has an increasing trend, with smaller values in the exponential case, without situations of non-monotonous behavior of the error norm (which causes the occurrence of negative EOC).

\begin{table}[!ht]
\centering
\caption{Computational order of convergence of the 1D GRW algorithm for Gaussian correlation of the $\ln(K)$ field.}
\label{tab:grw3}
\begin{tabular}{ |c|c|c|c|c|c|c|c|c|c|c|c|c| }
\hline\hline
$N$ & $\sigma^2$ & $\varepsilon_1$ & EOC & $\varepsilon_2$ & EOC &$\varepsilon_3$ & EOC & $\varepsilon_4$ & EOC & $\varepsilon_5$ \\
\hline
\multirow{3}{1em}{$10^2$} & $0.1$ & 5.74e-2 & 1.51 & 2.02e-2 & 1.79 & 5.83e-3 & 2.00 & 1.46e-3 & 2.30 & 2.96e-4\\
& $4$ & 3.63e-2 & 1.29 & 1.48e-2 & 1.61 & 4.86e-3 & 1.89 & 1.31e-3 & 2.26 & 2.74e-4\\
& $10$ & 2.59e-2 & 1.13 & 1.18e-2 & 1.47 & 4.25e-3 & 1.81 & 1.21e-3 & 2.23 & 2.58e-4\\
\hline\hline
\multirow{3}{1em}{$10^3$} & $0.1$ & 5.67e-2 & 1.43 & 2.11e-2 & 1.75 & 6.26e-3 & 1.99 & 1.58e-3 & 2.29 & 3.22e-4\\
& $4$ & 3.44e-2 & 1.38 & 1.32e-2 & 1.57 & 4.45e-3 & 1.83 & 1.25e-3 & 2.23 & 2.67e-4\\
& $10$ & 2.12e-2 & 1.15 & 9.53e-3 & 1.35 & 3.73e-3 & 1.69 & 1.16e-3 & 2.16 & 2.60e-4\\
\hline\hline
\multirow{3}{1em}{$10^4$} & $0.1$ & 5.12e-2 & 1.42 & 1.91e-2 & 1.78 & 5.58e-3 & 1.99 & 1.40e-3 & 2.30 & 2.84e-4\\
& $4$ & 2.48e-2 & 1.25 & 1.04e-2 & 1.57 & 3.51e-3 & 1.87 & 9.62e-4 & 2.25 & 2.02e-4\\
& $10$ & 1.80e-2 & 1.00 & 8.98e-3 & 1.39 & 3.43e-3 & 1.79 & 1.00e-3 & 2.21 & 2.16e-4\\
\hline\hline
\end{tabular}
\end{table}

\begin{table}[!ht]
\centering
\caption{Computational order of convergence of the 1D GRW algorithm for exponential correlation of the $\ln(K)$ field.}
\label{tab:grw4}
\begin{tabular}{ |c|c|c|c|c|c|c|c|c|c|c|c|c| }
\hline\hline
$N$ & $\sigma^2$ & $\varepsilon_1$ & EOC & $\varepsilon_2$ & EOC &$\varepsilon_3$ & EOC & $\varepsilon_4$ & EOC & $\varepsilon_5$ \\
\hline
\multirow{3}{1em}{$10^2$} & $0.1$ & 4.60e-2 & 1.26 & 1.92e-2 & 1.35 & 7.52e-3 & 1.44 & 2.77e-3 & 1.81 & 7.92e-4\\
& $4$ & 1.66e-2 & 1.04 & 8.09e-3 & 1.20 & 3.51e-3 & 1.35 & 1.38e-3 & 1.71 & 4.23e-4\\
& $10$ & 7.94e-3 & 1.04 & 3.87e-3 & 1.09 & 1.82e-3 & 1.20 & 7.94e-4 & 1.64 & 2.55e-4\\
\hline\hline
\multirow{3}{1em}{$10^3$} & $0.1$ & 4.66e-2 & 1.22 & 2.00e-2 & 1.31 & 8.09e-3 & 1.48 & 2.90e-3 & 1.84 & 8.09e-4\\
& $4$ & 2.14e-2 & 1.15 & 9.65e-3 & 1.16 & 4.33e-3 & 1.32 & 1.74e-3 & 1.81 & 4.97e-4\\
& $10$ & 1.07e-2 & 1.01 & 5.33e-3 & 0.87 & 2.91e-3 & 1.15 & 1.31e-3 & 1.76 & 3.86e-4\\
\hline\hline
\multirow{3}{1em}{$10^4$} & $0.1$ & 4.34e-2 & 1.20 & 1.89e-2 & 1.35 & 7.44e-3 & 1.52 & 2.60e-3 & 1.88 & 7.04e-4\\
& $4$  & 1.75e-2 & 1.22 & 7.53e-3 & 1.39 & 2.87e-3 & 1.55 & 9.83e-4 & 1.88 & 2.67e-4\\
& $10$ & 8.88e-3 & 1.19 & 3.88e-3 & 1.40 & 1.47e-3 & 1.40 & 5.56e-4 & 1.71 & 1.70e-4\\
\hline\hline
\end{tabular}
\end{table}

\subsection{Two-dimensional case}
\label{grw2D}

The 2D staggered scheme similar to (\ref{eq6.1}) reads as
\begin{eqnarray}\label{eq_grw3}
n(i,j,k+1)
&=&\left[1-r(i-1/2,j)-r(i+1/2,j)-r(i,j-1/2)-r(i,j+1/2)\right]n(i,j,k)\nonumber\\
& & + r(i-1/2,j)n(i-1,j,k)+r(i+1/2,j)n(i+1,j,k)\nonumber\\
& & + r(i,j-1/2)n(i,j-1,k)+r(i,j+1/2)n(i,j+1,k)-\lfloor \mathcal{N}f\Delta t\rfloor,
\end{eqnarray}
where the dimensionless parameters are defined for $\Delta x=\Delta y$ by
\begin{equation}
r(i,j)=\frac{K(i\Delta x,j\Delta x)a\Delta t}{(\Delta x)^{2}}. \label{eq_grw2}
\end{equation}
The contributions to lattice sites $(i,j)$ from neighboring sites summed up in (\ref{eq_grw3}) are obtained with the GRW algorithm which moves particles from sites $(l,m)$ to first-neighbor sites according to
\begin{eqnarray}\label{eq_grw4}
n(l,m,k)&=&\delta n(l,m|l,m,k)\nonumber\\
& &+\delta n(l-1,m|l,m,k)+\delta n(l+1,m|l,m,k)\nonumber\\
& &+\delta n(l,m-1|l,m,k)+\delta n(l,m+1|l,m,k).
\end{eqnarray}
For consistency with the (\ref{eq_grw3}), the stochastic averages of the quantities $\delta$ in (\ref{eq_grw4}) verify
\begin{eqnarray}\label{eq_grw5}
&&\overline{\delta n(l,m,k)}
=\left[1-r(l-1/2,m)-r(l+1/2,m)-r(l,m-1/2)-r(l,m+1/2)\right]\overline{n(l,m,k)}\nonumber\\
&&\overline{\delta n(l\mp 1,m|l,m,k)}=r(l\mp 1/2,m)\overline{n(l,m,k)}\nonumber\\
&&\overline{\delta n(l,m\mp 1|l,m,k)}=r(l,m\mp 1/2)\overline{n(l,m,k)}.
\end{eqnarray}
The parameters $r$ defined by (\ref{eq_grw2}) are jump probabilities which, as follows from (\ref{eq_grw5}), are now restricted by $0\le r\le 1/4$.

Similarly to the 1D case, we found that, even though it requires more iterations, the deterministic GRW is overall faster. Therefore, in the following we use deterministic GRW algorithms implemented in Matlab or C++ codes.

To solve the flow equation by the GRW algorithm (\ref{eq_grw2}-\ref{eq_grw5}), the boundary conditions (\ref{eq2.3}) and (\ref{eq2.4}) have to be completed by appropriate initial conditions. The homogeneous problem ($f=0$) is solved for an initial condition given by the mean slope of the pressure head (determined by the Dirichlet boundary conditions). For comparison with analytical manufactured solutions, the inhomogeneous problem ($f\neq 0$) is solved for an initial condition given by the analytical solution itself.

\begin{figure}[!ht]
\centering \begin{minipage}[t]{0.45\linewidth} \centering
\vspace*{0in}
\includegraphics[width=\linewidth]{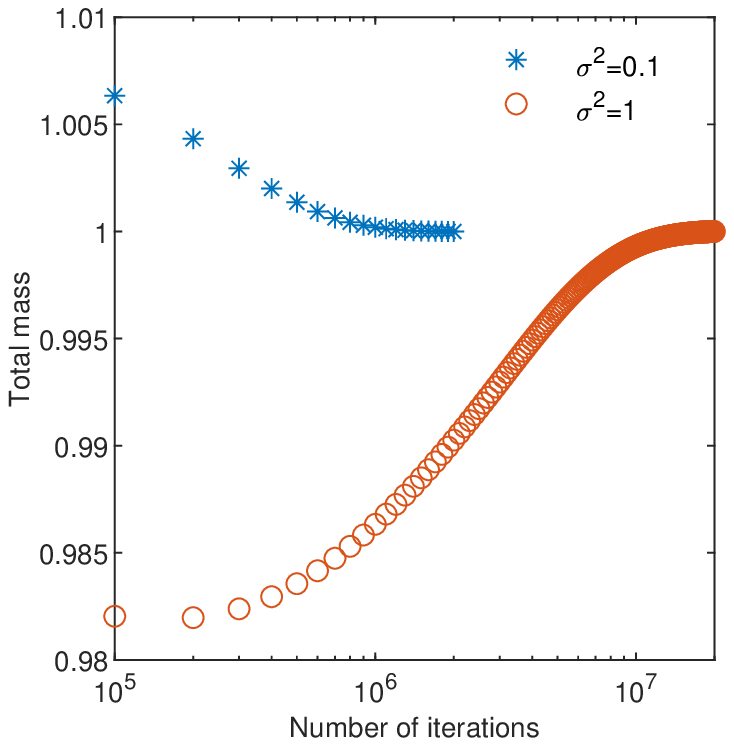}
\caption{\label{fig:conv_GRW_Gss}Convergence of the total mass of the GRW solution for $N=10^2$ modes and Gaussian correlation of the $\ln(K)$ field.}
\end{minipage}
\hspace{0.2cm} \centering \begin{minipage}[t]{0.45\linewidth}
\centering \vspace*{0in}
\includegraphics[width=\linewidth]{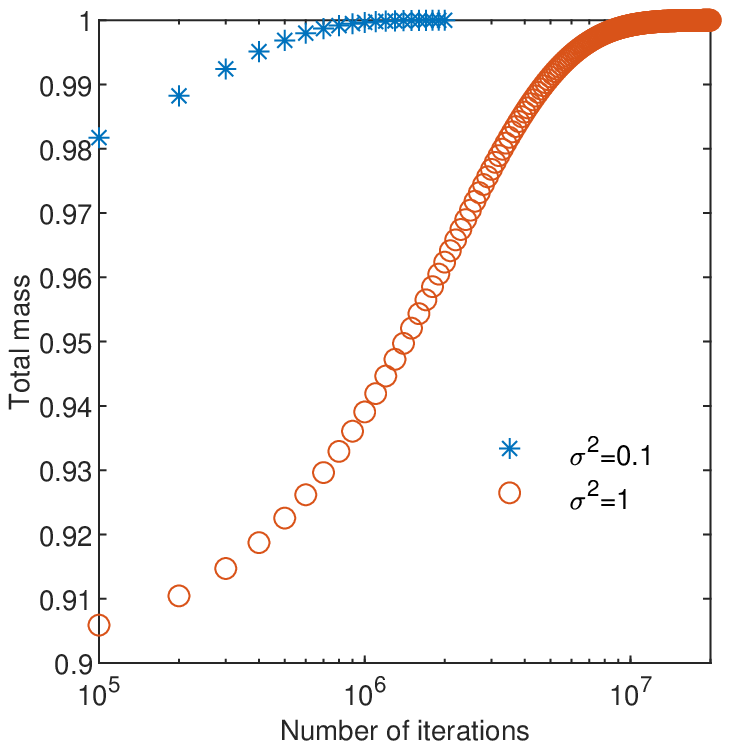}
\caption{\label{fig:conv_GRW_Exp}The same as in Fig.~\ref{fig:conv_GRW_Gss} for exponential correlation of the $\ln(K)$ field.}
\end{minipage}
\end{figure}

For an arbitrary initial condition, one obtains a transitory, non-stationary GRW solution which approaches the stationary solution after a number of time iterations depending on the  level of difficulty of the particular problem. The steady state is indicated by a constant number of particles or, in case of deterministic GRW, by the ``total mass'' defined as integral of the solution over the lattice, normalized by the final mass, as illustrated in Figs.~\ref{fig:conv_GRW_Gss}~and~\ref{fig:conv_GRW_Exp} for different solutions of the homogeneous problem.
The iterations needed to obtain stationary solutions require considerably large computing time (from 10 minutes for $2\cdot 10^6$ iterations to 2.5 hours for $2\cdot 10^7$ iterations). Faster computations can be designed by using a domain decomposition parallelization of each iteration.

Most of the computations were carried out with a Matlab implementation of the GRW code. For the comparisons with manufactured solutions in case of exponential correlation of the $\ln(K)$ field, which takes very long times, we use an executable mex-function of a C++ implementation of the flow solver. This procedure speeds up the computation by a factor of 1.5.

GRW solutions for code verification tests where computed on a coarse grid with $\Delta x=\Delta y=10^{-1}$ for Gaussian correlation and on the grid described in Section~\ref{problem}  with $\Delta x=\Delta y=2\cdot 10^{-2}$ for exponential correlation of the $\ln(K)$ field. The $L^2$ errors with respect to the manufactured solution (\ref{Flow:Eq5}) for Gaussian and exponential correlations are presented in Tables~\ref{tab:grw2_Gss}~and~\ref{tab:grw2_exp} below.  Since the time needed to reach the steady-state is considerably large, the GRW method is only tested for $N\leq 10^3$ and $\sigma^2\leq2$.

\begin{table}[!ht]
    \centering
    \captionof{table}{Numerical errors of 1D GRW solutions for \\
    Gaussian correlation of the $\ln(K)$ field.}
    \label{tab:grw2_Gss}
    \begin{tabular}{|| c || c | c | c | c | c | c | c || c ||}
\hline\hline
\diagbox[width=3em]{$N$}{$\sigma^2$} & 0.1 & 1 & 2  \\ \hline
$10^2$ &  3.16e-02 & 4.80e-02 & 1.35e-01 \\ \hline
$10^3$ &  6.81e-02 & 5.59e-01 & 1.78e+00 \\ \hline\hline
\end{tabular}
\end{table}

\begin{table}[!ht]
    \centering
    \captionof{table}{Numerical errors of 1D GRW solutions for \\
    exponential correlation of the $\ln(K)$ field.}
    \label{tab:grw2_exp}
    \begin{tabular}{|| c || c | c | c | c | c | c | c || c ||}
\hline\hline
\diagbox[width=3em]{$N$}{$\sigma^2$} & 0.1 & 1 & 2  \\ \hline
$10^2$ & 9.45e-02 & 7.57e-01 & 1.64e+00 \\ \hline
$10^3$ & 1.13e-01 & 1.28e+00 & 4.12e+00  \\ \hline\hline
\end{tabular}
\end{table}

As shown in Table~\ref{tab:grw2_Gss}, the GRW method solves the test problems in the Gaussian cases with $N=10^2$ with a good, or at least an acceptable accuracy. The increase of the number of modes to $N=10^3$ affects the accuracy of the GRW scheme, which for $\sigma^2=2$ produces errors larger that one (i.e., $50\%$ from the maximum values of the manufactured solution (\ref{Flow:Eq5})). The computing time for GRW solutions on the coarse grid ranges from a few minutes to a couple of hours.

GRW solutions in exponential cases can be obtained on the finer grid after considerable large numbers of iterations from $10^6$ to more than $10^8$, which require computing times from hours to days. Note that in the cases $N=10^3$ with $\sigma^2=1$ and $\sigma^2=2$ the GRW solutions are not yet strictly stationary and the corresponding $L^2$ values have to be regarded as lower bound errors. In these circumstances, accurate GRW solutions are feasible for $N=10^2$ and $\sigma^2\le 1$ or for $N=10^3$ and $\sigma^2=0.1$ (see Table \ref{tab:grw2_exp}).

Estimated orders of convergence are computed, similarly to the 1D case, from GRW solutions of the homogeneous problem obtained for successive refinements of the grid, from $\Delta x=\Delta y=10^{-1}$ to $\Delta x=\Delta y=3.125\cdot 10^{-3}$, according to formula~\ref{FDM_eqEOC}. EOC values are computed after the first temporal iteration of the GRW scheme, for $N=10^2$ and $N=10^3$.

One remarks that while in the Gaussian case, the EOC values are close to the theoretical convergence order 2 (Table~\ref{tab:Gauss2DgrwHom}), the the exponential case the order of convergence deteriorates to $1$ or $1.5$ in most cases (Table~\ref{tab:Exp2Dgrw3Hom}). Hence, the convergence behavior is influenced by the shape of the correlation function but there is no evidence for a dependence on $N$ and $\sigma^2$.
% EOC for 2D GRW:

\begin{table}[!ht]
\centering
\caption{Computational order of convergence of the GRW algorithm for Gaussian correlation of the $\ln(K)$ field.}
\label{tab:Gauss2DgrwHom}
\begin{tabular}{ |c|c|c|c|c|c|c|c|c|c|c|c|c| }
\hline\hline
$N$ & $\sigma^2$ & $\varepsilon_1$ & EOC & $\varepsilon_2$ & EOC & $\varepsilon_3$ & EOC & $\varepsilon_4$ & EOC & $\varepsilon_5$ \\
\hline
\multirow{3}{1.5em}{$10^2$} & $0.1$ & 2.80e-04 & 2.01 & 6.97e-05 & 2.02 & 1.72e-05 & 2.07 & 4.10e-06 & 2.32 & 8.20e-07\\
& $4$ & 2.78e-04 & 2.00 & 6.93e-05 & 2.02 & 1.71e-05 & 2.07 & 4.08e-06 & 2.32 & 8.15e-07\\
& $10$ & 2.84e-04 & 2.01 & 7.07e-05 & 2.01 & 1.75e-05 & 2.07 & 4.16e-06 & 2.32 & 8.32e-07\\
\hline\hline
\multirow{3}{1.5em}{$10^3$} & $0.1$ & 2.91e-04 & 2.01 & 7.25e-05 & 2.02 & 1.79e-05 & 2.07 & 4.27e-06 & 2.32 & 8.53e-07\\
& $4$ & 2.66e-04 & 2.00 & 6.63e-05 & 2.02 & 1.64e-05 & 2.07 & 3.90e-06 & 2.32 & 7.79e-07\\
& $10$ & 2.64e-04 & 2.00 & 6.58e-05 & 2.02 & 1.62e-05 & 2.07 & 3.87e-06 & 2.32 & 7.73e-07\\
\hline\hline
\end{tabular}
\end{table}

\begin{table}[!ht]
\centering
\caption{Computational order of convergence of the GRW algorithm for exponential correlation of the $\ln(K)$ field.}
\label{tab:Exp2Dgrw3Hom}
\begin{tabular}{ |c|c|c|c|c|c|c|c|c|c|c|c|c| }
\hline\hline
$N$ & $\sigma^2$ & $\varepsilon_1$ & EOC & $\varepsilon_2$ & EOC & $\varepsilon_3$ & EOC & $\varepsilon_4$ & EOC & $\varepsilon_5$ \\
\hline
\multirow{3}{1.5em}{$10^2$} & $0.1$ & 7.22e-04 & 1.49 & 2.57e-04 & 1.10 & 1.20e-04 & 1.01 & 5.97e-05 & 1.69 & 1.85e-05\\
& $4$ & 4.58e-04 & 2.34 & 9.07e-05 & 1.66 & 2.88e-05 & 1.18 & 1.27e-05 & 1.84 & 3.54e-06\\
& $10$ & 4.34e-04 & 2.62 & 7.06e-05 & 1.86 & 1.94e-05 & 0.96 & 1.00e-05 & 1.77 & 2.94e-06\\
\hline\hline
\multirow{3}{1.5em}{$10^3$} & $0.1$ & 4.04e-04 & 1.64 & 1.30e-04 & 1.40 & 4.91e-05 & 1.56 & 1.67e-05 & 1.74 & 4.99e-06\\
& $4$ & 3.90e-04 & 2.06 & 9.38e-05 & 2.15 & 2.12e-05 & 1.53 & 7.34e-06 & 2.10 & 1.71e-06\\
& $10$ & 4.72e-04 & 2.11 & 1.09e-04 & 2.54 & 1.88e-05 & 1.65 & 5.98e-06 & 2.12 & 1.38e-06\\
\hline\hline
\end{tabular}
\end{table}

\section{Statistical inferences}
\label{stat}

Ensembles of numerical solutions of the flow problem are often used for the purpose of validation of the numerical schemes through comparisons with first-order approximation results (e.g., \cite{Bellinetal1992,deDreuzyetal2007,SalandinandFiorotto1998,Srzicetal2013}). The latter are obtained by perturbation expansions truncated at the first order in $\sigma^2$, equivalent to a linearization of the flow problem (\ref{eq2.1}-\ref{eq2.4}) (e.g., \cite{Dagan1989,GelharandAxness1983}). Here, we consider first-order approximations of the velocity field numerically generated with a Kraichnan procedure, based on proportionality of the spectral densities of the random velocity and that of the $\ln(K)$ field \cite[Appendix A]{Suciuetal2016}. The number of random periodic modes used in the Kraichnan generators of the hydraulic conductivity and of the velocity field (in case of linear approximation) is set to $N=100$. Validation tests of the five schemes are performed for $\ln(K)$ fields with Gaussian and exponential correlations and a fixed small variance $\sigma^2=0.1$. The deviation from the predictions of the linear approximation is further investigated with the FDM, FEM, DGM, and GRW schemes for increasing $\sigma^2$ of 0.5, 1, 1.5, and 2. The number of realizations, in all cases investigated here, is fixed to $R=100$. Numerical investigations indicate that the chosen values of the parameters $N$ and $R$ ensure unbiased statistical estimations of the dispersion of a passive scalar transported over tens of correlation lengths in heterogeneous aquifers characterized by a small variance $\sigma^2=0.1$ of the $\ln(K)$ field \cite{Eberhardetal2007}.

\begin{figure}[h]%[tb]
\centering
\includegraphics[width=\linewidth]{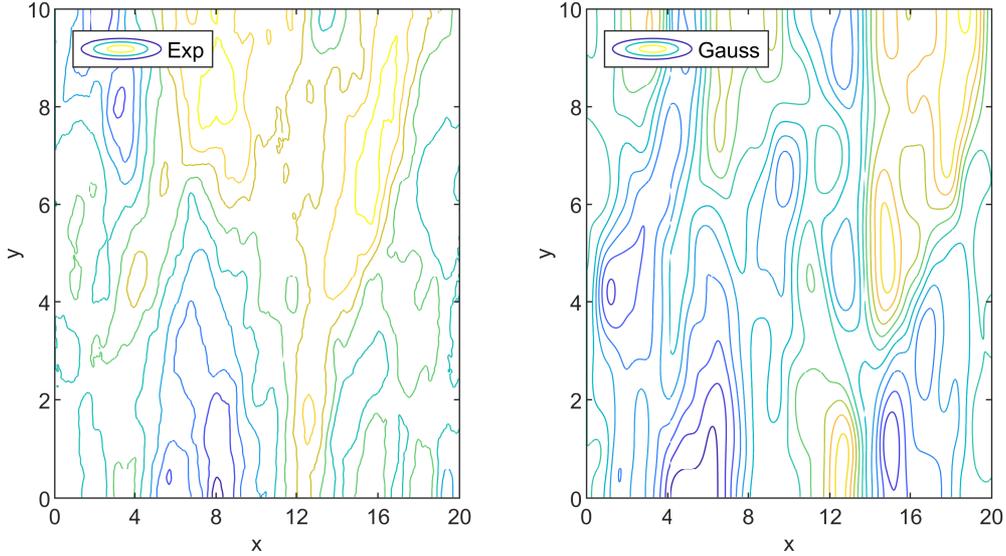}
\caption{\label{fig:statistics_contours_h} Contours of the hydraulic head fluctuations $\Delta h(x,y)$ computed with exponential
and Gaussian correlation of the $\ln(K)$ field.}
\end{figure}

\begin{figure}[h]%[tb]
\centering
\includegraphics[width=\linewidth]{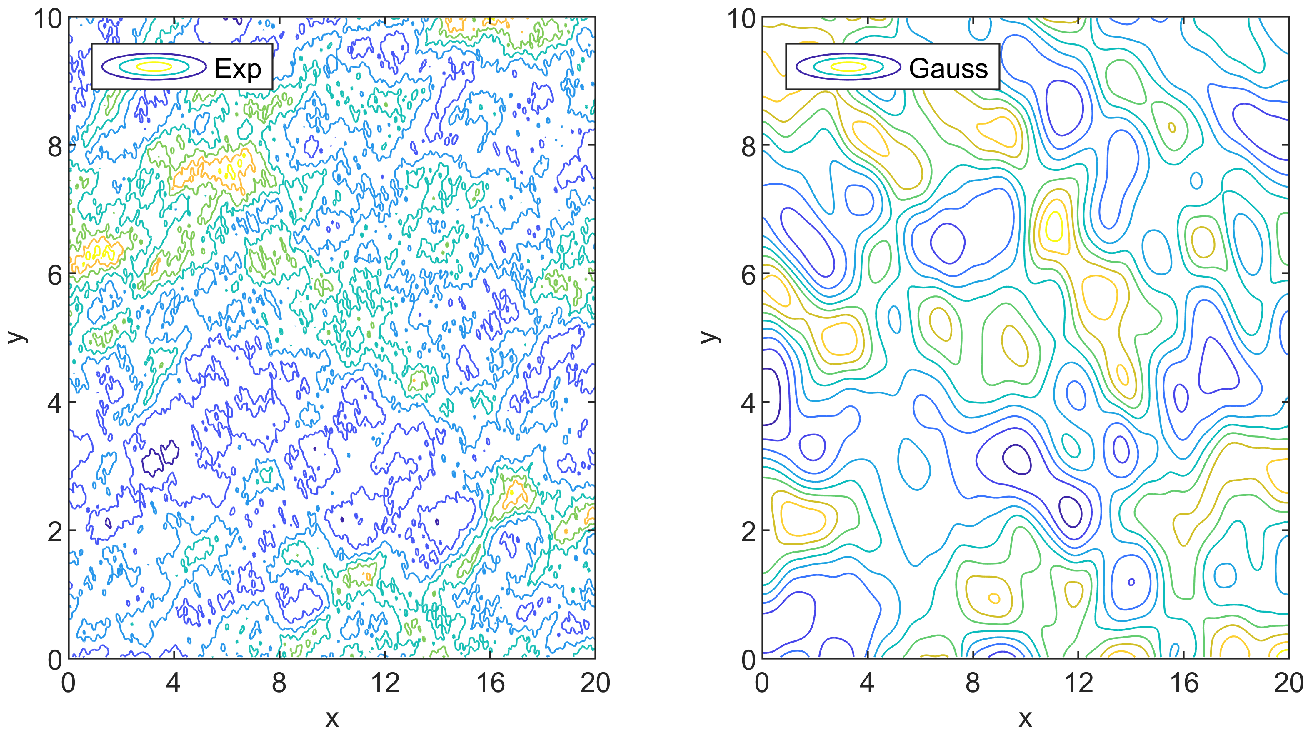}
\caption{\label{fig:statistics_contours_VxVy} Contours of absolute values of velocity
fields computed with exponential and Gaussian correlation of the $\ln(K)$ field.}
\end{figure}

The homogenous problem (\ref{eq2.1}-\ref{eq2.4}) is solved in the same two-dimensional domain, with $L_x=20$ and $L_y=10$, as that used for code verification tests in the previous section. The ensemble of realizations were computed for space steps $\Delta x=\Delta y=0.02$, with the FDM scheme and, to reduce the computational effort, for larger steps $\Delta x=\Delta y=0.05$, with the FEM and DGM schemes, and  $\Delta x=\Delta y=0.1$, with the GRW scheme. Constrained by the maximum Matlab array size allowed on the computer used to obtain CSM solutions, we use a limited number of 160 Chebyshev collocation points in both $x$- and $y$-directions, even though, as indicated by Fig.~\ref{fig:Chebyshev_coeff_2d_Gauss_homogen}, a larger number of points in $x$-direction is required to obtain accurate solutions. Sample solutions obtained with FDM are illustrated in Figs.~\ref{fig:statistics_contours_h}--\ref{fig:statistics_contours_VxVy}. FEM, DGM, and GRW solutions, not shown here, have a similar appearance.

Instead, among the realizations of the CSM solution for exponential correlation of the $\ln(K)$ field there also are unacceptable solutions with extremely large and irregular variations. Therefore, we selected an ensemble of 100 valid realizations by disregarding the ``bad'' solutions after the visual inspection of the larger initial ensemble of solutions, as illustrated in Appendix~\ref{appendixB}. The issue of ``bad realizations'' is common in situations where, due to computational constraints, numerical simulations have to be conducted on relatively coarse grids (we already encountered this issue during the preparation of the transport simulations presented in \cite{Raduetal2011,Suciuetal2013}).

Monte Carlo inferences of the mean and variance are obtained by computing averages over the ensembles of $R$ realizations followed by spatial averages \cite[Appendix B3]{Suciuetal2006}. Error bounds of the ensemble-space averages are given by standard deviations of the ensemble averages at grid points, estimated by spatial averages. The spatial averaging domain is an inner region of the computational domain unaffected by the boundary conditions approximately delimited by visual inspection of the variances estimated on longitudinal and transverse sections through the computation domain presented in Appendix~\ref{appendixC} (see also \cite{Bellinetal1992,KurbanmuradovandSabelfeld2010}.

Theoretical investigations indicate that the head variance is much smaller than that of the hydraulic conductivity. Spectral perturbation approaches for flow in two- and three-dimensions predict variances of the order $\sigma^{2}_{h}\sim (\sigma\lambda J)^2$ \cite{Ababouetal1989,Bakretal1978,Mizelletal1982}. The estimated variances presented in Table~\ref{tab:stat_h_1stOrder} agree with the spectral theory, which for our numerical parameters predict $\sigma^{2}_{h}\sim 10^{-4}$.

The first-order approximate relations between the variance of the
velocity components, $\sigma^{2}_{V_{x}}$ and $\sigma^{2}_{V_{y}}$,
and the variance $\sigma^2$ of the $\ln(K)$ field are given by
$\sigma^{2}_{V_{x}}=\frac{3}{8}\sigma^2U$ and
$\sigma^{2}_{V_{y}}=\frac{1}{8}\sigma^2U$, respectively \cite[Eq.
(3.7.27)]{Dagan1989}. Note that these relations do not depend on the shape of the correlation function of the $\ln(K)$ field. Since, as  mentioned in Section~\ref{problem}, for the purpose of statistical inferences we use dimensionless velocities, $U=1$. Then, with the smallest variance of the
$\ln(K)$ field considered in this study, $\sigma^2=0.1$, the velocity variances predicted by the linear theory are $\sigma^{2}_{V_{x}}=0.0375$ and
$\sigma^{2}_{V_{y}}=0.0125$, respectively.

The estimated mean values and variances of the velocity components are presented in Tables~\ref{tab:stat_1stOrder_Gss}~and~\ref{tab:stat_1stOrder_Exp}. FDM, FEM, DGM, and GRW results are generally close to theoretical first-order predictions within a range of less than 10\% in case of FDM, FEM, DGM, and GRW approaches, excepting FEM estimated variance of the transverse velocity in the Gaussian case which deviates with about 30\%. CSM results show deviations of about 30\%, for the variance of the transverse velocity in the Gaussian case, and deviations even larger than 100\% for both velocity variances, in the exponential case. These differences can be again attributed to the limited number of collocation points used in the CSM approach.

\begin{figure}[!ht]
\centering \begin{minipage}[t]{0.45\linewidth} \centering
\vspace*{0in}
\includegraphics[width=\linewidth]{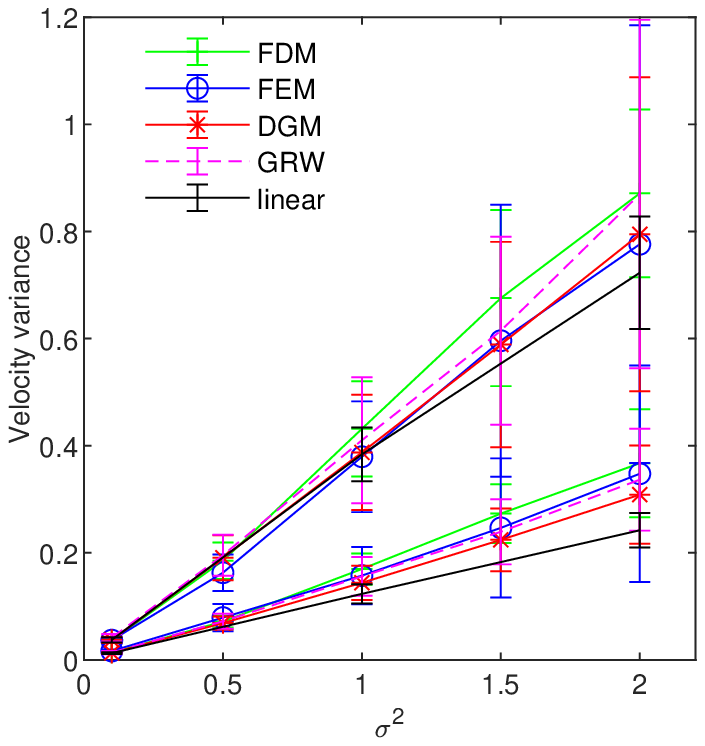}
\caption{\label{fig:vel_variance_Gauss}Comparison of the velocity variances resulted from FDM, FEM, DGM, GRW, and linear approximations simulations for Gaussian correlation of the $\ln(K)$ field. Upper branches correspond to longitudinal components and lower branches correspond to transverse components.}
\end{minipage}
\hspace{0.2cm} \centering \begin{minipage}[t]{0.45\linewidth}
\centering \vspace*{0in}
\includegraphics[width=\linewidth]{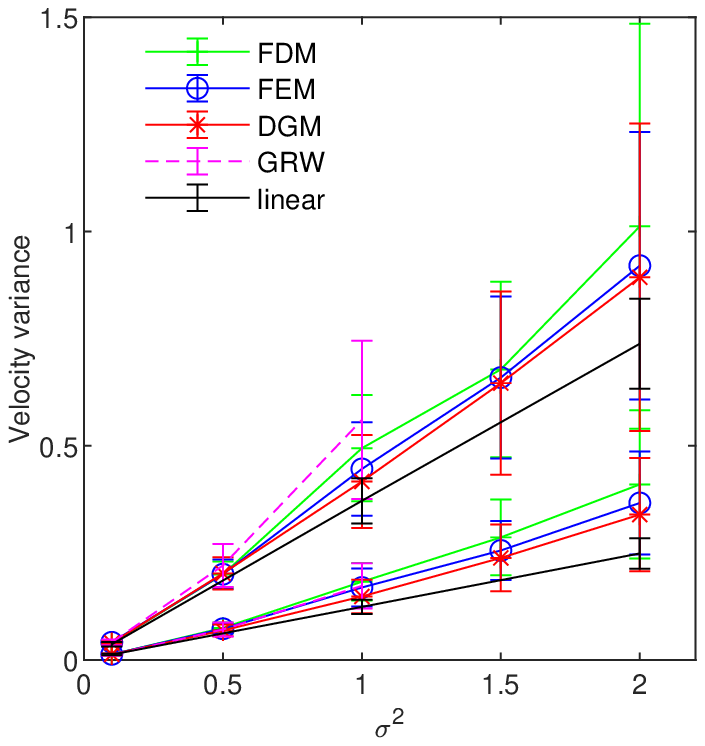}
\caption{\label{fig:vel_variance_Exp}The same as in Fig.~\ref{fig:vel_variance_Gauss}, for exponential correlation of the $\ln(K)$ field. GRW results are provided only for $\sigma^2\leq 1$.}
\end{minipage}
\end{figure}

The results obtained with the FDM, FEM, DGM, and GRW approaches for larger $\sigma^2$ are shown in Figs.~\ref{fig:vel_variance_Gauss}~and~\ref{fig:vel_variance_Exp}. We remark that the velocity variances progressively increase above the trend predicted by the linear theory, similarly to the Monte Carlo results presented in the literature (e.g., \cite{Bellinetal1992,KurbanmuradovandSabelfeld2010,SalandinandFiorotto1998}).
The differences between the results obtained with the three methods remain within the estimated error bounds determined by the present numerical setup (number of realizations, resolution, and dimension of the computational domain). For $\sigma^2\leq 0.5$ the estimated variances practically coincide with the first-order predictions. Within this range, velocity fields can be quite well approximated by the linearized solution of the flow problem and simulations of solute transport in groundwater can be carried out with Kraichnan generated velocity fields at lower computational costs \cite{Suciuetal2006,Suciu2014,Suciuetal2016}.

We conclude this section with a remark about the GRW results. The ensembles of GRW simulations were computed for $10^6$ iterations in all cases, excepting the Gaussian case with $\sigma^2=2$, where the number of iterations was set to $5\cdot 10^6$. As indicated by Figs.~\ref{fig:conv_GRW_Gss}~and~\ref{fig:conv_GRW_Exp}, as well as by Supplementary materials S3 and S4, these numbers of iterations are not large enough to ensure the convergence of the solution for homogeneous problems. Surprisingly, the variance estimates from Tables~\ref{tab:stat_1stOrder_Gss}-\ref{tab:stat_1stOrder_Exp} deviate with maximum 10\% from the predictions of the first order theory for small $\sigma^2$. For larger $\sigma^2$, the GRW results are also close to those obtained with the FDM, FEM, and DGM schemes for $\sigma^2\leq 2$ in Gaussian case (Fig.~\ref{fig:vel_variance_Gauss}) and for $\sigma^2\leq 0.5$ in exponential case (Fig.~\ref{fig:vel_variance_Exp}).

\begin{table} [h]
\centering
    \captionof{table}{Variance of hydraulic head, $\sigma^{2}_{h}$, estimated by Monte Carlo simulations for $\sigma^2=0.1$.}
    \label{tab:stat_h_1stOrder}
\begin{tabular}{|c | c | c | c|}
  \hline
  & Gaussian correlation & Exponential correlation \\
  \hline
  FDM  & 3.47e-04 $\pm$ 3.77e-05 & 5.91e-04 $\pm$ 5.79e-05 \\
  FEM  & 6.05e-04 $\pm$ 1.06e-04 & 5.52e-04 $\pm$ 8.45e-05 \\
  DGM  & 2.48e-04 $\pm$ 1.12e-04 & 4.69e-04 $\pm$ 2.44e-04 \\
  CSM  & 2.07e-04 $\pm$ 1.46e-04 & 5.64e-04 $\pm$ 4.57e-04 \\
  GRW  & 3.81e-04 $\pm$ 6.06e-05 & 5.69e-04 $\pm$ 1.08e-04 \\
  \hline
\end{tabular}
\end{table}

\begin{table} [h]
\centering
    \captionof{table}{Gaussian correlation: Mean values and variances of the velocity components estimated by Monte Carlo simulations for $\sigma^2=0.1$.}
    \label{tab:stat_1stOrder_Gss}
\begin{tabular}{| c | c | c | c | c |}
  \hline
  & $\langle V_x\rangle$ & $\langle V_y\rangle$ & $\sigma^{2}_{V_{x}}$ & $\sigma^{2}_{V_{y}}$ \\
  \hline
  linear  & 1.00 $\pm$ 2.07e-02 &  1.56e-03 $\pm$ 1.13e-02 & 3.75e-02 $\pm$ 4.96e-03 & 1.27e-02 $\pm$ 1.61e-03 \\
  FDM     & 1.01 $\pm$ 1.08e-01 & -5.32e-03 $\pm$ 8.37e-03 & 3.66e-02 $\pm$ 4.52e-03 & 1.34e-02 $\pm$ 1.22e-03 \\
  FEM     & 1.00 $\pm$ 2.13e-02 & -1.05e-02 $\pm$ 1.49e-02 & 3.70e-02 $\pm$ 4.51e-03 & 1.67e-02 $\pm$ 2.94e-03 \\
  DGM     & 1.01 $\pm$ 9.40e-02 & -3.20e-03 $\pm$ 9.80e-03 & 3.76e-02 $\pm$ 5.91e-03 & 1.32e-02 $\pm$ 1.73e-03 \\
  CSM     & 1.00 $\pm$ 2.27e-02 &  4.46e-04 $\pm$ 9.17e-03 & 3.96e-02 $\pm$ 9.11e-03 & 8.03e-03 $\pm$ 5.36e-03 \\
  GRW     & 1.00 $\pm$ 1.96e-02 &  -9.19e-04 $\pm$ 1.21e-02 & 4.14e-02 $\pm$ 6.80e-03 & 1.38e-02 $\pm$ 3.00e-03 \\
  \hline
\end{tabular}
\end{table}

\vspace{-0.3cm}
\begin{table} [h]
\centering
    \captionof{table}{Exponential correlation: Mean values and variances of the velocity components estimated by Monte Carlo simulations for $\sigma^2=0.1$.}
    \label{tab:stat_1stOrder_Exp}
\begin{tabular}{| c | c | c|  c | c |}
  \hline
  & $\langle V_x\rangle$ & $\langle V_y\rangle$ & $\sigma^{2}_{V_{x}}$ & $\sigma^{2}_{V_{y}}$ \\
  \hline
  linear  & 0.99 $\pm$ 1.81e-02 & 2.60e-03 $\pm$ 1.04e-02 & 3.66e-02 $\pm$ 5.17e-03 & 1.25e-02 $\pm$ 1.78e-03 \\
  FDM     & 1.00 $\pm$ 1.06e-01 & 2.20e-04 $\pm$ 1.03e-02 & 3.97e-02 $\pm$ 4.04e-03 & 1.27e-02 $\pm$ 8.97e-04 \\
  FEM     & 1.00 $\pm$ 1.56e-02 & 7.26e-04 $\pm$ 9.46e-03 & 4.14e-02 $\pm$ 6.48e-03 & 1.25e-02 $\pm$ 2.05e-03 \\
  DGM     & 1.01 $\pm$ 9.35e-02 & 1.66e-03 $\pm$ 1.10e-02 & 3.96e-02 $\pm$ 4.71e-03 & 1.30e-02 $\pm$ 1.86e-03 \\
  CSM     & 0.99 $\pm$ 3.07e-02 & 1.63e-03 $\pm$ 1.83e-02 & 8.65e-02 $\pm$ 4.17e-02 & 3.61e-02 $\pm$ 2.55e-02 \\
  GRW     & 1.00 $\pm$ 2.00e-02 & -1.60e-03 $\pm$ 1.09e-02 & 4.08e-02 $\pm$ 6.70e-03 & 1.19e-02 $\pm$ 1.80e-03 \\
  \hline
\end{tabular}
\end{table}

\section{Discussion and conclusions}
\label{concl}

The numerical approaches tested in the present benchmark study passed the validation test consisting of comparisons of the numerical solutions obtained for $\sigma^2=0.1$ and $N=100$ with theoretical results provided by first-order perturbation approaches. For larger variances, up to $\sigma^2=2$, the statistical inferences performed by averaging over Monte Carlo ensembles of solutions given by the FDM, FEM, and DGM schemes were close to those obtained by simulations with similar numerical setup reported in the literature. However, these encouraging results are not a guarantee for the accuracy of the numerical method in case of higher values of $\sigma^2$ and $N$, needed to solve the flow problem for highly heterogeneous aquifers and large spatial scales. This is indicated, for instance, by the situation of the GRW ensembles of solutions for Gaussian correlation of the $\ln(K)$ field which provide satisfactory statistical estimates even though they are not yet convergent (see Section~\ref{stat}) and cannot be used for single-realization approaches in subsurface hydrology.

While for Gaussian correlation of the $\ln(K)$ field the test problems can be accurately solved (with the only exception of the 2D FDM solutions for $N=10^3$ and $\sigma^2\geq 6$ and that of the 2D GRW solution for $N=10^3$ and $\sigma^2=2$), this is not the case for exponentially correlated fields. In fact, the most important result from a practical point of view of our investigations is that in case of exponential correlation and large values of the parameters $\sigma^2$ and $N$ the flow problem is not computationally tractable for all the numerical methods considered in this study. That means that all the schemes fail to solve one or more test problems by using reasonably large computing resources, corresponding to a resolution of the numerical solution which is one order of magnitude higher than recommended in the literature (see e.g., \cite{Ababouetal1989,deDreuzyetal2007}). This conclusion is consistent with the
numerical results presented by \cite{Gotovacetal2009}, which show that for high resolution of the exponentially correlated $\ln(K)$ field with $\sigma^2>2$ the required discretization of the head solution exceeds the limit of the available computing resources.

The numerical approaches compared in this study cannot be ranked from most effective to least effective. As discussed in the following, each numerical scheme has its advantages and limitations in solving problems for fixed realizations of the hydraulic conductivity or in Monte Carlo simulations.

FDM schemes are easily implementable and require a small amount of computing time to solve the whole problem, from about one minute, to solve 1D problems, to about 7 minutes, for 2D problems. The time needed to solve the linear system of equations for 2D problems, which allows a fair comparison with FEM and DGM approaches, was about 3 seconds in all cases, with the exception of exponential correlation with parameter pair $(N=10^4, \sigma^2=10)$, where it was of about 5 seconds. The verification of the FDM codes by comparisons with 1D manufactured solutions indicates a good accuracy for the entire range of parameters $\sigma^2$ and $N$ in case of Gaussian correlation of the $\ln(K)$ field, and for all values of $\sigma^2$ and $N\leq 10^3$ in exponential case. An acceptable accuracy of the 2D solutions is obtained in case of Gaussian correlation for all parameters excepting values $\sigma^2\geq 6$ and $N=10^3$. Instead, in case of exponential correlation, solutions with acceptable accuracy can be obtained only for $\sigma^2=0.1$ and $N\leq 10^3$. The estimated order of convergence of the 1D FDM solutions is close to the theoretical order 2 only for Gaussian correlation of the $\ln(K)$ field with $\sigma^2\leq 4$. For larger variances, in case of Gaussian correlation, and for all combinations of parameters, in case of exponential correlation, the EOC values for 1D solutions deviate from the theoretical order and show a non-monotonous oscillatory behavior. The 2D FDM solutions converge with EOC values close to 2, for Gaussian correlation, but, for exponential correlation, EOC oscillates and takes negative values. Since the computing time is relatively small, the 2D FDM scheme is well suited for Monte Carlo simulations of flows in aquifers with low to moderate heterogeneities modeled by $\ln(K)$ fields with Gaussian correlation structure.

FEM schemes are more elaborated, with higher accuracy, but require larger computing times for the assembly of the linear system. In terms of solving the linear system, it takes less than 0.1 seconds in the 1D test case and around 15 seconds in the 2D case, a few times more than for the FDM scheme. The comparisons with manufactured solutions show that the 1D FEM solutions for Gaussian correlation of the $\ln(K)$ field, and for all the combinations of parameters $N$ and $\sigma^2$, are accurate, with errors between $10^{-6}$ and $10^{-2}$, for both $\mathbb{P}_1$ and $\mathbb{P}_2$ approximations of $K$ and $f$. For exponential correlation, the 1D FEM solutions have the same accuracy as in Gaussian case if $N\leq 10^3$, but for $N=10^4$ the errors  increase with $\sigma^2$ from one to $10^7$ even when the $\mathbb{P}_3$ approximation is used. In 2D, for Gaussian correlation, accurate FEM solutions with errors between $10^{-3}$ and $10^{-2}$ can be obtained with the quadratic approximation $\mathbb{P}_2$. The 2D FEM solutions for exponential correlation obtained with the quadratic approximation $\mathbb{P}_2$ are still fairly accurate for $N=10^3$ and $\sigma^2\leq 1$, as well as for $N=10^4$ and $\sigma^2=0.1$. The convergence of the 1D FEM solutions for Gaussian correlation of the $\ln(K)$ field, for all pairs of parameters excepting $(N=10^4, \sigma^2=10)$, is indicated by EOC values close to 2, but in case of exponential correlation one obtains oscillating errors and a few negative EOC values, similarly to the 1D FDM solutions. In the 2D case, the error norm decreases monotonously, with EOC values close to the theoretical order of convergence 2, for Gaussian correlation, and smaller EOC values for exponential correlation. Though the computing time is larger than for FDM, the FEM scheme can be used in Monte Carlo simulations of flow in highly heterogeneous aquifers with Gaussian correlated $\ln(K)$ fields (for all parameters $\sigma^2$ and $N$ considered in this study), and in case of exponentially correlated fields up to the limit of tractability (specified by parameters $\sigma^2$ and $N$ for which one obtains accurate solutions).

The 2D DGM scheme is even more complex than classical FEM schemes. Consequently, the computing time required by the pre-processing step of generating the computational grid and by the assembly of the linear system of equation is about six times larger than for the 2D FEM scheme used in this study. The time to solve the system of equations is between 42 and 46 seconds, that is, around three times more than that of the FEM scheme. The code verification tests, for Gaussian correlation and all combinations of parameters $\sigma^2$ and $N$, show that the errors with respect to the analytical manufactured solutions are of the order $10^{-3}$, that is up to one order of magnitude smaller than for the FEM solutions. In the exponential case, acceptable errors of orders between $10^{-2}$ and $10^{-1}$ can be obtained for ($N=10^2$, $\sigma^2\leq 2$), ($N=10^3$, $\sigma^2\leq 1$), and ($N=10^4$, $\sigma^2=0.1$). Similarly to FEM estimates, EOC values are close to theoretical order 2 of convergence in the Gaussian case and become generally smaller than 2 in the exponential case. DGM scheme also can be used in Monte Carlo simulations for the range of $\sigma^2$ and $N$ parameters, identified at the verification stage, which ensure the accuracy of the solution.

The error of the numerical solution obtained by the discretization schemes is composed of the approximation error of the method and the arithmetic error due to floating point roundoffs. For the 1D FDM and FEM solutions obtained in case of Gaussian correlation of the $\ln(K)$ field with small $\sigma^2$ and $N$, the errors with respect to the manufactured solutions are of the order $\Delta x^2$. They are thus consistent with the theoretical order of convergence and indicate that the corresponding systems of equations were solved with negligible errors. However, this is no longer the case for exponential correlation and larger parameter values. A possible explanation of the large errors produced in these cases are the arithmetic errors in the linear solver (in addition to the discretization errors) stemming from high condition numbers of the stiffness matrices and large residuals of the solution. For instance, in 2D cases, the condition numbers given by the {\it condest} Matlab function are comparable across different discretizations and increase with $\sigma^2$ from $10^6$ up to $10^{12}$, for FDM and FEM, and up to $10^{13}$ for DGM, for both Gaussian and exponential correlations and for all the three values of $N$. In the Gaussian cases, the computed residuals show small variations around a value of $10^{-11}$ (with larger values of order $10^{-8}$ for FDM). However, for exponential correlation, the residuals increase with increasing $\sigma^2$ and $N$ from $10^{-11}$ to $10^{-6}$, for the FEM scheme, and to $10^{-7}$ for the DGM scheme (and again, with larger values between $10^{-8}$ and $10^{-3}$, for the FDM scheme). The product of condition number and residual is an upper bound for the arithmetic error \cite{Kelley1995}. This error bound increases in case of exponential correlation from $10^{-5}$ to $10^6$, for the FEM and DGM schemes, and from $10^{-2}$ to $10^9$, for the FDM scheme.

In exponential cases of large $N$ and $\sigma^2$, the coefficients $K$ contain large numbers of periodic modes with amplitudes spanning over several orders of magnitude. The high resolution necessary to prevent the occurrence of ill-conditioned matrices and large residuals leads to extremely large degrees of freedom of the linear system of equations which render impractical, from the point of view of computational resources, the use of direct flow solvers such as FDM, FEM, and DGM considered in this study. A possible remedy could be using affordable discretization levels and various preconditioning techniques \cite{KnabnerandAngermann2003}. An alternative could be provided by numerical upscaling techniques developed for problems with rough coefficients, which capture the effect of small scales on large scales without resolving the small scale details \cite{Caloetal2011,Houetal1999,Kornhuberetal2018}.

CSM schemes verify the manufactured analytical solutions with very small errors between $10^{-9}$ and $10^{-14}$ in both 1D and 2D cases. This high precision is rendered possible by the use of the analytical expressions of the derivatives of the $K$ field given by the Kraichnan algorithm. The convergence is demonstrated by the decrease of Chebyshev coefficients towards the roundoff plateau. The non-homogeneous flow problem can be solved with relatively small numbers of collocation points (less than 100 in both spatial directions) which are enough for the spectral representation of the smooth manufactured solutions chosen for the purpose of code verification. But in case of highly oscillating solutions of the homogeneous problem for exponential correlation of the $\ln(K)$ field, a much larger number of points is required, which increases the dimension of the problem beyond the limit of available memory (on a computer with 32 GB RAM). Due to this constraint, the maximum number of collocation points in each direction was of 160, which do not ensure reliable solutions for any realization of $\ln(K)$. Therefore the  ensemble of valid solutions for the hydraulic head  and velocity components used in Monte Carlo assessments had to be selected after visual inspection of solution quality (see Appendix~\ref{appendixB}).

The GSM scheme has been proposed to avoid the use of the analytical derivative of the hydraulic conductivity $K$ and has been used only in solving the 1D non-homogeneous problem for comparisons with the manufactured solutions. The errors are again very small, between $10^{-7}$ and $10^{-12}$, in the Gaussian case, but extremely large in the exponential case. Some tests made in the attempt to improve the precision indicate that exceedingly large computing resources would be necessary to render the problem computationally tractable.

The GRW scheme is a transitory scheme using computational particles, accurate and unconditionally stable. The drawback is the large number of time iterations needed to reach the stationary state corresponding to the steady-state solution of the flow problem. In the 1D Gaussian case, approximations with satisfactory accuracy of the manufactured solution can be obtained in a about one hour. For the largest couple of parameters ($N=10^4$, $\sigma^2=10$) in the 1D exponential case, the amount of computing time is of tens of hours to reduce the error to a value smaller than 10 (from errors of order $10^7$ or $10^8$ obtained with the FDM and FEM approaches) and of about 7 days to obtain errors of order $10^{-2}$. The 2D GRW scheme solves the code verification tests with errors smaller than one for parameter ranges $(N=10^2\; , \sigma^2\leq 2)$, $(N=10^3\; , \sigma^2\leq 1)$, in the Gaussian case, and for $(N=10^2\; , \sigma^2\leq 1)$, $(N=10^3\; , \sigma^2=0.1)$, in the exponential case. The computing time increases up to hours in the Gaussian case and up to days in the exponential case. Since the benchmark problems are not computable in reasonably large times, they are practically computationally intractable for the GRW scheme as well.

As we have seen, the numerical schemes tested so far on benchmark problems have their own strengths which make them useful in specific applications, depending on the objective pursued and the issue to be solved. FDM schemes are appropriate for fast computations of Monte Carlo ensembles in case of moderate aquifer heterogeneity, FEM and DGM schemes are appropriate to cope with higher heterogeneity in either single-realization or Monte Carlo approaches, GRW is rather appropriate for single-realization problems with highly heterogeneous coefficients, and CSM/GSM approaches are useful in solving problems with heterogeneous coefficients given by smooth functions. We have also seen that all schemes face the computational tractability issue in case of an exponential structure of the log-conductivity field. It is therefore recommendable that, unless well documented by experimental data for a specific problem, the exponential correlation model should be avoided. Instead, Gaussian models can be used to account for the effective dispersion associated with the integral scale of the correlation function \cite{Dagan1989,Trefryetal2003}, as  well as in flow and transport simulations for fractal aquifer heterogeneity \cite{DiFedericoandNeuman1997,Suciuetal2015}.

\section*{Acknowledgements}
The authors are grateful to Dr. Maria Cr\u{a}ciun, for her help in testing the random field generators, and to Dr. Emil C\u{a}tina\c{s} for carefully reading the manuscript. The work of Nicolae Suciu was supported by the Deutsche Forschungsgemeinschaft (DFG) -- Germany under Grant SU 415/4-1 \enquote{Integrated global random walk model for reactive transport in groundwater adapted to measurement spatio-temporal scales}.
Andreas Rupp acknowledges the financial support of the DFG Grants RU 2179 \enquote{MAD Soil - Microaggregates: Formation and turnover of the structural building blocks of soils} and EXC 2181 \enquote{STRUCTURES: A unifying approach to emergent phenomena in the physical world, mathematics, and complex data}.

\appendix
\section{Manufactured solutions}
\label{appendixA}
%
%\setcounter{figure}{0}
%\subsection{...}\label{appendixA1}

For the 1D case, we have selected the following
manufactured solution :
\begin{equation}\label{Flow:Eq1}
    h(x) = 3 + \sin(x) \ , \text{ with } x \in [0,L] \ .
\end{equation}
This leads to the following Dirichlet boundary conditions :
\begin{equation}\label{Flow:Eq2}
\begin{split}
\begin{cases}
    & h(0) = 3 , \\
    & h(L) = 3 + \sin(L)  .
\end{cases}
\end{split}
\end{equation}
The isotropic hydraulic conductivity $K$ is computed by fixing
$y=1$ in  (\ref{eq2.5}) as follows:
\begin{equation}\label{Flow:Eq3}
K(x) = C_1 \exp \left(C_2
\sum_{i=1}^{N} \cos \left( \phi_i + 2 \pi \left( k_{i,1} x  + k_{i,2}
\right) \right) \right),
\end{equation}
where we use the shorthand notations $C_1=\langle K \rangle \exp\left(-\dfrac{\sigma^2}{2}\right)$ and $C_2=\sigma\sqrt{\dfrac{2}{N}}$.
Finally, after inserting (\ref{Flow:Eq1}) and (\ref{Flow:Eq3}) in Eq.~(\ref{eq2.2}), one obtains a source term $-f$, where $f$ has the following form:
\begin{equation}\label{Flow:Eq4}
\begin{split}
f(x) &= C_1\exp\left(C_2 \sum_{i=1}^{N} \cos \left(
\phi_i + 2 \pi \left( k_{i,1} x  + k_{i,2} \right) \right) \right)\\
&\cdot\left(C_2 \sum_{i=1}^{N} (-2 \pi) k_{i,1}
\sin \left( \phi_i + 2 \pi \left( k_{i,1} x + k_{i,2} \right) \right)
\cos(x) - \sin(x)\right).
\end{split}
\end{equation}

For the 2D case, we consider the following smooth manufactured solution :
\begin{equation}\label{Flow:Eq5}
h(x,y) = 1 + \sin(2 x + y) \ , \text{ with } x \in [0,L_x] \text{ and } y \in [0,L_y] \ ,
\end{equation}
along with the Dirichlet and Neumann boundary conditions :
\begin{equation}\label{Flow:Eq6}
    \begin{split}
        \begin{cases}
            & h(0,y) = 1 + \sin(y) ,\;\; \forall y\in[0,L_y],\\
            & h(L_x,y) = 1 + \sin(2L_x + y) ,\;\; \forall y\in[0,L_y],\\
            & \dfrac{\partial h}{\partial y} (x,0) = \cos(2 x ) ,\;\; \forall x\in[0,L_x],\\
            & \dfrac{\partial h}{\partial y} (x,L_y) = \cos(2 x + L_y)\;\; \forall x\in[0,L_x] .
        \end{cases}
    \end{split}
\end{equation}
The function $K$ is now given by
\begin{equation}\label{Flow:Eq7}
K(x,y) = C_1 \exp \left( C_2 \sum_{i=1}^{N} \cos \left( \phi_i + 2 \pi \left( k_{i,1} x  + k_{i,2}y \right) \right) \right).
\end{equation}

Similarly to the 1D case, after inserting (\ref{Flow:Eq5}) and (\ref{Flow:Eq7}) in Eq.~(\ref{eq2.2}), one obtains a source term $-f$, where $f$ is given by
\begin{equation}\label{Flow:Eq8}
\begin{split}
f(x,y)
&= 2C_{1}C_{2}\sum _{i=1}^{N}-2 \pi k_{i,1}\sin \left( \phi_i+2 \left( xk_{i,1}+yk_{i,2} \right) \pi \right)\\
&\cdot \exp\left(C_2 \sum _{i=1}^{N}\cos \left( \phi_i+2\left(xk_{i,1}+yk_{i,2} \right) \pi \right) \right)\cos \left( 2x+y \right) \\
&-5C_{1} \exp\left(C_2 \sum _{i=1}^{N}\cos \left( \phi_i +2 \left( xk_{i,1}+yk_{i,2} \right) \pi \right) \right)\sin
\left( 2x+y \right) \\
&+C_{1}C_{2} \sum _{i=1}^{N}-2 \pi k_{j,2} \sin \left( \phi_i+2 \left( xk_{i,1}+yk_{i,2} \right) \pi \right) \\
&\cdot \exp\left(C_2 \sum_{i=1}^{N}\cos \left( \phi_i+2\left( xk_{i,1}+ yk_{i,2}\right) \pi \right)\right)\cos \left( 2x+y \right).
\end{split}
\end{equation}

\section{Discarding bad  realizations of the hydraulic head}
\label{appendixB}

\begin{figure}[p]
\centering
\includegraphics[width=\linewidth]{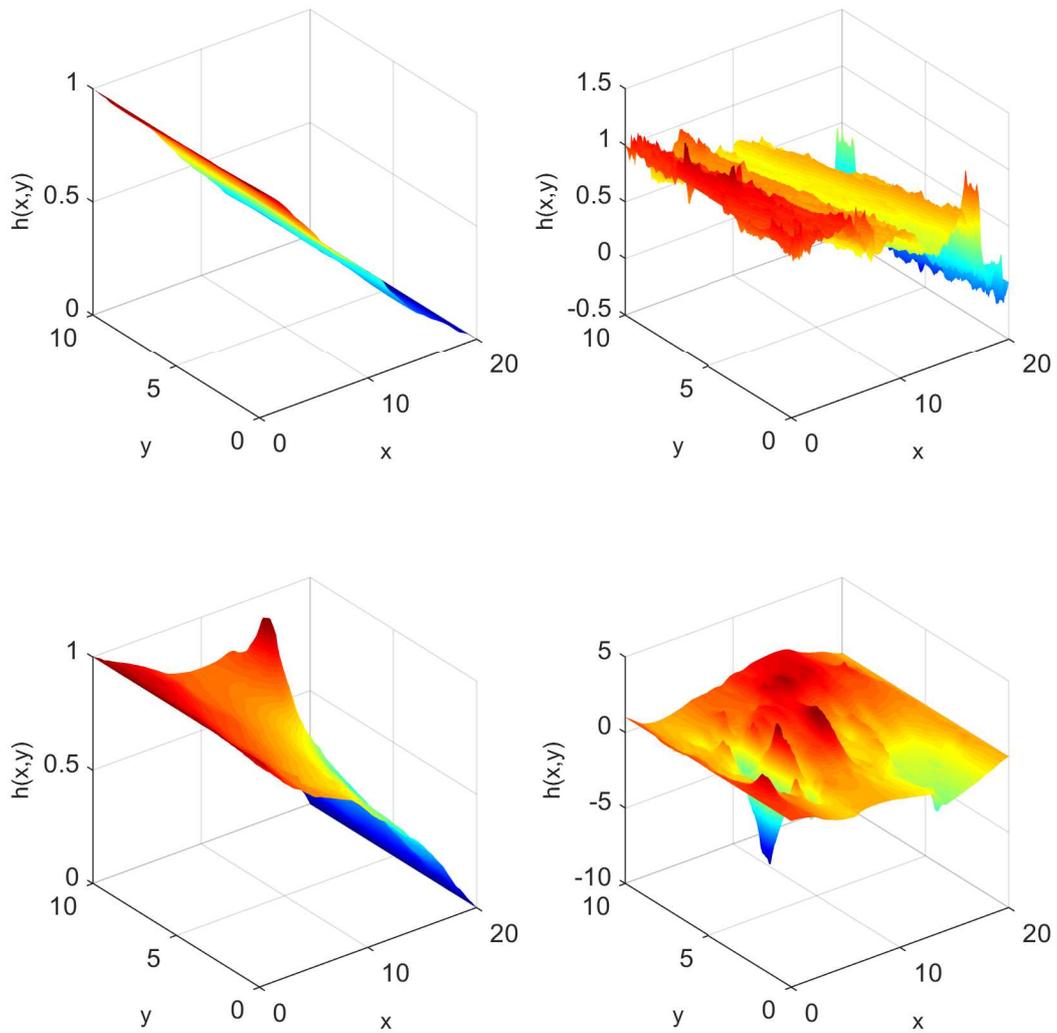}
\caption{\label{fig:h}Realizations of the hydraulic head  solution $h(x,y)$ computed with exponential correlation of the $\ln(K)$ field.}
\end{figure}

\begin{figure}[p]
\centering
\includegraphics[width=\linewidth]{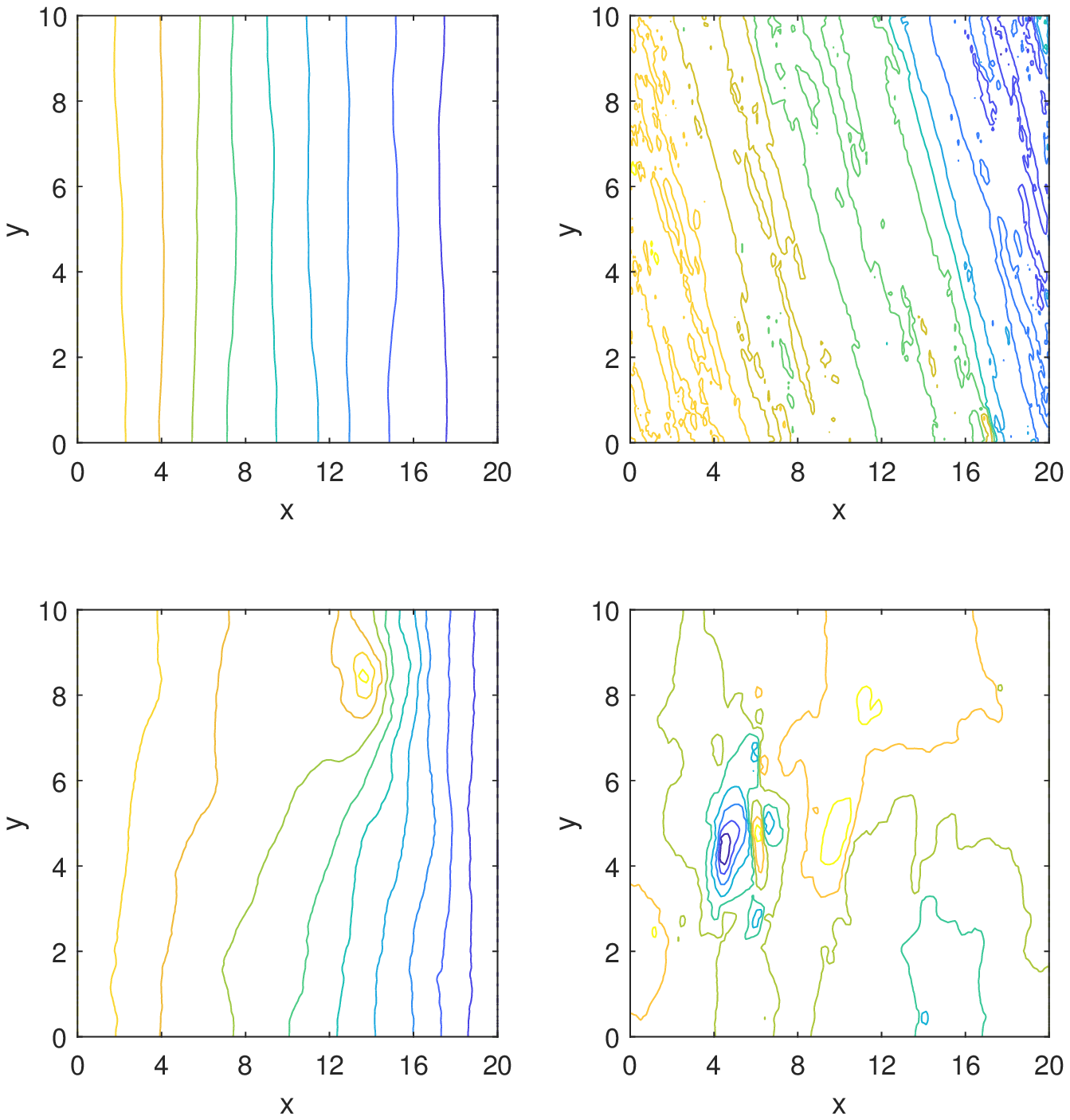}
\caption{\label{fig:contours_h}Contours of the hydraulic head solutions for exponential correlation of the $\ln(K)$ field presented in from Fig.~\ref{fig:h}.}
\end{figure}

In case of collocation spectral method using 160 collocation points in both $x$- and $y$-directions an ensemble of ``good'' realizations of the hydraulic head  solution (see top-links plots in Figs.~\ref{fig:h}--\ref{fig:contours_h}) is selected by discarding all realizations with extremely large variations (Fig.~\ref{fig:h}) and closed contour lines (Fig.~\ref{fig:contours_h}).

%\newpage

\section{Influence of boundary conditions on the head and velocity variances}
\label{appendixC}

\begin{figure}[p]
\centering \begin{minipage}[t]{0.45\linewidth} \centering
\vspace*{0in}
\includegraphics[width=\linewidth]{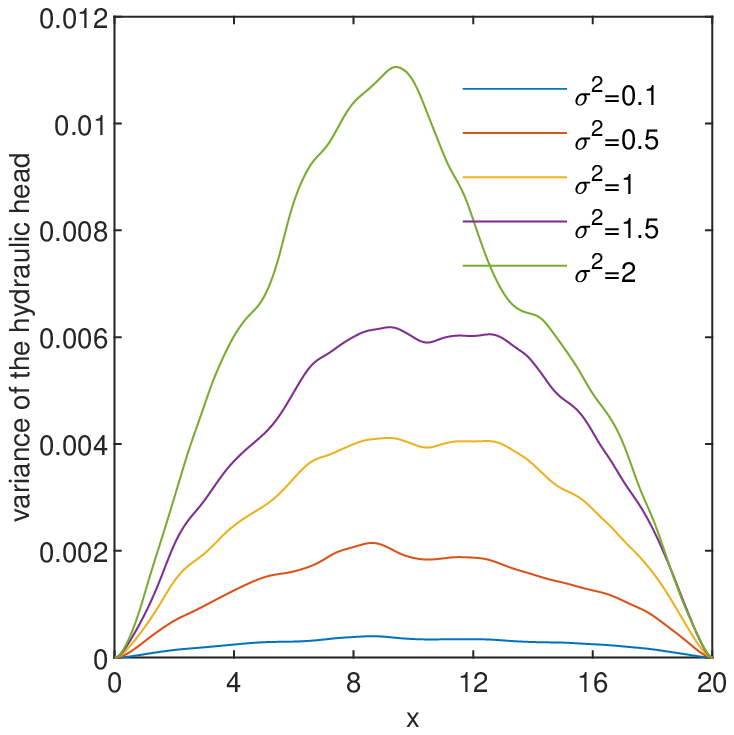}
\caption{\label{fig:FDM_BC_Dirichlet_h_Gss}FDM, Gaussian correlation: Influence of inflow/outflow Dirichlet boundary conditions on the variance of the hydraulic head.}
\end{minipage}
\hspace{0.2cm} \centering \begin{minipage}[t]{0.45\linewidth}
\centering \vspace*{0in}
\includegraphics[width=\linewidth]{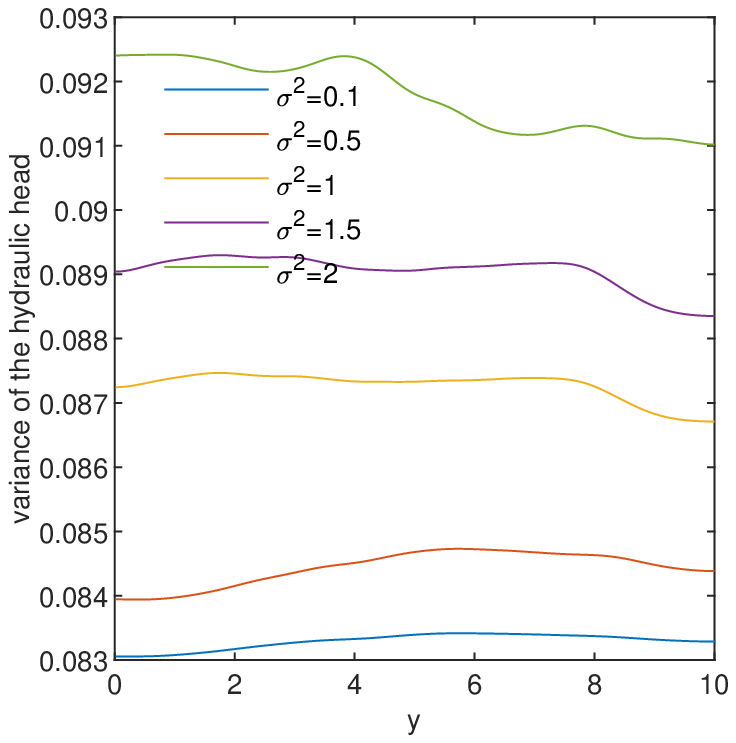}
\caption{\label{fig:FDM_BC_Neumann_h_Gss}FDM, Gaussian correlation: Influence
of top/bottom Neumann boundary conditions on the variance of the hydraulic head.}
\end{minipage}
\end{figure}

Figures~\ref{fig:FDM_BC_Dirichlet_h_Gss}--\ref{fig:GRW_BC_Neumann_h_Gss} show variances of the hydraulic head computed by finite difference (FDM), finite element (FEM), discontinuous Galerkin (DGM), and global random walk (GRW) approaches for increasing variance $\sigma^2$ of the $\ln(K)$ fields with Gaussian and exponential correlation. The variances are estimated along midlines $x=L_y/2$ and $y=L_x/2$ of the computation domain by averages over the ensemble of $R=100$ realizations followed by spatial averages over the $y$- and $x$-directions, respectively.

The influence of the Dirichlet inflow/outflow and Neumann no-flow boundary conditions on velocity variance is assessed in a similar way by ensemble and space averages. Figures~\ref{fig:FDM_statistics_X_Gauss}--\ref{fig:GRW_statistics_Y_Gauss} show that a plateau of the spatial variation of the variance (required for reliable statistical estimations of the mean and of the variance) can be more or less identified in an interior sub-domain delimitated by 1 to 4 dimensionless length units away from boundaries, but only for variances of the $\ln(K)$ field up to $\sigma^2=1.5$.

%\subsection{Finite difference method}

\begin{figure}[p]
\centering \begin{minipage}[t]{0.45\linewidth} \centering
\vspace*{0in}
\includegraphics[width=\linewidth]{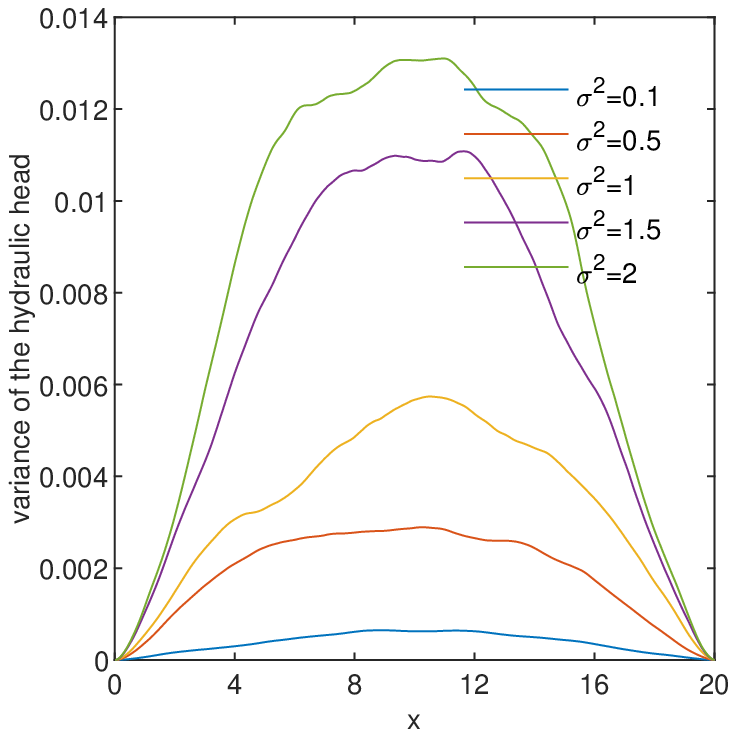}
\caption{\label{fig:FDM_BC_Dirichlet_h_Exp}FDM, Exponential correlation: Influence
of inflow/outflow Dirichlet boundary conditions on the variance of the hydraulic head.}
\end{minipage}
\hspace{0.2cm} \centering \begin{minipage}[t]{0.45\linewidth}
\centering \vspace*{0in}
\includegraphics[width=\linewidth]{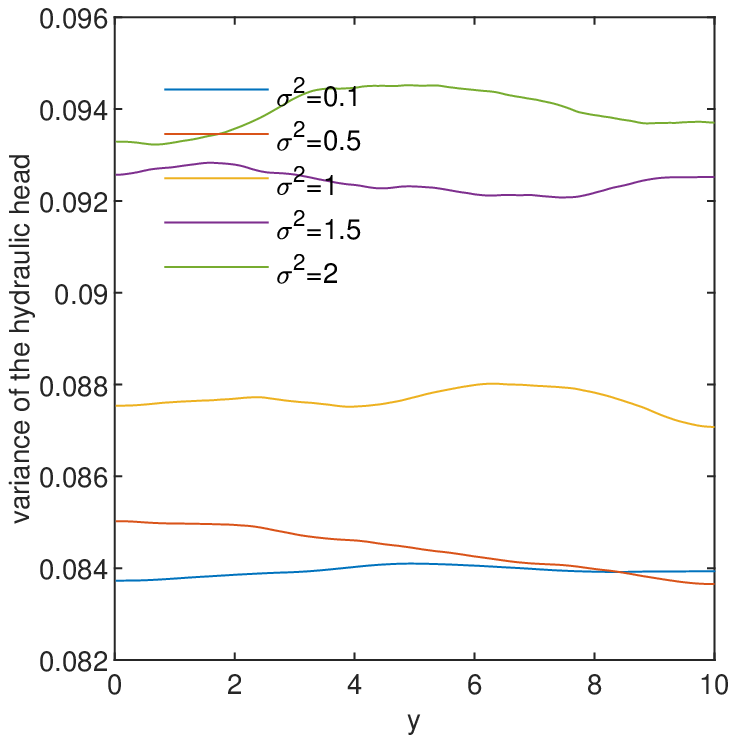}
\caption{\label{fig:FDM_BC_Neumann_h_Exp}FDM, Exponential correlation: Influence of top/bottom Neumann boundary conditions on the variance of the hydraulic head.}
\end{minipage}
\end{figure}

%%\newpage
%%\subsection{Finite element method}
\begin{figure}[p]
\centering \begin{minipage}[t]{0.45\linewidth} \centering
\vspace*{0in}
\includegraphics[width=\linewidth]{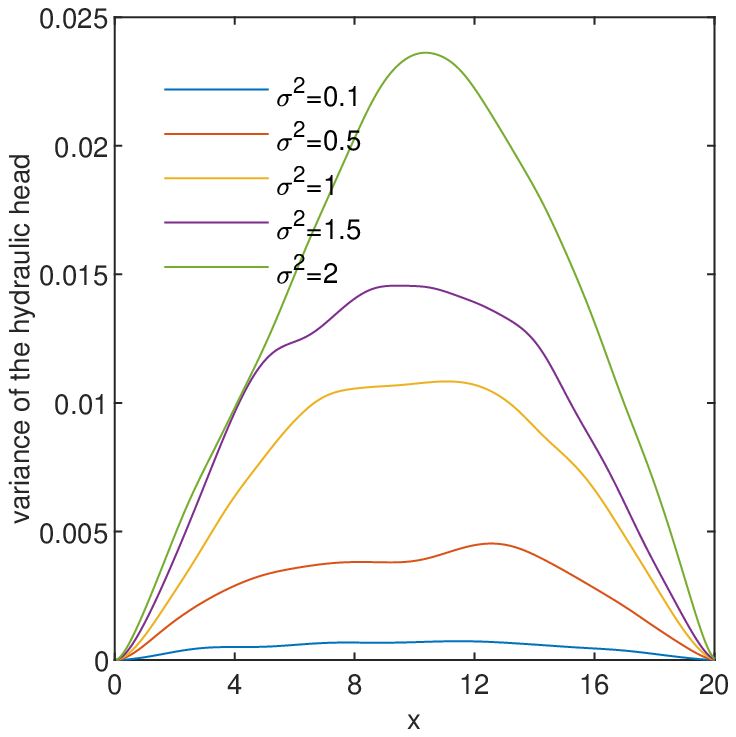}
\caption{\label{fig:FEM_BC_Dirichlet_h_Gss}FEM, Gaussian correlation: Influence
of inflow/outflow Dirichlet boundary conditions on the variance of the hydraulic head.}
\end{minipage}
\hspace{0.2cm} \centering \begin{minipage}[t]{0.45\linewidth}
\centering \vspace*{0in}
\includegraphics[width=\linewidth]{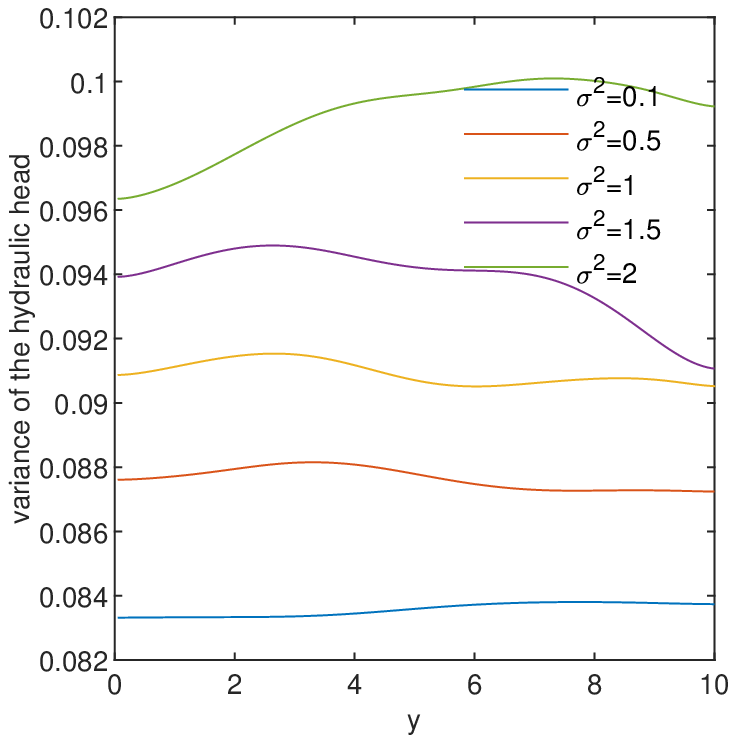}
\caption{\label{fig:FEM_BC_Neumann_h_Gss}FEM, Gaussian correlation: Influence
of top/bottom Neumann boundary conditions on the variance of the hydraulic head.}
\end{minipage}
\end{figure}

\begin{figure}[p]
\centering \begin{minipage}[t]{0.45\linewidth} \centering
\vspace*{0in}
\includegraphics[width=\linewidth]{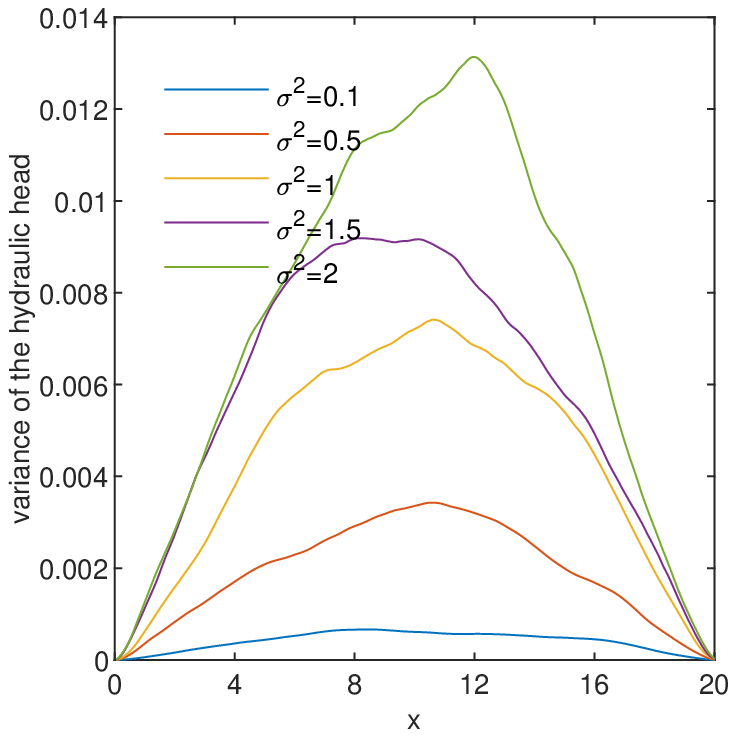}
\caption{\label{fig:FEM_BC_Dirichlet_h_Exp}FEM, Exponential correlation: Influence
of inflow/outflow Dirichlet boundary conditions on the variance of the hydraulic head.}
\end{minipage}
\hspace{0.2cm} \centering \begin{minipage}[t]{0.45\linewidth}
\centering \vspace*{0in}
\includegraphics[width=\linewidth]{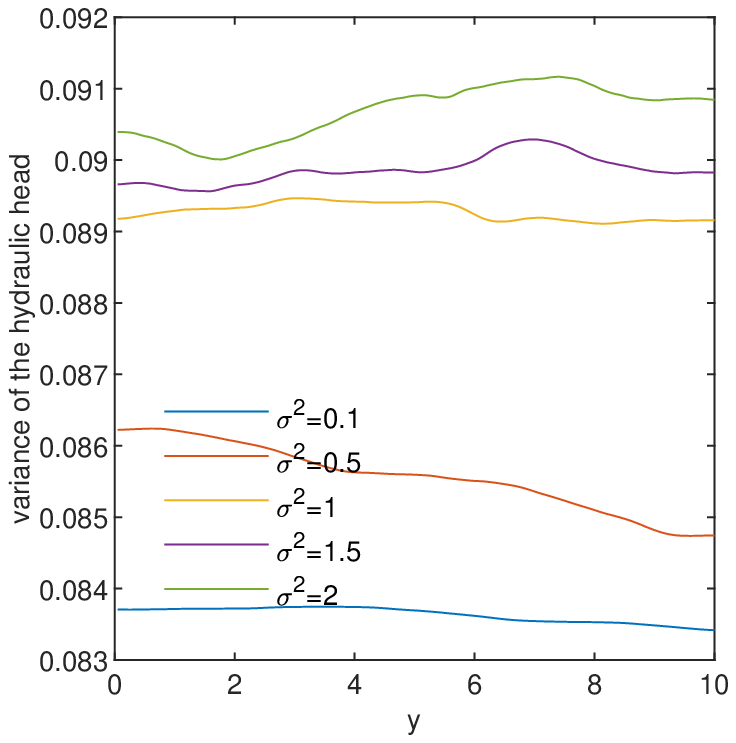}
\caption{\label{fig:FEM_BC_Neumann_h_Exp}FEM, Exponential correlation: Influence
of top/bottom Neumann boundary conditions on the variance of the hydraulic head.}
\end{minipage}
\end{figure}

%\newpage
%\subsection{Discontinuous Galerkin method}

\begin{figure}[p]
\centering \begin{minipage}[t]{0.45\linewidth} \centering
\vspace*{0in}
\includegraphics[width=\linewidth]{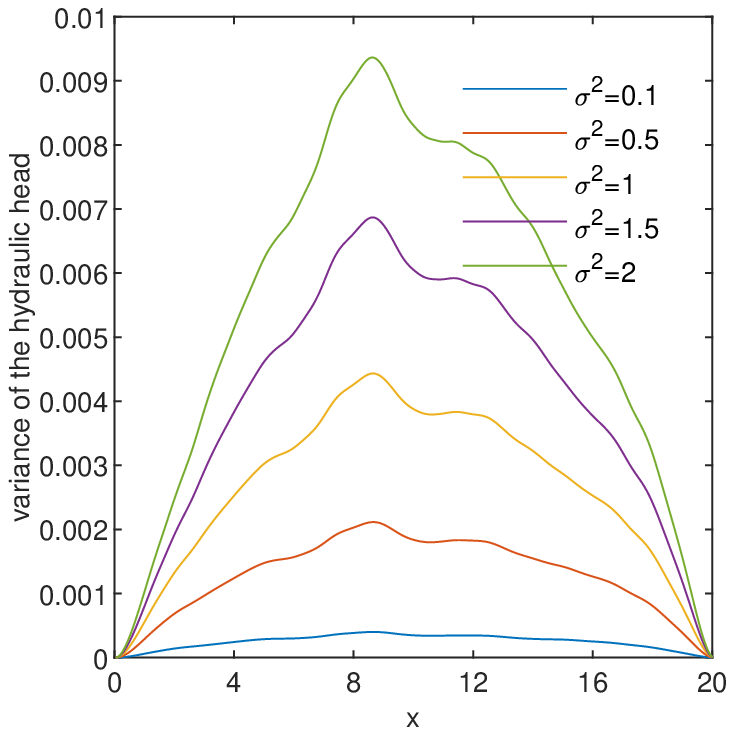}
\caption{\label{fig:DGM_BC_Dirichlet_h_Gss}DGM, Gaussian correlation: Influence of inflow/outflow Dirichlet boundary conditions on the variance of the hydraulic head.}
\end{minipage}
\hspace{0.2cm} \centering \begin{minipage}[t]{0.45\linewidth}
\centering \vspace*{0in}
\includegraphics[width=\linewidth]{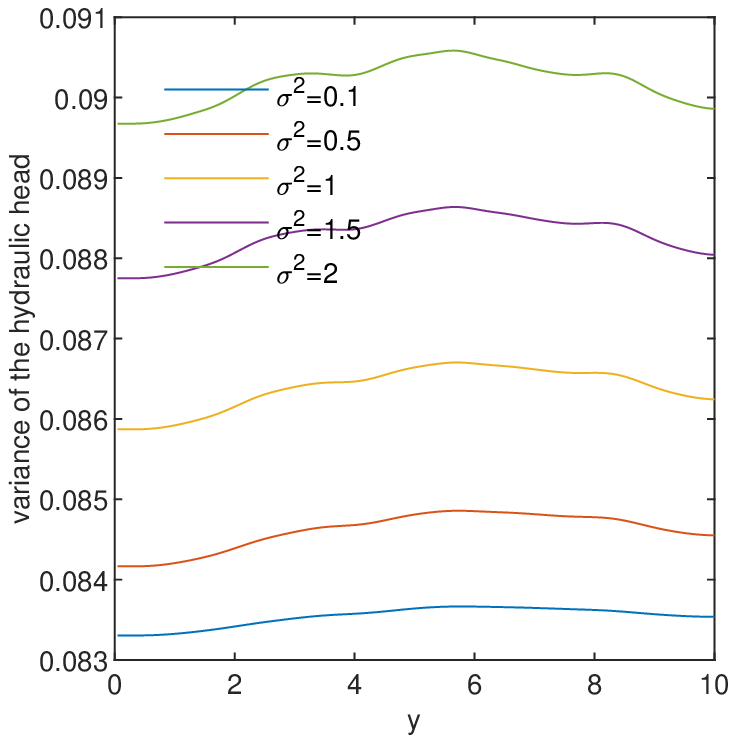}
\caption{\label{fig:DGM_BC_Neumann_h_Gss}DGM, Gaussian correlation: Influence
of top/bottom Neumann boundary conditions on the variance of the hydraulic head.}
\end{minipage}
\end{figure}

\begin{figure}[p]
\centering \begin{minipage}[t]{0.45\linewidth} \centering
\vspace*{0in}
\includegraphics[width=\linewidth]{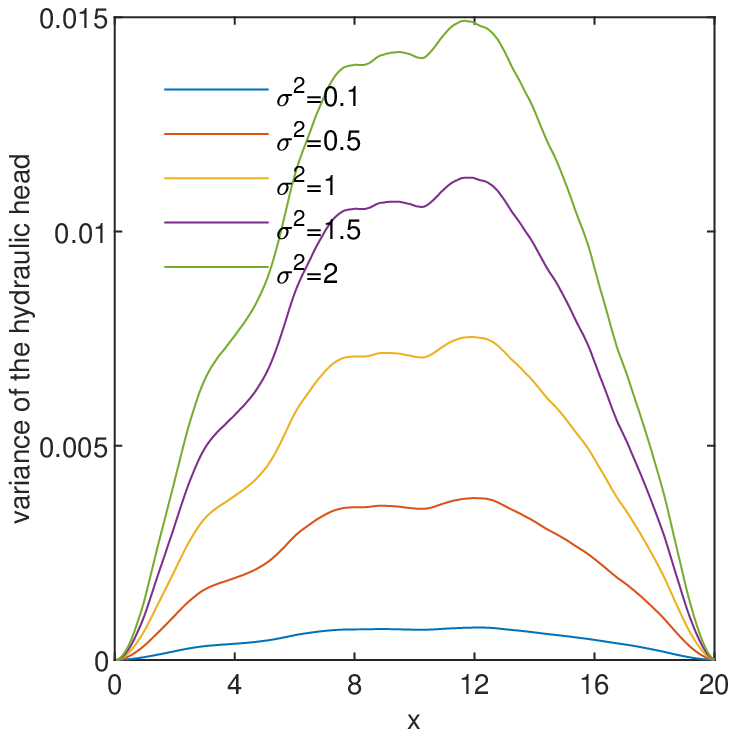}
\caption{\label{fig:DGM_BC_Dirichlet_h_Exp}DGM, Exponential correlation: Influence
of inflow/outflow Dirichlet boundary conditions on the variance of the hydraulic head.}
\end{minipage}
\hspace{0.2cm} \centering \begin{minipage}[t]{0.45\linewidth}
\centering \vspace*{0in}
\includegraphics[width=\linewidth]{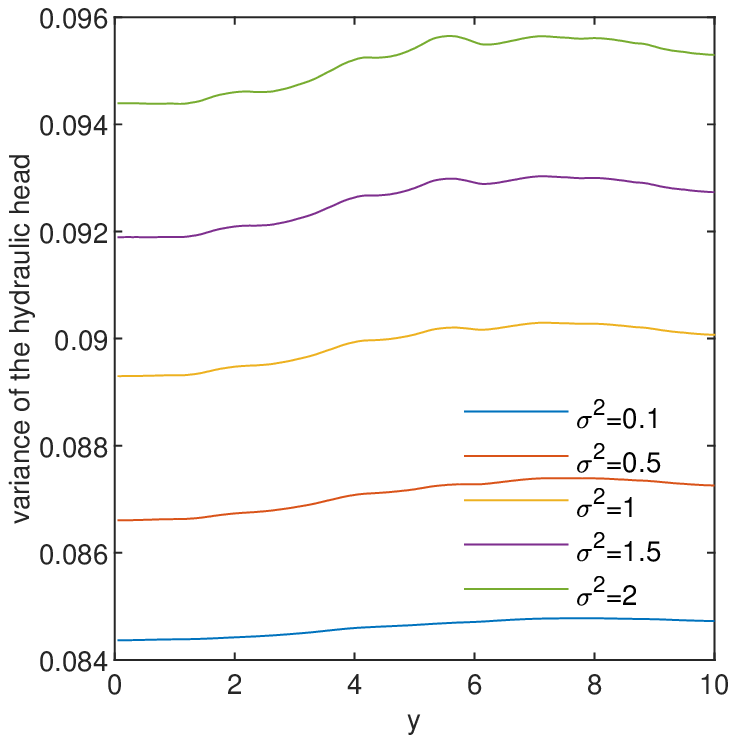}
\caption{\label{fig:DGM_BC_Neumann_h_Exp}DGM, Exponential correlation: Influence of top/bottom Neumann boundary conditions on the variance of the hydraulic head.}
\end{minipage}
\end{figure}

%\subsection{Global random walk method}

\begin{figure}[p]
\centering \begin{minipage}[t]{0.45\linewidth} \centering
\vspace*{0in}
\includegraphics[width=\linewidth]{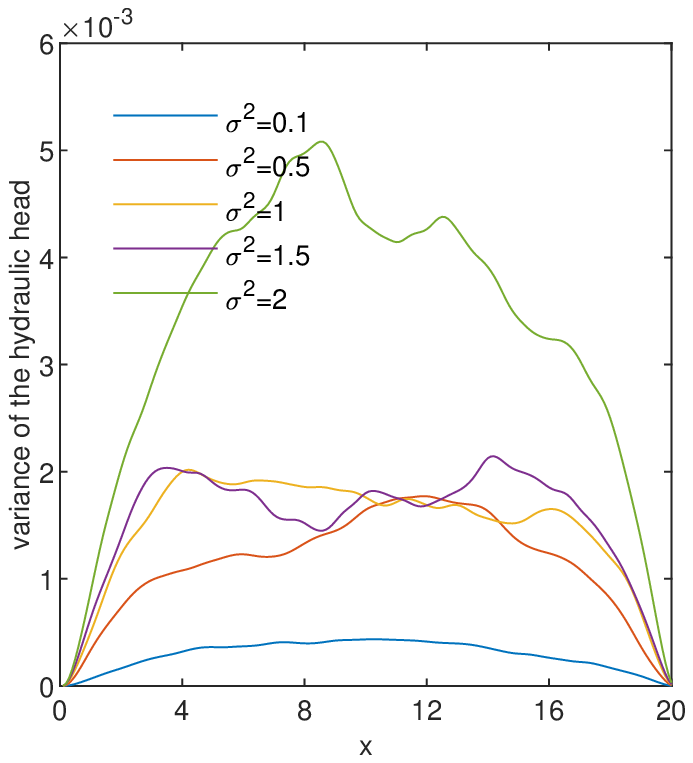}
\caption{\label{fig:GRW_BC_Dirichlet_h_Gss}GRW, Gaussian correlation: Influence
of inflow/outflow Dirichlet boundary conditions on the variance of the hydraulic head.}
\end{minipage}
\hspace{0.2cm} \centering \begin{minipage}[t]{0.45\linewidth}
\centering \vspace*{0in}
\includegraphics[width=\linewidth]{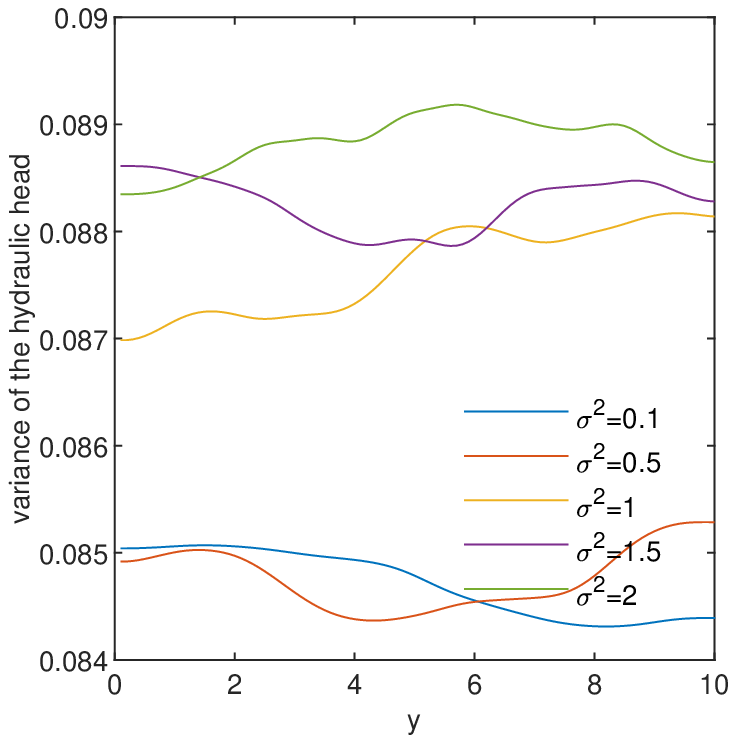}
\caption{\label{fig:GRW_BC_Neumann_h_Gss}GRW, Gaussian correlation: Influence
of top/bottom Neumann boundary conditions on the variance of the hydraulic head.}
\end{minipage}
\end{figure}

\begin{figure}[p]
\centering \begin{minipage}[t]{0.45\linewidth} \centering
\vspace*{0in}
\includegraphics[width=\linewidth]{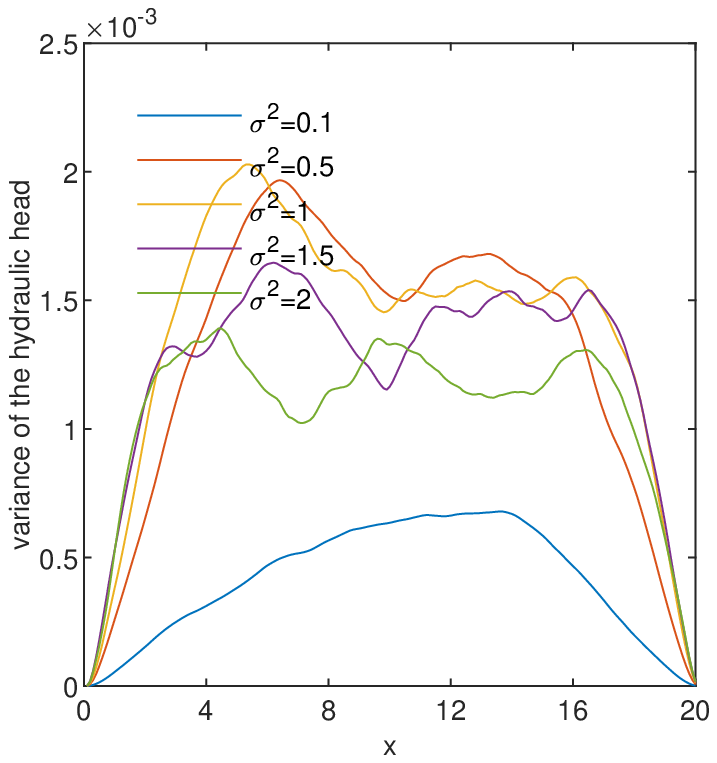}
\caption{\label{fig:GRW_BC_Dirichlet_h_Gss}GRW, exponential correlation: Influence
of inflow/outflow Dirichlet boundary conditions on the variance of the hydraulic head.}
\end{minipage}
\hspace{0.2cm} \centering \begin{minipage}[t]{0.45\linewidth}
\centering \vspace*{0in}
\includegraphics[width=\linewidth]{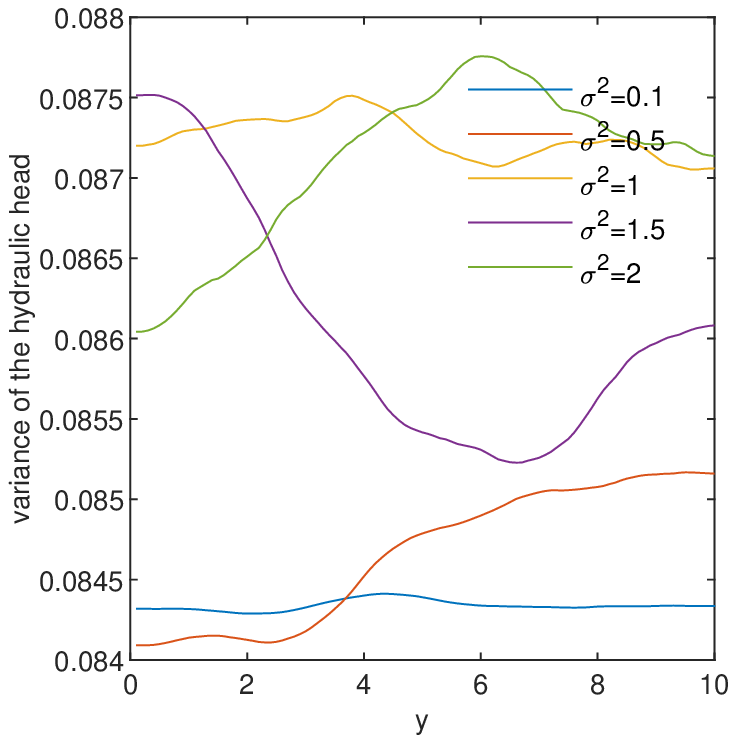}
\caption{\label{fig:GRW_BC_Neumann_h_Gss}GRW, exponential correlation: Influence
of top/bottom Neumann boundary conditions on the variance of the hydraulic head.}
\end{minipage}
\end{figure}

%\section{Influence of boundary conditions on velocity variance}
%\subsection{Finite difference method}

\begin{figure}[p]
\centering
\includegraphics[width=\linewidth]{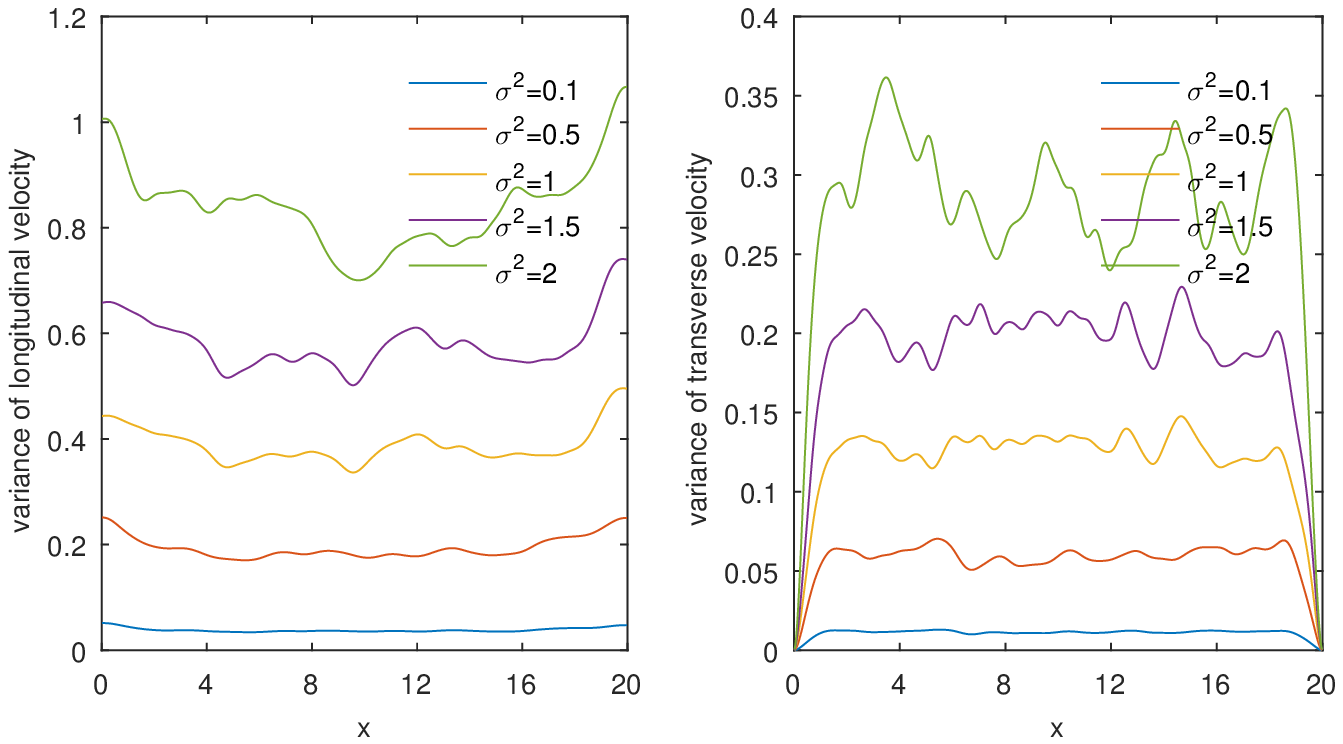}
\caption{\label{fig:FDM_statistics_X_Gauss}FDM, Gaussian correlation: Influence of inflow/outflow Dirichlet boundary conditions on velocity
variances.}
\end{figure}

\begin{figure}[p]
\centering
\includegraphics[width=\linewidth]{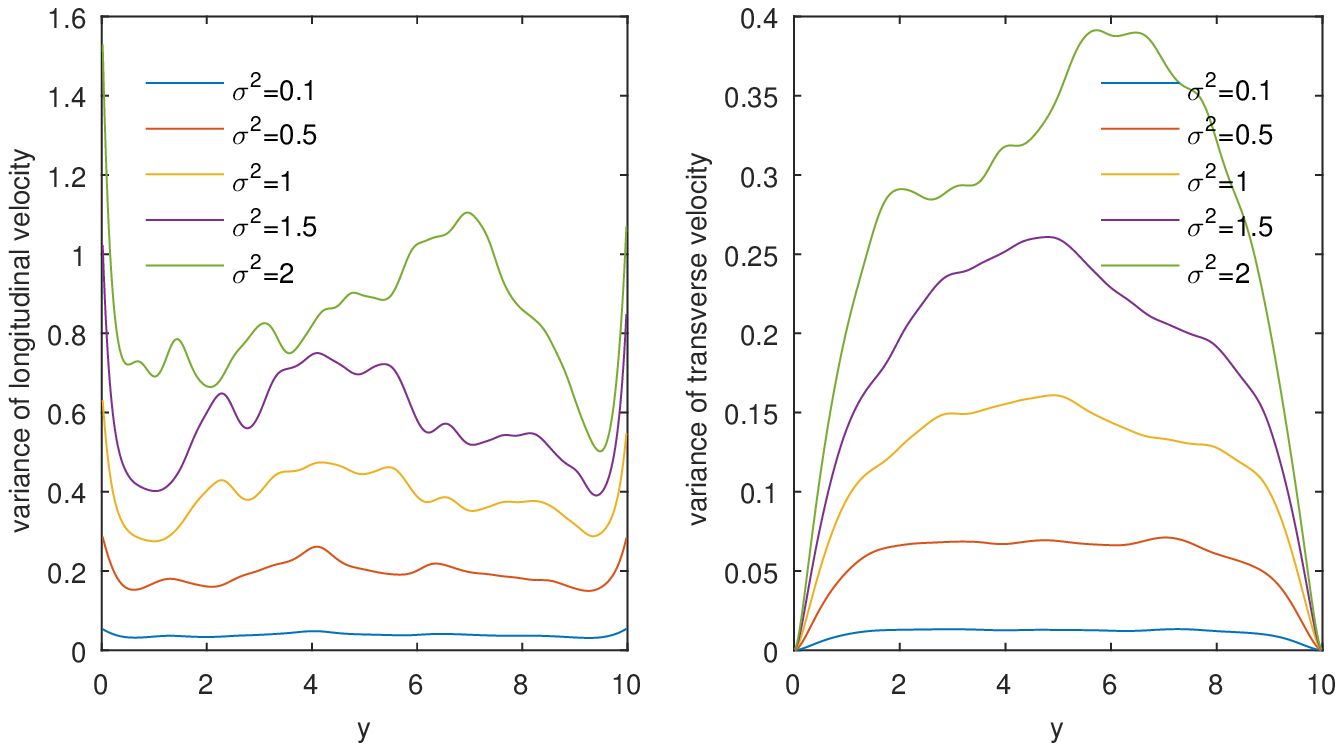}
\caption{\label{fig:FDM_statistics_Y_Gauss}FDM, Gaussian correlation: Influence of top/bottom Neumann no-flow boundary conditions on velocity
variances.}
\end{figure}

%\newpage
\begin{figure}[p]
\centering
\includegraphics[width=\linewidth]{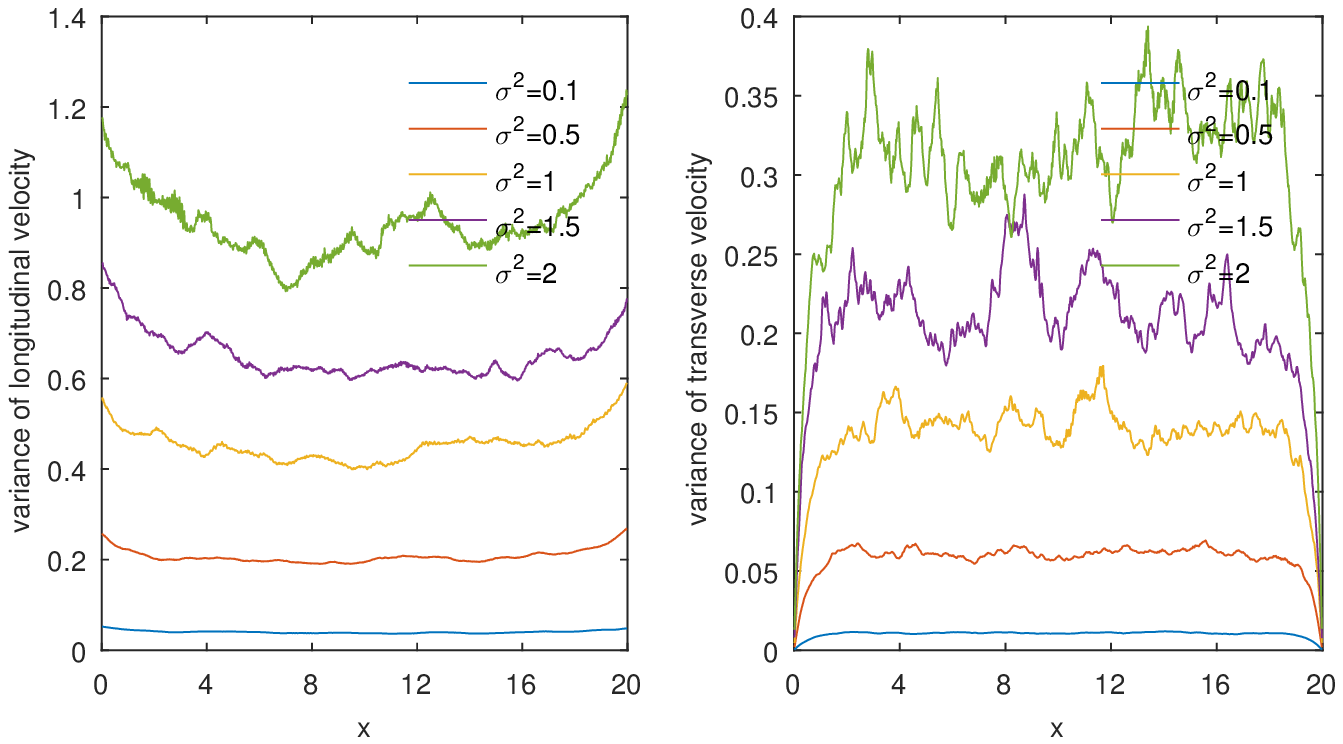}
\caption{\label{fig:FDM_statistics_X_Exp}FDM, exponential correlation: Influence of inflow/outflow Dirichlet boundary conditions on velocity
variances.}
\end{figure}

\begin{figure}[p]
\centering
\includegraphics[width=\linewidth]{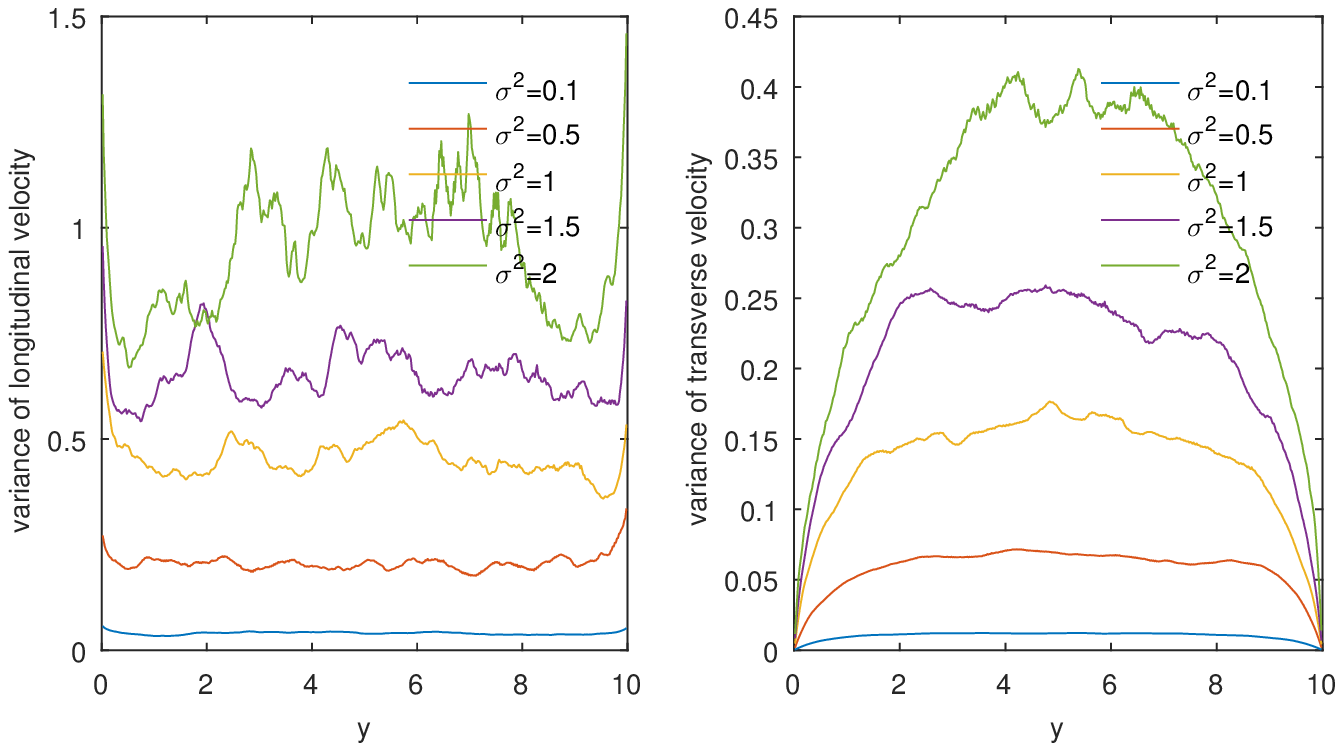}
\caption{\label{fig:FDM_statistics_Y_Exp}FDM, exponential correlation: Influence of top/bottom Neumann no-flow boundary conditions on velocity
variances.}
\end{figure}

%%\newpage
%%\subsection{Finite element method}

\begin{figure}[p]
\centering
\includegraphics[width=\linewidth]{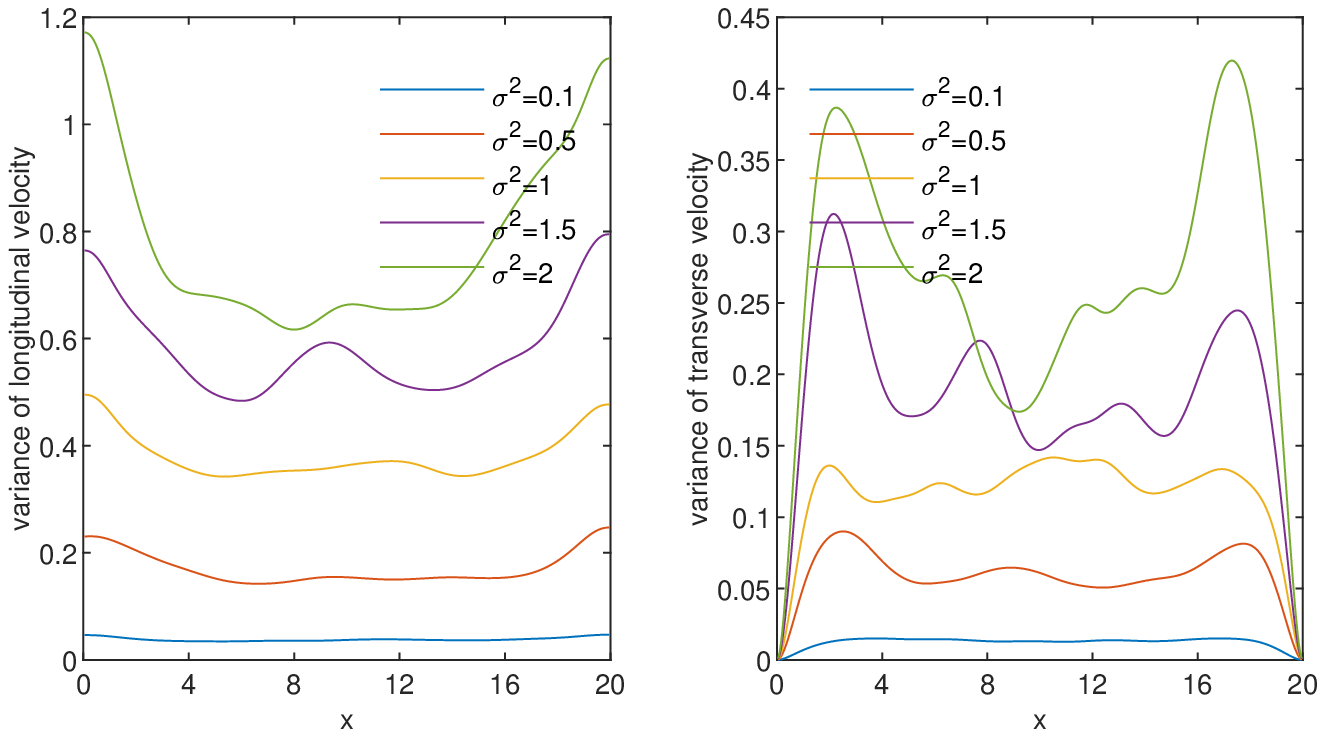}
\caption{\label{fig:FEM_statistics_X_Gauss}FEM, Gaussian correlation: Influence
of inflow/outflow Dirichlet boundary conditions on velocity
variances.}
\end{figure}

\begin{figure}[p]
\centering
\includegraphics[width=\linewidth]{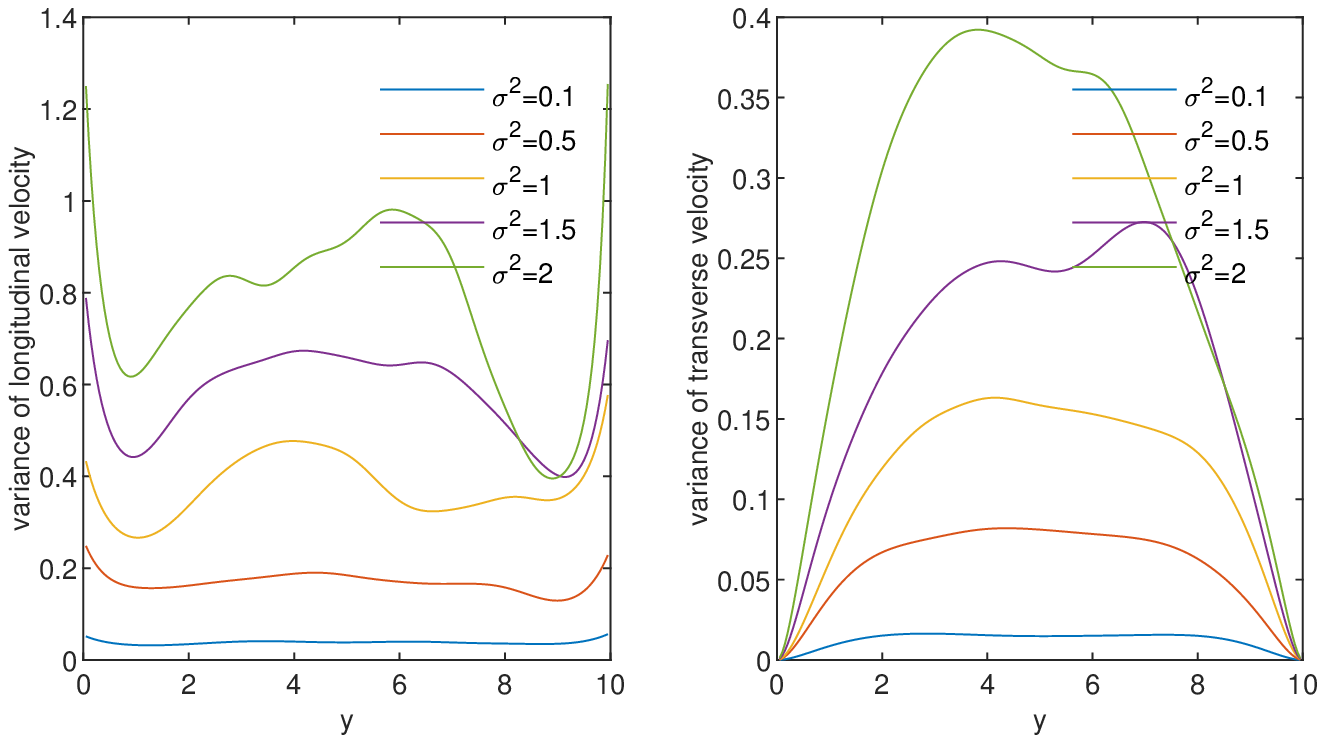}
\caption{\label{fig:FEM_statistics_Y_Gauss}FEM, Gaussian correlation: Influence
of top/bottom Neumann no-flow boundary conditions on velocity
variances.}
\end{figure}

%\newpage
\begin{figure}[p]
\centering
\includegraphics[width=\linewidth]{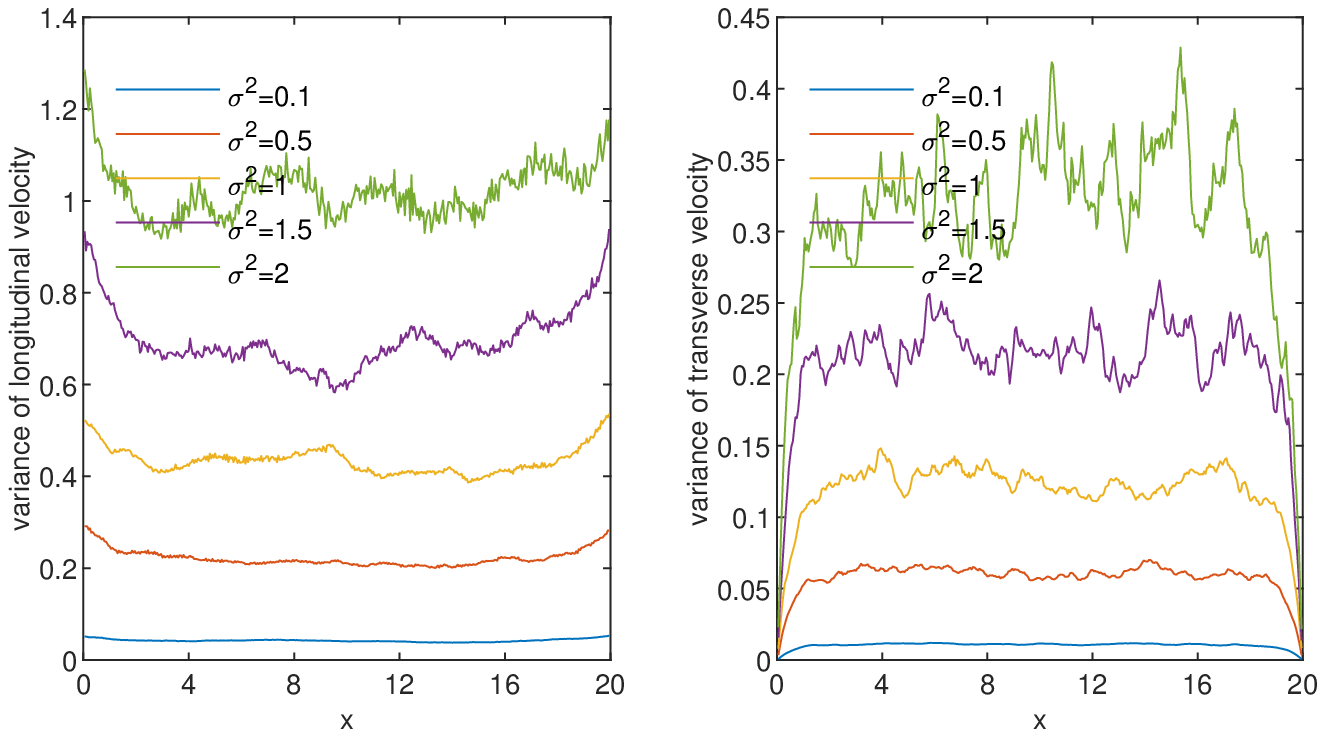}
\caption{\label{fig:FEM_statistics_X_Exp}FEM, exponential correlation: Influence
of inflow/outflow Dirichlet boundary conditions on velocity
variances.}
\end{figure}

\begin{figure}[p]
\centering
\includegraphics[width=\linewidth]{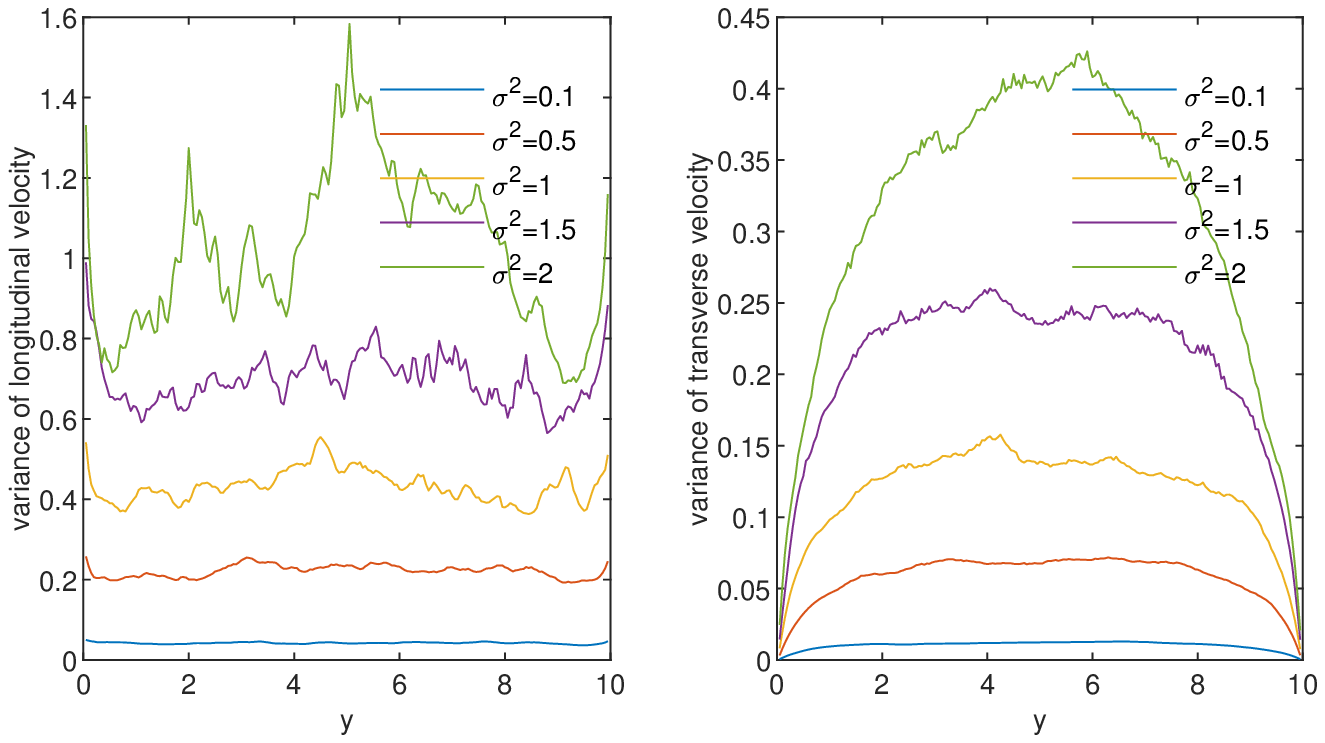}
\caption{\label{fig:FEM_statistics_Y_Exp}FEM, exponential correlation: Influence
of top/bottom Neumann no-flow boundary conditions on velocity
variances.}
\end{figure}

%\newpage
%\subsection{Discontinuous Galerkin method}

\begin{figure}[p]
\centering
\includegraphics[width=\linewidth]{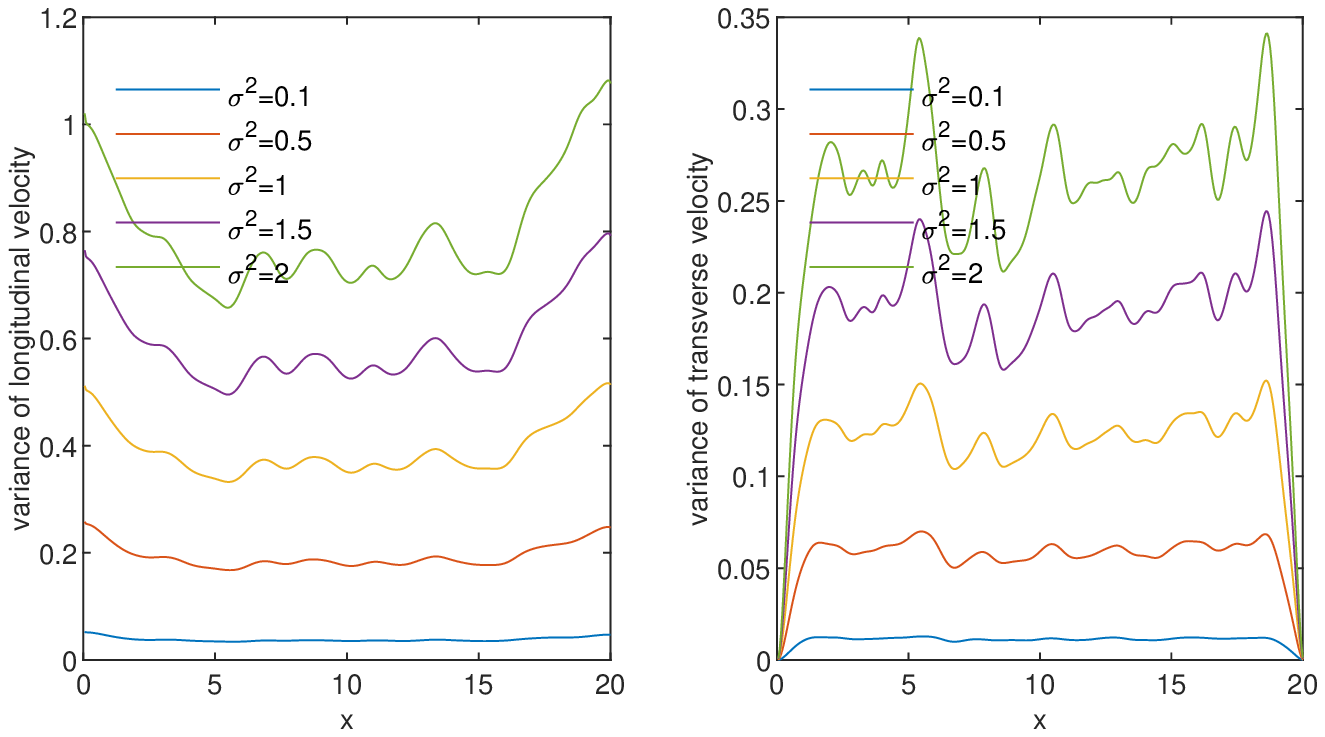}
\caption{\label{fig:DGM_statistics_X_Gauss}DGM, Gaussian correlation: Influence of inflow/outflow Dirichlet boundary conditions on velocity
variances.}
\end{figure}

\begin{figure}[p]
\centering
\includegraphics[width=\linewidth]{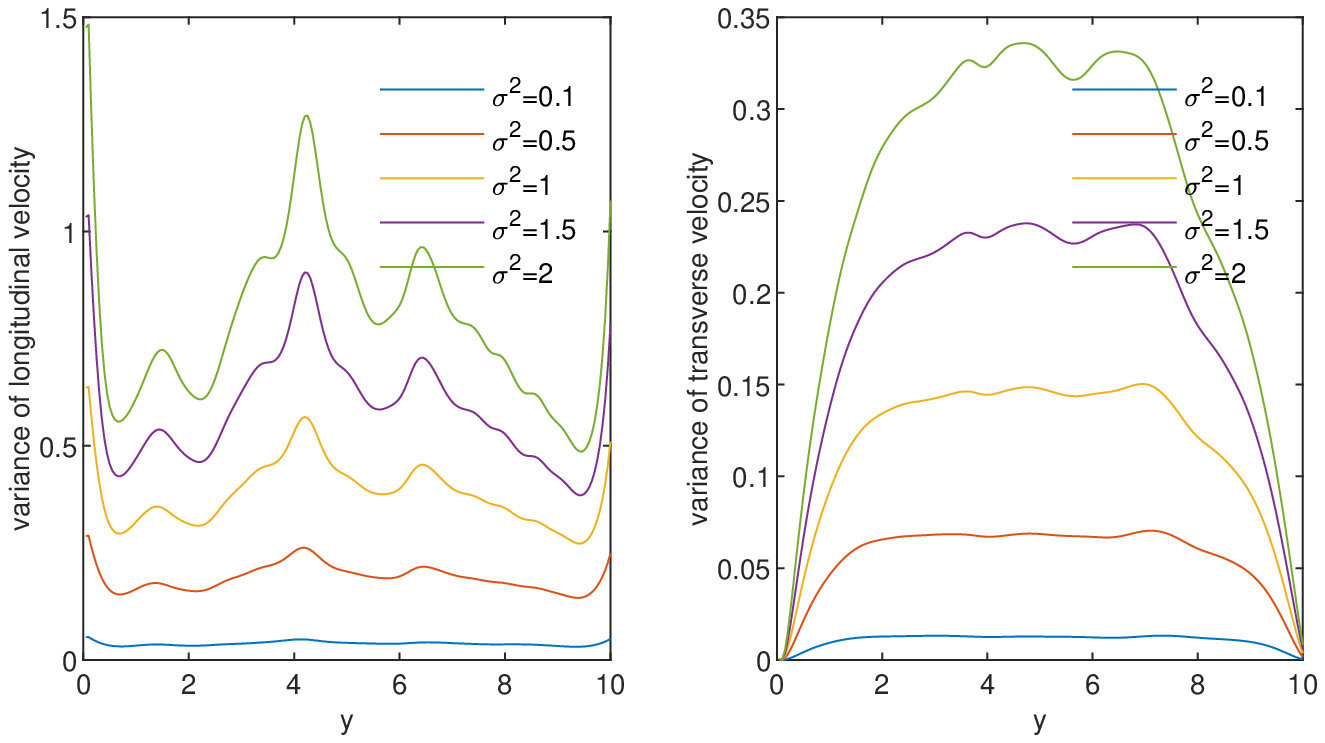}
\caption{\label{fig:DGM_statistics_Y_Gauss}DGM, Gaussian correlation: Influence of top/bottom Neumann no-flow boundary conditions on velocity
variances.}
\end{figure}

%\newpage
\begin{figure}[p]
\centering
\includegraphics[width=\linewidth]{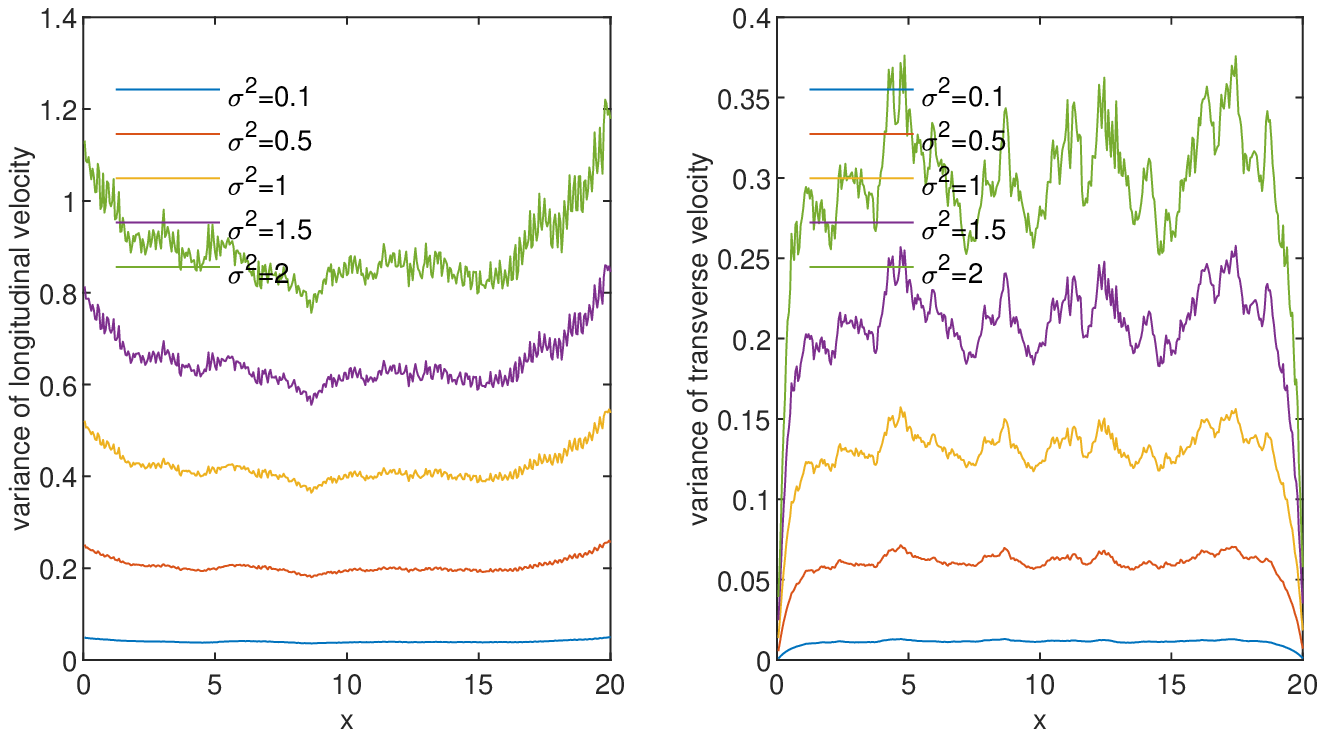}
\caption{\label{fig:DGM_statistics_X_Exp}DGM, exponential correlation: Influence of inflow/outflow Dirichlet boundary conditions on velocity
variances.}
\end{figure}

\begin{figure}[p]
\centering
\includegraphics[width=\linewidth]{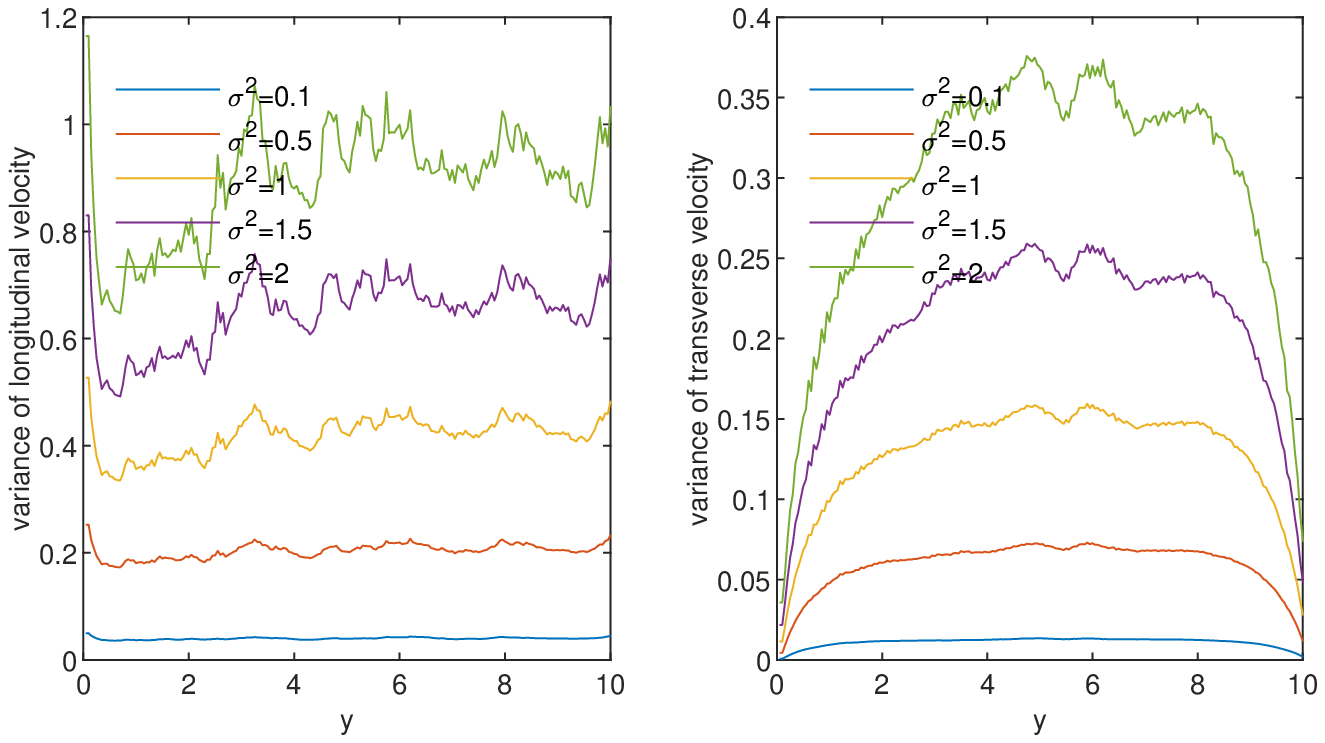}
\caption{\label{fig:DGM_statistics_Y_Exp}DGM, exponential correlation: Influence of top/bottom Neumann no-flow boundary conditions on velocity
variances.}
\end{figure}

%\subsection{Global random walk method}

\begin{figure}[p]
\centering
\includegraphics[width=\linewidth]{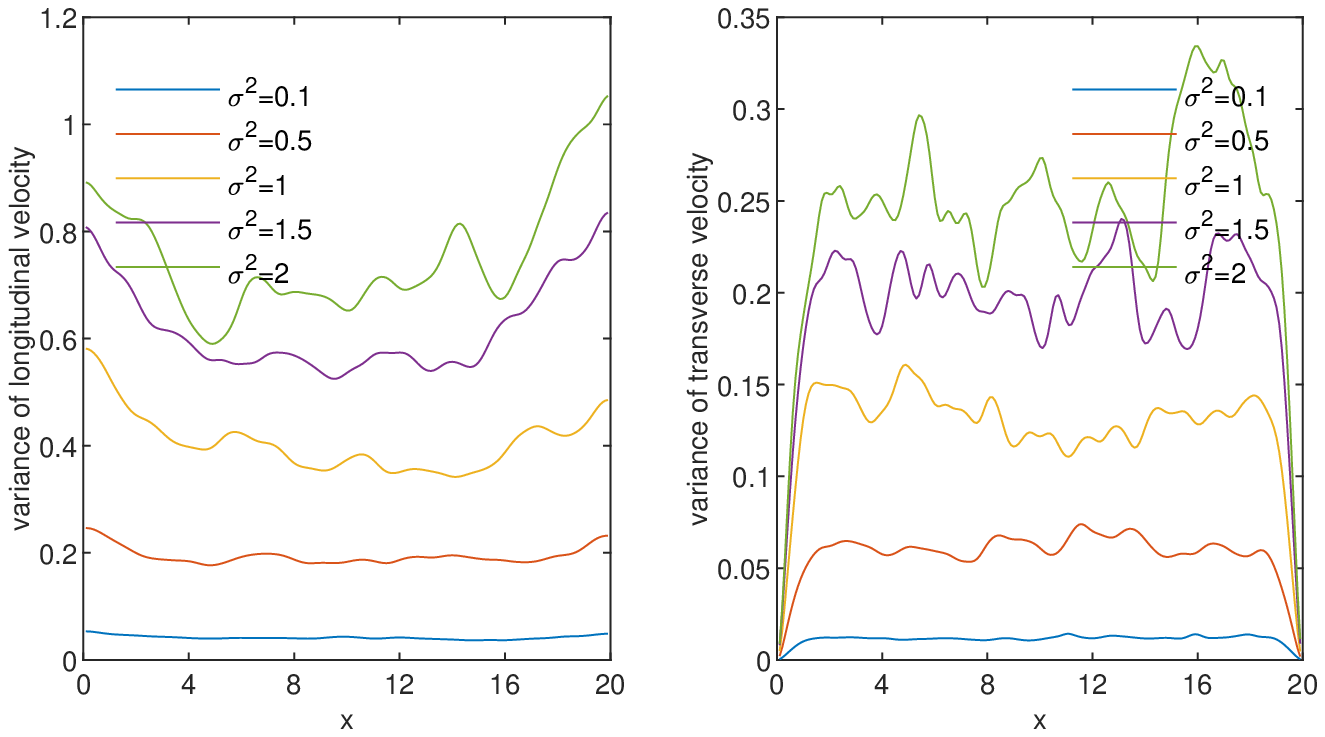}
\caption{\label{fig:GRW_statistics_X_Gauss}GRW, Gaussian correlation: Influence
of inflow/outflow Dirichlet boundary conditions on velocity
variances.}
\end{figure}

\begin{figure}[p]
\centering
\includegraphics[width=\linewidth]{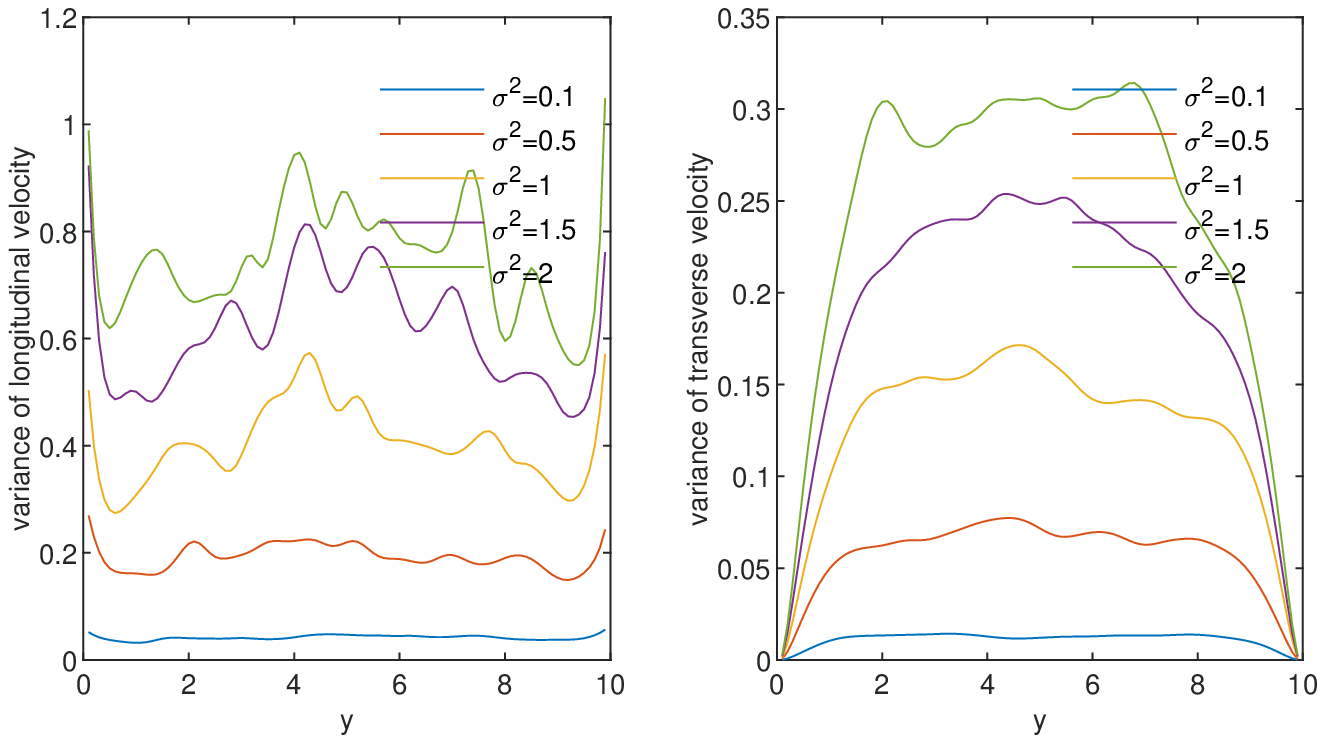}
\caption{\label{fig:GRW_statistics_Y_Gauss}GRW, Gaussian correlation: Influence
of top/bottom Neumann no-flow boundary conditions on velocity
variances.}
\end{figure}

\begin{figure}[p]
\centering
\includegraphics[width=\linewidth]{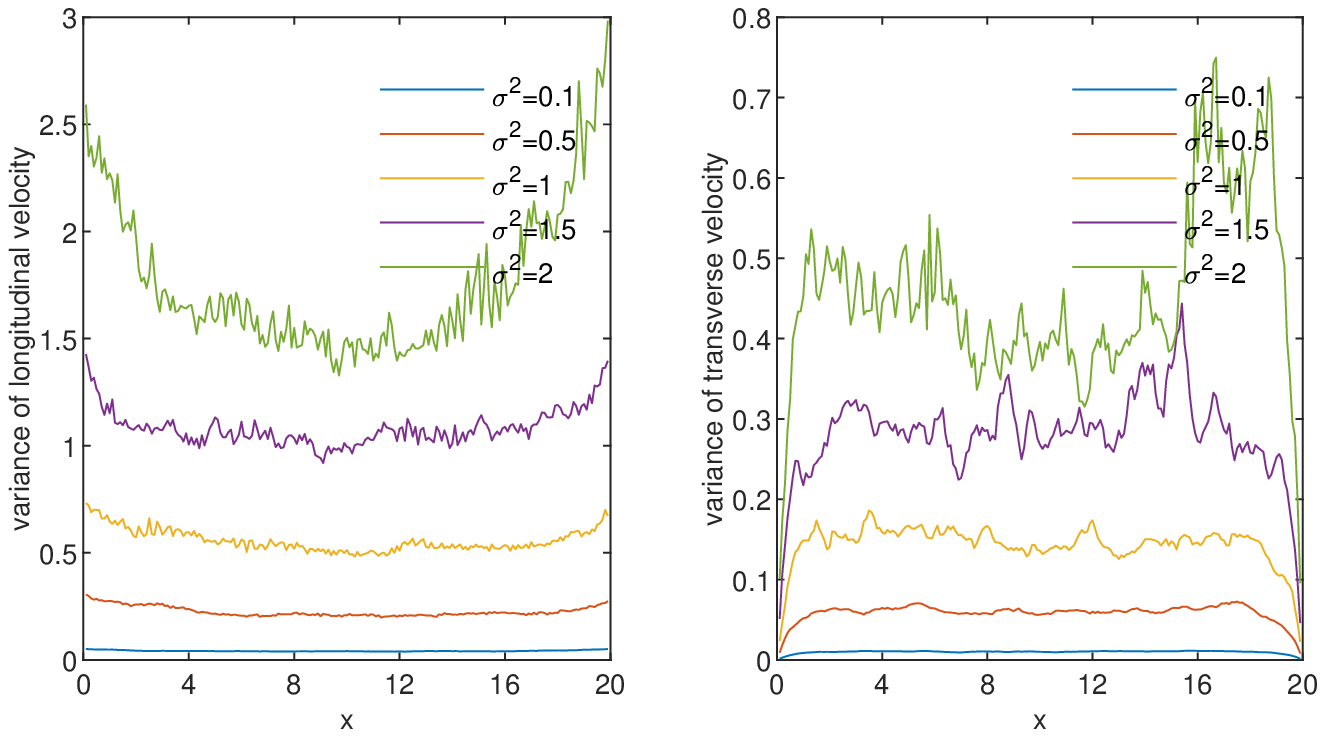}
\caption{\label{fig:GRW_statistics_X_Gauss}GRW, exponential correlation: Influence
of inflow/outflow Dirichlet boundary conditions on velocity
variances.}
\end{figure}

\begin{figure}[p]
\centering
\includegraphics[width=\linewidth]{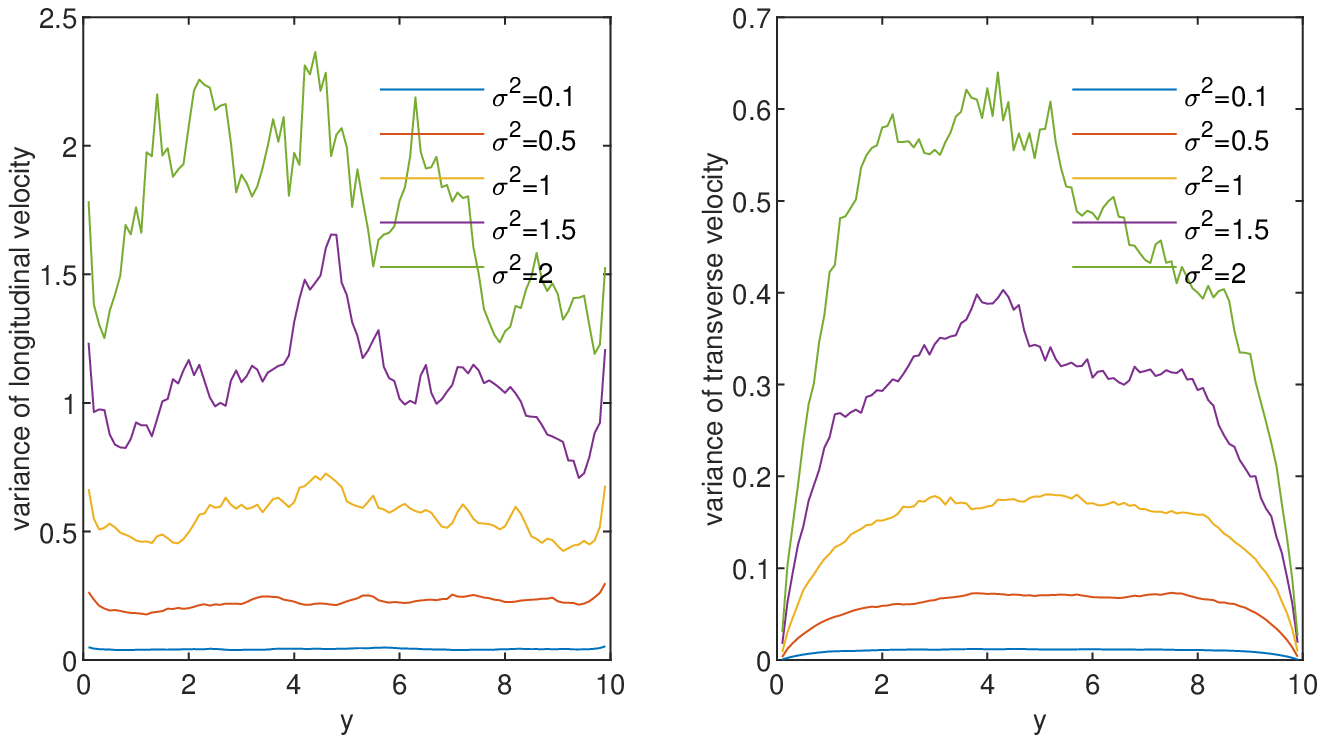}
\caption{\label{fig:GRW_statistics_Y_Gauss}GRW, exponential correlation: Influence
of top/bottom Neumann no-flow boundary conditions on velocity
variances.}
\end{figure}

%\section{References}
%\label{ref}

\newpage

\end{document}